\documentclass[usenatbib,babel]{aa}
\usepackage[varg]{txfonts}

\usepackage[english,english]{babel}
\usepackage{amsmath}
\usepackage{amssymb,amsfonts,textcomp}
\usepackage{array}
\usepackage{supertabular}
\usepackage{hhline}
\usepackage{hyperref}
\usepackage{balance}
\usepackage[usenames]{color}
\hypersetup{dvips, colorlinks=true, linkcolor=black, citecolor=black, filecolor=black, urlcolor=black}

\def\gtrsim{\lower.5ex\hbox{$\; \buildrel > \over \sim \;$}}
\usepackage{graphicx}

\newcommand{\hagn}{\mbox{{Horizon-AGN\,\,}}}

\def \apjl {{\it ApJ Let.}}
\def \apjs {{\it ApJ Sup.}}

\def \aap {{\it A\&A}}

\definecolor{grey}{rgb}{0.75,0.75,0.75}
\definecolor{Orange}{rgb}{1.0,0.5,0.15}
\definecolor{brown}{rgb}{0.7,0.25,0.0}
\definecolor{pink}{rgb}{1.0,0.5,0.5}
\definecolor{darkerred}{rgb}{0.8,0,0}
\definecolor{darkerblue}{rgb}{0,0,0.8}
\definecolor{Blue}{rgb}{0,0.08,0.65}
\definecolor{Red}{rgb}{0.65,0.08,0.05}
\definecolor{Green}{rgb}{0.15,0.45,0.25}

\begin{document}

\title{Caught in the rhythm II: }
\subtitle{Competitive alignments of satellites with their inner halo and  central galaxy.}

\author{C. Welker  
	\inst{1,2,3}
	\and
	C. Power \inst{1,2}
	\and 
	C. Pichon \inst{3,4}	
	\and
	Y. Dubois \inst{3}
	\and 
	J. Devriendt \inst{5,6}
	\and 
	S. Codis \inst{3}
	}
	
\institute{  ICRAR, The University of Western Australia, Crawley, Perth, 6009, WA, Australia
		  \and
		  ARC Centre of Excellence for All-sky Astrophysics (CAASTRO)
		\and
		CNRS and UPMC Univ. Paris 06, UMR 7095, Institut d'Astrophysique de Paris,
                98 bis Boulevard Arago, F-75014 Paris, France
                \and
                Korea Institute of Advanced Studies (KIAS) 85 Hoegiro, Dongdaemun-gu, Seoul, 02455, Republic of Korea
                \and
                Sub-department of Astrophysics, University of Oxford, Keble Road, Oxford, OX1 3RH, United Kingdom
                \and
                Observatoire de Lyon, UMR 5574, 9 avenue Charles Andr\'e, Saint Genis Laval 69561, France
                }

\date{Accepted . Received ; in original form }

\abstract{
    The anisotropic distribution of satellites around the central galaxy of their host halo is well-documented, both observationally and in simulations of the $\Lambda$CDM model. However the relative impact of baryons and dark matter in shaping the gravitational potential that drives this distribution is still  debated.
	 Using the cosmological hydrodynamics simulation Horizon-AGN,  the angular distribution of satellite galaxies  with respect to their central counterpart and  their halo is quantified, with a special focus on the redshift range  0.3 to 0.8.
    Haloes and their galaxies are identified and their kinematics computed using dark matter and stellar particles respectively. The cosmic web is extracted as a network of contiguous segments tracing ridges of the density field. 
   The relative tendency of the central-to-satellite separation vector to align with the spin and minor-axis of both its central counterpart and its host halo is investigated, treating separately its core and outskirts.
    On scales smaller than one virial radius, satellites with masses greater than $5\%$ of their central's  tend to cluster more strongly in the plane of the central, rather than merely tracing the shape of their host halo. This is explained i) by the increased isotropy of inner haloes and ii) by the radial decrease of their triaxiality acquired through their inside-out assembly in vorticity-rich flows along the cosmic web. 
    However, while the effect of  torques from the central galaxy decreases with distance, halo triaxiality increases, impacting more and more  the satellite's distribution until it becomes comparable to that of the central, just outside one virial radius. Above this scale, the filamentary infall from the cosmic web also impacts the satellites distribution, dominating above two virial radii in the unrelaxed outskirts of the halo. 
     The central's morphology also plays a governing role: the  alignment w.r.t. the central plane is four times stronger in haloes hosting stellar discs than in spheroids. While the average inner ellipticity within one virial radius is comparable for both  types, it indirectly impacts their  tendency to trace their halo's shape in the mass range where centrals and haloes' minor axis are significantly aligned. For $10^{12.5}\, M_{\odot}<M_{\rm 0}\!< 10^{13.5}\,M_{\odot}$, the satellite's alignment around haloes' minor axis is $40\%$ stronger for disc hosts than it is for spheroid hosts. Nonetheless, the impact of the galactic plane on satellites decreases for lower satellite-to-central mass ratios and higher halo ellipticities, suggesting that these results might not hold for dwarf satellites of the  Local group. The  orientation of the Milky-Way's satellites traces their nearby cosmic filament, and their level of coplanarity is consistent with systems of    similar mass and cosmic locations found in Horizon-AGN.
 However, the strong impact of  galactic planes in massive groups and clusters bounds the likelihood of finding a relaxed region where satellites can be  used to infer halo shape.  The minor-to-major axis ratios for stacked haloes with mass $M_{\rm 0}> 10^{13.5}\, M_{\odot}$ is  underestimated by $10\%$. This error can  soar quickly to $30-40\%$ for individual halo measurements.
}


 \titlerunning{Competitive alignments of satellites with their inner halo and  central galaxy.}
 
\maketitle



\section{Introduction}

The complex intertwined interactions between satellite galaxies, their centrals, and the dark matter
haloes they orbit within have been studied extensively in both observations and numerical simulations in
recent years \citep[e.g.][]{Brainerd05,Aubert04,Sales09,Ibata13}. Both the angular distribution and orientation
of satellites with respect to their central galaxies and host haloes should encode information about
the structure of the gravitational potential in which they orbit and the larger scale cosmic web from
which they were accreted. This has important implications for how one interprets a range of observational
measurements, for example, estimates of halo shapes, and the clustering of satellites in
planes around galaxies. Cosmological $N$-body simulations, tracking the self-consistent growth of structure
from early times to the present day, predict that anisotropic satellite distributions should be a natural
byproduct of hierarchical assembly, but how these predictions will be modified by the presence of baryons
remains a topic of active debate. For example, dissipation by baryons in the centres of haloes will lead
to both a contraction and a flattening of the gravitational potential in which satellites orbit, but one does
not expect any impact on how satellites are accreted from the cosmic web.This paper  quantifies in detail
how the presence of baryons influence the spatial distribution and orientation of satellites around the
centrals and in their host haloes.

One of the first observational studies to consider the angular distributions of satellite galaxies around
their centrals dates back to \citet{Holmberg69}, who reported that satellites followed a prefentially polar
alignment (the so-called ``Holmberg effect''; see also \citealt{Zaritsky97}). The reality  of
such an effect has long been debated \citep[e.g.][]{Hawley75,Phillips15}, and more recent
studies, based on large galaxy surveys such as Sloan Digital Sky Survey, report that satellite galaxies
lie within the plane of their central galaxy~\citep{Brainerd05,Yang06,Abazajian09,Sales09,Wang10,Nierenberg12, smithetal15,Huang17}. These trends are in broad agreement with the results of $N$-body
\citep{Aubert04, Zentner05} and hydrodynamical cosmological simulations \citep{Dong14}. Interestingly, a
number of studies report spatially thin and kinematically cold planes of satellites around the Milky Way
and Andromeda ~\citep{Bahl14, Pawlowskietal14, Gilletetal15} whose physical origin is more challenging to
understand.

These broadbrush observational results can be understood if satellites follow dynamically relaxed orbital
distributions within their haloes. However, \citet{Wang05} and \citet{Agustsson10} highlighted that the
observed
concentration of satellites in the rotation plane of their host halo implies that it is triaxial
\citep{BarnesE87,Warren92,Yoshida2000,Meneghetti01,Jing02}, which is a consequence of the anisotropic
hierarchical assembly of structure via the cosmic web. Massive haloes are built via multiple mergers between
objects drifting along the cosmic web whose orbital angular momentum -- preferentially perpendicular to
filaments along which they flow -- is partially transferred to the intrinsic angular momentum of the
remnant, resulting in massive haloes having a spin preferentially perpendicular to their nearby filament
\citep{vanhaarlem93,Tormen97,Aubert04,Sousbie08,Bailin08,Paz11,Codis12,Zhang13,Dubois14, Welker14}. This
means that the elongation of both halo and filamentary infall should align, and the tendency of satellites
to orbit in a specific plane is a combination of not only the host's triaxiality but also of the continued
infall of satellites along the cosmic web \citep{Aubert04,Knebe04,Wang05,Zentner05}. Observations of the
planes of satellites of M31 or the Milky Way~\citep{Ibata13, Libeskind15} as well as the detection of
alignments in the SDSS by \cite{Paz08} \citep[see also][]{Tempel15} strongly support this claim.  	
	
 \citet[][hereafter Paper I]{Welker17} found strong evidence for another distinct alignment trend
that becomes evident at small galaxy-satellite separations, and is a likely signature of the dissipative
nature of baryons. As a
satellite's orbit takes it deeper into its host halo's potential well, the orbit bends into the plane of their
central galaxy; this occurs even when the plane is
strongly misaligned with the nearby filament and the host halo shape on larger scales. This is an important result as this location of satellites with respect to the orientation of the central galaxy contributes to the "one-halo" term in intrinsic alignment models, and must therefore be accounted for to properly model the impact of intrinsic alignments on the extraction of small scale information, as was already pointed out in previous work \citep{Chisari15, Chisari17}.

 This suggests
that gravitational torques from the central galaxy, especially when it is disc-like, influences the
dynamical evolution of satellites as they orbit at small radii, and therefore impact the reliability of
halo shapes derived from distributions of satellites. It is also possible that central galaxies remain
statistically well aligned with the inner core of their host dark matter halo, in which case satellites
simply follow the geometry determined by the inner parts of the dark matter halo, which matches that of
the central galaxy. This paper uses the cosmological hydrodynamics simulation
Horizon-AGN~\citep{Dubois14} to explore these scenarios and to quantitatively establish the relative
importance of baryons - principally stars - and dark matter in determining the alignments, especially within
the Virial radius. It also aims to assess the reliability of halo axis ratios derived from their distribution of
satellites. 
	
This paper is structured as follows: after a short review of the numerical setup and methods used in Section~\ref{section:virtual}, the radial evolution of DM haloes shape and inertial twist is explored, and related  to the inside-out build up of haloes within the cosmic web in Section \ref{section:shapes}.   Section \ref{section:DMvsbaryons}  tests for alignments of satellites with the  inner and outer parts of the halo and explicitly shows that the tidal influence of the central is dominant in the inner parts of halo, which arises because of the stronger anisotropy of the stellar material compared to the DM core.  Section \ref{section:halogal}  quantifies the intrinsic misalignments between centrals and their DM halo  on increasing scales, in terms of both minor axis and spin, to estimate possible couplings between both signals. Section~\ref{section:virialshell}  quantifies such coupling within the Virial radius in terms of central morphology and cosmic web orientation. Implications for  local planes of satellites can be found in Section~\ref{section:MW}. Finally Section~\ref{section:conclusion} summarises the main results. Complementary analysis of the effects of various parameters -- such as the central mass, the overall halo shape and the satellite-to-central mass ratio -- on satellite alignments can be found in Appendices. Most importantly, Appendix~\ref{section:ellipticity} makes predictions for axis ratios derived from populations of satellites tracing the local shape of their halo, and consequently estimates the error on such measurements induced by central alignments through comparison in Horizon-AGN.
\section{Numerical methods \& definitions}
\label{section:virtual}
\subsection{The Horizon-AGN simulation}
\label{section:HAGN}

The details of the Horizon-AGN\footnote{\url{www.horizon-simulation.org}} simulation can be found in~\cite{Dubois14}. This simulation is run in a 
$L_{\rm box} = 100\,  h^{-1} \rm Mpc $ cube with a $\Lambda$CDM cosmology compatible with the WMAP-7 data~\citep{komatsuetal11}. 
The mass resolution is $M_{\rm  DM, res}=8\times10^7 \, \rm M_\odot$ for DM, and $M_{\rm gas,res}=1\times 10^7 \, \rm M_\odot$ for the initial gas resolution. It is run with the {\sc ramses} code~\citep{teyssier02} on a grid  that adaptively refines down to $\Delta x=1$ proper kpc, with refinement triggered if the number of DM particles in a cell becomes greater than 8, or if the total baryonic mass reaches 8 times the initial baryonic mass resolution in a cell.

It includes elaborate sub-grid physics including: uniform UV-heating of the gas after redshift $z_{\rm  reion} = 10$ following~\cite{haardt96}, cooling down to $10^4\, \rm K$ through H and He collisions, star formation (in regions of gas number density above $n_0=0.1\, \rm H\, cm^{-3}$) following a Schmidt law, feedback from stellar winds, supernovae type Ia and type II  with mass, energy and metal release and AGN feedback with heating or jet mode when the accretion rate is respectively above and below one per cent of Eddington \citep[see][for details]{duboisetal12}.

\subsection{Galaxies and haloes: matching and kinematics}
\label{section:structures}

 More details and discussion on the selection procedure can be found in Paper I.  Only  a brief summary is provided here. Galaxies and haloes are identified from star/DM particles with the AdaptaHOP halo finder~\citep{Aubert04,tweedetal09} with the same parameters than in~\cite{Dubois14}. It typically selects objects with masses larger than  $1.7 \times 10^8 \, \rm M_\odot$. Catalogues with up to $\sim 180 \, 000$ galaxies are produced for each redshift output analysed in this paper ($0.3 < z < 0.8$). The cut $M_{\rm g}>10^9 \, \rm M_\odot$ is imposed on central galaxies. Pair results are stacked over the whole range of redshifts.

Each galaxy is matched with its closest main halo (not its sub-halo).   The central galaxy is identified as the most massive galaxy contained within a sphere of radius $R=0.25 \, R_{\rm vir}$ with $R_{\rm vir}$ the Virial radius of the halo. This cut in both mass and distance is chosen so as to mimick identification of brightest cluster galaxies (BCGs) in observations, around which most of the alignment trends are evaluated. Unless otherwise specified, sub-haloes are not considered as able to host central galaxies. The focus here is on satellite galaxies situated within a sphere of radius  $R=3 \, R_{\rm vir}$ around the centre of the halo.

Further cuts in distance to the centre of the halo can be performed afterwards and will be specified in each case. Let us stress the fact that above 2 $R_{\rm vir}$,  galaxies neighbouring a halo are not necessarily bound to this halo -which is itself mostly unrelaxed- and can therefore also be centrals of another halo. Since our aim is  precisely to show the continuity between unrelaxed and relaxed distributions of satellites in haloes, i.e. between the motion of galaxies in the cosmic web and the orientation of satellites in the outer and inner parts of the haloes they enter, this analysis considers a $2-3R_{\rm vir}$ bin to investigate the continuity of some alignment trends from extra- to intra-halo scales.
The  kinematics and inertia tensors for galaxies and haloes from stellar and DM particles respectively are computed. As for galaxies, they are computed for stellar material within the half mass radius of the galaxy - defined as the radius that contains half the mass of the galaxy, noted $M_{\rm g}$. Note that alignments within galactic planes computed from all the stars are then expected to be even stronger \citep{Chisari15}, but using the half mass radius is more directly comparable to observations. For haloes they are computed on various scales, namely:
\begin{itemize}
\item on all DM bound particles identified by HaloMaker, excising satellites (sub-haloes).
\item for DM material within 5 spheres of increasing radius centred around the halo's centre of mass. The 5 radii cuts correspond to: $R=0.25 \, R_{\rm vir}, 0.5 \, R_{\rm vir}, \, R_{\rm vir}, 2 \, R_{\rm vir}, 3 \, R_{\rm vir}$.
\end{itemize}

In this latter case, extra care is needed to prevent particles in sub-haloes from seriously impacting the overall calculation of the diffuse halo shape. Numerous technics have been used in the past to correctly estimate the overall shape of the halo \citep{Power03,Hayashi07,Despali16,VegaFerrero17}, from strict exclusion of satellite material to fitting ellipsoids to iso-potential contours in place of density contours \citep{Hayashi07}. Since the gravitational potential is less sensitive to local over-densities, this latter method is naturally less sensitive to sub-haloes and does not require somewhat arbitrary exclusion of sub-haloes. However, using it makes satellite alignments around haloes less comparable with alignments around central discs and along cosmic filaments, which are identified as purely density features.  

This paper therefore  focusses  on the inertia tensor computed directly on particles and    sub-haloes are dealt with by excising density spikes within $R=1 \, R_{\rm vir}^{\rm sub}$ (with $R_{\rm vir}^{\rm sub}$ the Virial radius of the sub-halo) and smoothing out the density in the excised regions.   An  NFW profile is first fit to each host halo to derive the expected density in each artificially emptied sub-volume, hence the expected number of missing particles $N_{\rm miss}$.  The volume is the populated by generating  a random set a $N_{\rm miss}$ positions within the empty region. Note that no  artificial velocities is assigned to such particles, hence  the angular momentum is not computed with this method of satellite extraction.

 A distinct  method is adopted to compute the angular momentum in all shells:  density spikes produced by sub-haloes from the calculation are excised, but without trying to exclude the DM fluff from the outskirts of the sub-halo that are in the process of diluting into the host halo. More precisely,   DM particles are excluded within $R=0.5 \, R_{\rm vir}^{\rm sub}$ around each sub-halo, $R_{\rm vir}^{\rm sub}$ being the Virial radius of a given sub-halo. Note that inertia tensors computed with this latter method were found to be completed consistent with the ones computed by the first method. Using one method or the other does not impact qualitatively the results presented in this study, and the quantitative impact is limited to a few $\%$ error, within poissonian error bars.

For all quantities, the subscript {\it g} is used for central galaxies, {\it 0} for their host halo, {\it s} for satellites.
 The angular momentum of a galaxy is measured with respect to the centre of mass of the galaxy:
 	\begin{eqnarray}
	\mathbf{L_{\rm g}}=\sum_{i} m_{i}(\mathbf{r_{i}} -\mathbf{r_{\rm g}} ) \times (\mathbf{v_{i}} -\mathbf{v_{\rm g}} )\, ,
	\end{eqnarray}
	with $\mathbf{r_{i}}$, $m_{i}$ and $\mathbf{v_{i}} $ the position, mass and velocity of particle {\it i}, $\mathbf{r_{\rm g}}$ the position of the centre of mass of the galaxy
	and $\mathbf{v_{\rm g}}$ its centre of mass velocity.  The angular momentum of haloes is computed with the same definition.  All measurements are carried out for satellites located within a sphere of radius $3\, R_{\rm vir}$ around the host halo of the central, further cuts being applied throughout the paper to focus on inner parts of the halo.
	 The inertia tensor of each object is computed from its particles masses ($m_{l}$) and positions ($\mathbf{x^{l}}=\mathbf{r_{l}} -\mathbf{r_{\rm g}}$) (in its barycentric coordinate system):
	  	\begin{eqnarray}
 	I_{ij} = \sum_{l} m_{l}\Big(\delta_{ij} (\sum_{k}x^{l}_{k} x^{l}_{k})-x^{l}_{i} x^{l}_{j}\Big)\, ,
 	\end{eqnarray}	
where $\delta_{ij}$ is the Kronecker symbol. The lengths of the semi-principal axes (with $c<b<a$) are derived from the moments of inertia tensor. For haloes, it was checked that computing the inertia tensor within a given radius iteratively, by first selecting all DM material within the given radius, computing the bounding ellipsoid of similar circularised radius then repeating the operation on material within the bounding ellispoid did not significantly impact the values obtained by omitting the second step.

An alternative proxy to determine the shape of a structure is the
reduced inertia tensor, which weighs each particle by the inverse square distance
to the centre. This method typically weighs up the innermost parts of the structure probed. However as  a radial decomposition of the halo was adopted and as the focus is on the ability of satellites to trace the overall shape of their halo,  the simple inertia tensor was retained. Note that although radial misalignments of the halo dark matter component shape will induce variations between the two proxys, this should not qualitatively affect the results. More details on the variations between the simple and reduced inertia tensors can be found in \cite{Chisari17}.

The shapes of structures with small number of particles
can be biased due to insufficient resolution. Following
the criteria set in \cite{Chisari15} our analysis is restricted to galaxies and halo parts (radial bins) 
resolved with more than 300 particles (including in their innermost core $R<0.25\, R_{\rm vir}$). Additionally,   the slope of halo density profiles should be resolved enough across our radial bins to ensure the gravitational potential exerted on satellites is realistic enough. haloes with $M_{\rm 0}<10^{11.5}\, M_{\odot}$ (i.e. defined by less than 4000 particles as a whole by the halo finder) are therefore excluded from our sample, as well as central galaxies with $M_{\rm 0}<10^{9}\, M_{\odot}$.
Moreover, innermost radial bins where the resolution might drop below 1000 particles for less massive haloes despite these cuts  are systematically indicated and usually plotted as dashed lines in our analysis.

\begin{figure}
\center \includegraphics[width=0.95\columnwidth]{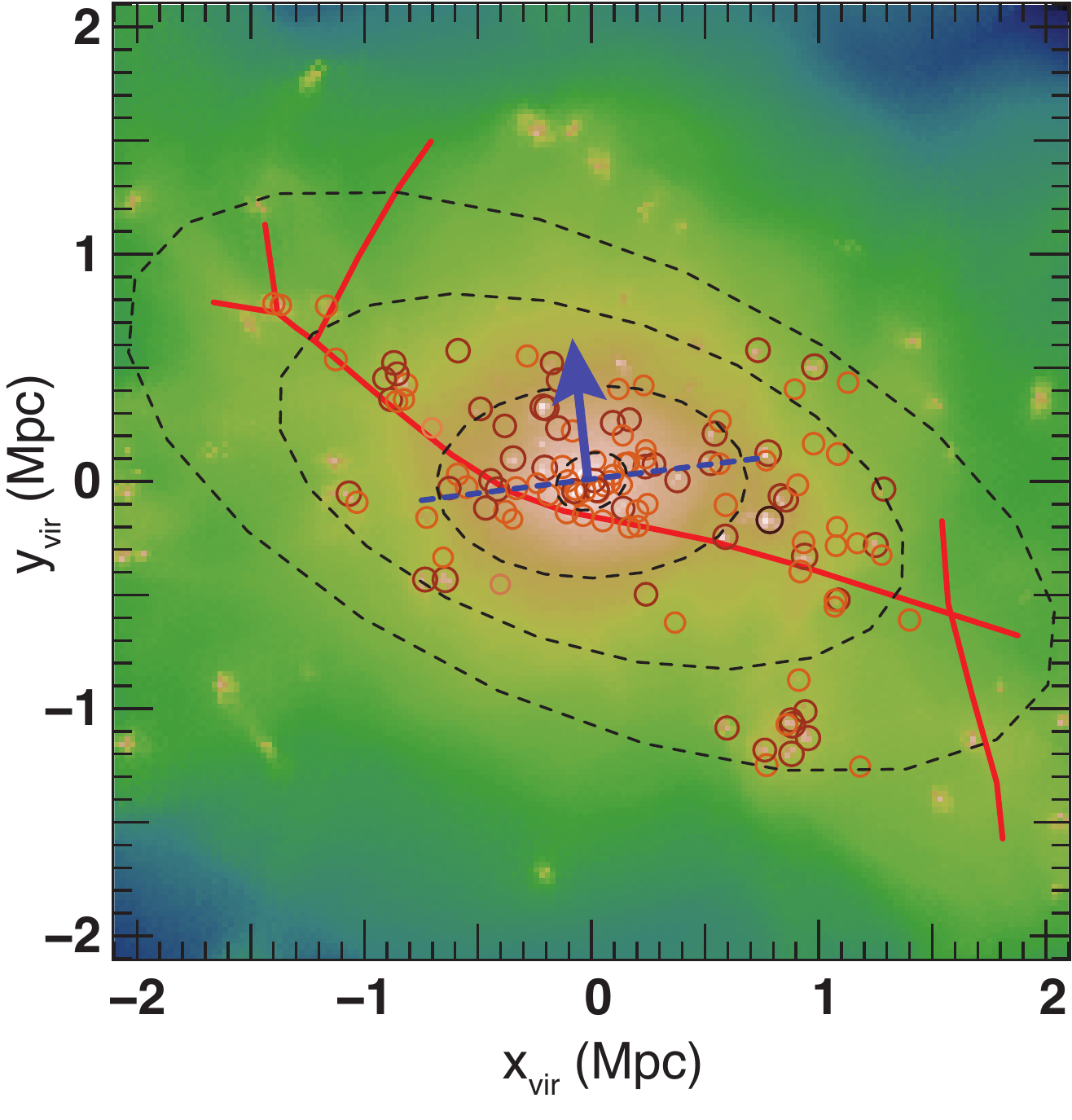}
 \caption{Gas density map of a massive halo ($10^{13.9} \, M_{\odot}$) at $z=0.3$ in Horizon-AGN with 121 satellites within $3\, R_{\rm vir}$. All quantities are projected along the minor axis of the DM halo, computed from DM particles within $R_{\rm vir}$. Dashed black lines indicate the projection of the inertial ellipsoids computed from DM particles within 0.25, 1, 2 and 3 $\, R_{\rm vir}$ respectively. Orange circles indicate the location of satellites (with darker shades for more massive galaxies, from $10^{9}$ to $10^{12}\,M_{\odot}$). The dashed blue line shows the orientation of the galactic plane, with the central's spin indicated as a blue arrow. Solid red lines indicate the spine of the skeleton intersecting the halo within $3.5\, R_{\rm vir}$. The angular distribution of satellites displays a complex set of correlations with  the cosmic web,
the inner structure of the halo  and the central galaxy.}
\label{fig:halo}
\end{figure}

	The {\it central separation vector}, or position vector of each satellite in the rest frame of its central galaxy is defined as $\mathbf{r_{\rm gs}}= \mathbf{r}_{s}-\mathbf{r_{\rm g}}$ with $\mathbf{r}_{s}$ the position of the satellite. Its norm is the separation $R_{\rm gs} = || \mathbf{r}_{\rm gs} || $. Similarly,   the {\it halo separation vector} and its norm is defined  for each satellite $\mathbf{r}_{\rm 0s}$ and    $R_{\rm 0s}$, i.e its position computed in the rest frame of its host halo. The distance from the halo  to make radial cuts is always $R_{\rm 0s}$, and each separation vector is used to quantify alignments around the specific structure it is centred on. In practice, since the centre of mass of haloes is identified iteratively through embedded density grids of increasing resolution, there is little deviation from one vector to the other: the median relative angle between the two is $\approx 0.4^{\rm o}$ and the median relative length difference is $1\%$. Choosing only one of the two to quantify all alignments has little impact on the results when excluding extreme outliers which correspond to major galaxy mergers.

\begin{table}
\renewcommand{\arraystretch}{1.2}
{ \center
\begin{tabular}{|l|c|r|}
\hline
{\bf Angle} & {\bf definition} \\
 \hline
 halo minor axis - filament & $\nu_{\rm 0}=\cos\alpha_{\rm 0}$ \\
 \hline
halo major axis -filament & $\nu_{\rm 1}=\cos\alpha_{\rm 1}$\\
 \hline
halo spin -filament & $\nu_{\rm 0s}=\cos\alpha_{\rm 0s}$\\
 \hline
central spin -filament & $\nu_{\rm c}=\cos\alpha_{\rm c}$\\
 \hline
Satellite separation vector - central minor axis & $\mu_{\rm c}=\cos \theta_{\rm c}$ \\
\hline
Satellite separation vector -halo minor axis & $\mu_{\rm 0}=\cos\theta_{\rm 0}$\\
\hline
halo spin - central spin & $\gamma_{\rm 0c}=\cos \beta_{\rm 0c}$\\
\hline
halo minor axis - central minor axis & $\kappa=\cos \theta_{\rm 0c}$\\
\hline
\end{tabular}
\center\caption{Definitions of the different angles used in this work. All these angles are made scale dependent by selecting the maximal sphere (or shell) in which DM particles and satellites are to be found for the computation.}
\label{angles}
}
\end{table}

\begin{figure}
\center \includegraphics[width=0.75\columnwidth]{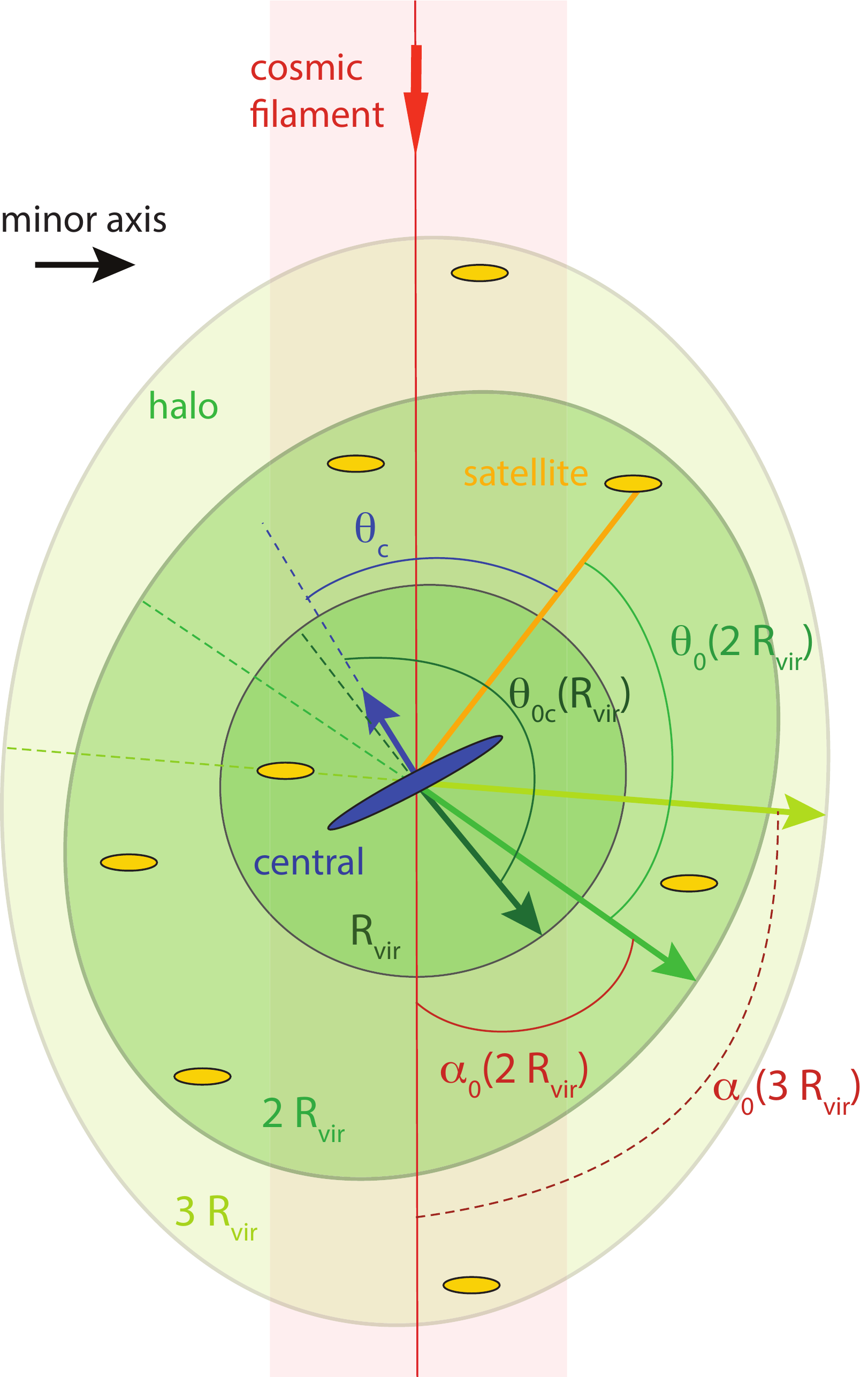}
 \caption{Sketch of a halo with radially varying shape -- as traced by the inertia tensor --  (green shades), hosting a central galaxy (in blue), a population of satellites (in orange) and connected to one cosmic filament (in red). The relative angles between the different components identified in this work are indicated with appropriate colours. Only the definition of angles calculated around the minor axis of the halo and the central galaxy are given here. Complementary definitions can also be made using the spin or the major axis instead of the minor axis (See Table.~\ref{angles} for details). The angles around the halo's minor axis depend on the scale on which the halo shape is computed.   Each of these angles are computed on five increasing scales: 0.25, 0.5, 1, 2 and $3 \, R_{\rm vir}$. Only one instance of each angle is represented on this sketch for readability.}
\label{fig:angle}
\end{figure}

Table~\ref{angles} summarises the definitions of all the angles used in this paper to follow alignment trends and an illustration of the most important ones can be found in Fig.~\ref{fig:angle}. All these angles depend in principle on the scale selected by the maximal sphere (or shell) in which DM particles and satellites are to be found for the computation. This study will be typically following the evolution of the PDF of $\mu_{\rm 0}=\cos\theta_{\rm 0}(r/r_{\rm vir})$, where the shape of the halo is computed using all DM particles within normalised distance $r/r_{\rm vir}$ from the halo centre of mass, and only satellites within  normalised distance $r/r_{\rm vir}$ are considered. Unless specified otherwise, indicated error bars are $1\sigma$ poissonian error bars.

\subsection{Characterization of the cosmic web}

In order to quantify the orientation of galaxies relative to the cosmic web,  a geometric three-dimensional ridge extractor called the {\sc skeleton}~\citep{sousbie09} is used. It is computed from the full volume DM density distribution 
sampled on a $512^3$ cartesian grid. This density distribution is smoothed with a gaussian kernel of length $3 \, h^{-1}$ comoving $\rm Mpc$. The orientation and distribution of galaxies are then measured relative to the direction of 
	the closest filament segment. Note that these filaments are defined as ridge lines of the density field and therefore have no thickness. The closest filament of a given galaxy is thus simply the segment whose euclidian distance to the galaxy 
	is the smallest. All central galaxies in the sample are separated from their nearest filament by less than 1 Mpc, and the vast majority of them by less than 0.5 Mpc (the peak of the galaxy distance-to-filament 
	distribution lies around 0.2 Mpc). In comparison, the virial radii of the dark matter haloes analysed in this work range from 50 kpc for lowest mass haloes to around 1.2 Mpc for most massive clusters ( $10^{14.5} \,M_{\odot}$) with a median around 300 kpc, a value typical of Milky-Way type haloes.
	
	As an example, the projected gas density map of a $10^{13.9}\, M_{\odot}$ halo in Horizon-AGN is shown on Fig.~\ref{fig:halo} with its local network of cosmic filaments overlaid as solid red lines, The projected inertial ellipsoids computed from its DM particles within 0.25, 1, 2 and 3 $R_{\rm vir}$ are overlaid as dashed black lines, the satellites are circled in various shades of orange with darker shades indicating higher masses (from $10^{9}$ to $10^{12}\,M_{\odot}$), the central galactic plane is shown in blue, with the spin of the central galaxy indicated as a blue arrow. The discrepancy between the outer and the inner angular distributions of satellites is already visible on this example.

\subsection{The effect of grid-locking}

A common caveat of cartesian based Poisson solvers is the numerical anisotropy that arises in
the force calculation. On smallest mass scales, this can lead to spurious alignments of spins with the cartesian grid. This effect was explicitly tested in Horizon-AGN in \cite{Dubois14}. The main results are:
 the spins of  less massive galaxies, $M_{\rm g}<5.10^9 \, M_\odot$, show some preferential spin alignments with the grid while no obvious
alignment is seen for the high-mass galaxies. While low-mass galaxies are preferentially locked with the grid because they are composed of very few grid elements, this effect disappears for more massive galaxies due to the
larger number of resolution elements to describe those objects. Cosmic filaments were found not to be subject to grid-locking, coherently with their large-scale nature. 

In the present study, the threshold chosen for central galaxies and the scales considered ensure that most of our results are not subject to grid-locking. A detailed analysis of those effects in Horizon-AGN can also be found in \cite{Chisari15}.


\section{Alignments of satellites with  halo shape \& CW}
\label{section:shapes}

This section quantifies the degree of alignment of satellite to  the inner and outer parts of their host halo and  to their embedding  cosmic filament, and relate these findings to the inside-out build-up of dark haloes across cosmic time.

\subsection{Radial evolution of satellite alignments with halo shape.}

\begin{figure}
\center \includegraphics[width=0.85\columnwidth]{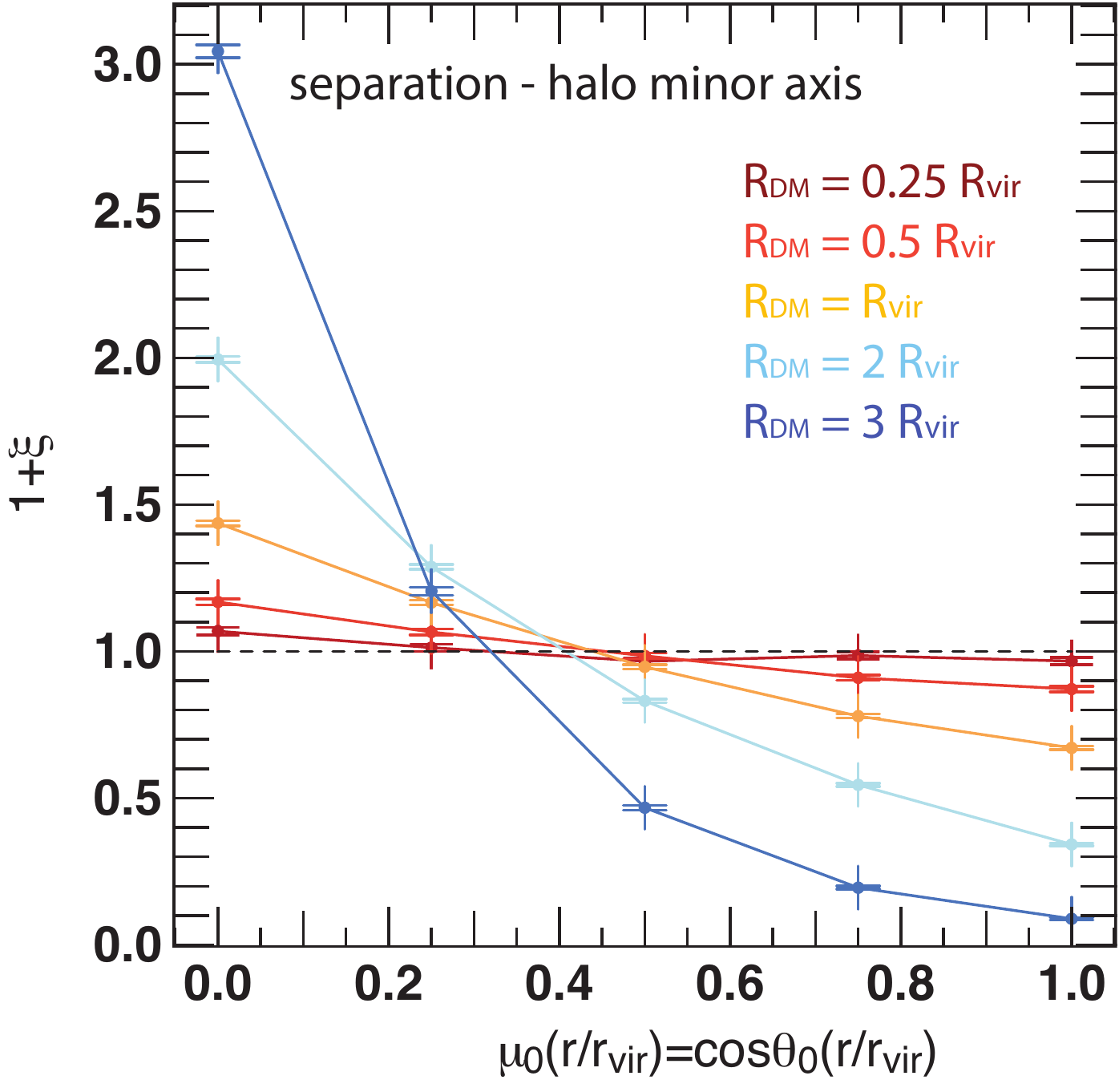}
 \caption{PDF of $\mu_{0}(r/r_{\rm vir})$, the cosine of $\theta_{0}(r/r_{\rm vir})$, the angle between the minor axis of the DM halo for material within a sphere of radius $r=R_{\rm DM}$ around its centre and the satellite separation vector, for all satellites within 5 radial shells from $0<R_{\rm 0s}<0.25 \, R_{\rm vir}$ (dark red) to $2 \, R_{\rm vir}<R_{\rm 0s}<3 \, R_{\rm vir}$ (navy blue). Satellites in the halo outskirts lie orthogonally to the minor axis of their host.}
\label{fig:satshell}
\end{figure}

Let us first quantify the degree of orthogonality between the satellite separation vectors and the minor axis of their host halo on various scales, and  compare it to the corresponding signal found around central galaxies\footnote{Note that an alternative way of testing the ability of satellites to trace their host's shape would be to estimate the alignment of satellites along the major axis of their host. This gives similar results but is less comparable to alignments in the central galactic plane (which are tested through  orthogonality to the galaxy's minor axis).}.  
Fig.~\ref{fig:satshell} displays the PDF of $\mu_{0}(r/r_{\rm vir})$, the cosine of $\theta_{0}(r/r_{\rm vir})$, the angle between the minor axis of the DM halo computed for all DM material within a sphere of radius $r=R_{\rm DM}$, with $R_{\rm DM}=0.25$, 0.5, 1, 2 or $3\, R_{\rm vir}$ around its centre of mass and the satellite separation vector. To obtain each curve, results are stacked for all satellite-host pairs identified in the considered $R_{\rm DM}$ {\it shell}. For instance, the light red curve corresponds to $R_{\rm DM}=0.5\, R_{\rm vir}$ and is obtained using the minor axis of the DM material within $0.5\; R_{\rm vir}$ of the halo centre of mass and taking into account all satellites with $0.25\; R_{\rm vir}<R_{\rm 0s}<0.5\; R_{\rm vir}$. 

One can clearly notice that the tendency for satellites to lie orthogonally to their host's halo minor axis is strongly scale dependent. While on the outskirts of the halo ($R_{\rm DM}>R_{\rm vir}$, $2\,R_{\rm vir}$ and $3\,R_{\rm vir}$ respectively) satellites show a high degree of orthogonality relative to their host's minor axis (excess probability of $\xi(0) =0.45$, $1.0$ and $2.1$ respectively), this trend decreases sharply when the inner parts of the halo are probed (and is barely detected below $R_{\rm DM}=0.25\;R_{\rm vir}$, with $\xi(0)<0.08 $). For comparison this means that $53\%$ of satellites within the $2\,R_{\rm vir}<R_{\rm 0s}<3\,R_{\rm vir}$ shell are misaligned by more than $75^{\rm o}$ with their host minor axis, while only $25\%$ are expected to do so in a uniform random distribution. For the innermost shell, $R_{\rm 0s}<0.25\,R_{\rm vir}$, this fraction drops to $26\%$, close to random.
%

Satellites are therefore more orthogonal to their host's minor axis in its outskirts than within its virial radius. This might indicate that outer shells of the halo are more elongated than its inner parts, possibly due to their lower state and relaxation and stronger connection with the filamentary infall.

\begin{figure*}
\center \includegraphics[width=1.6\columnwidth]{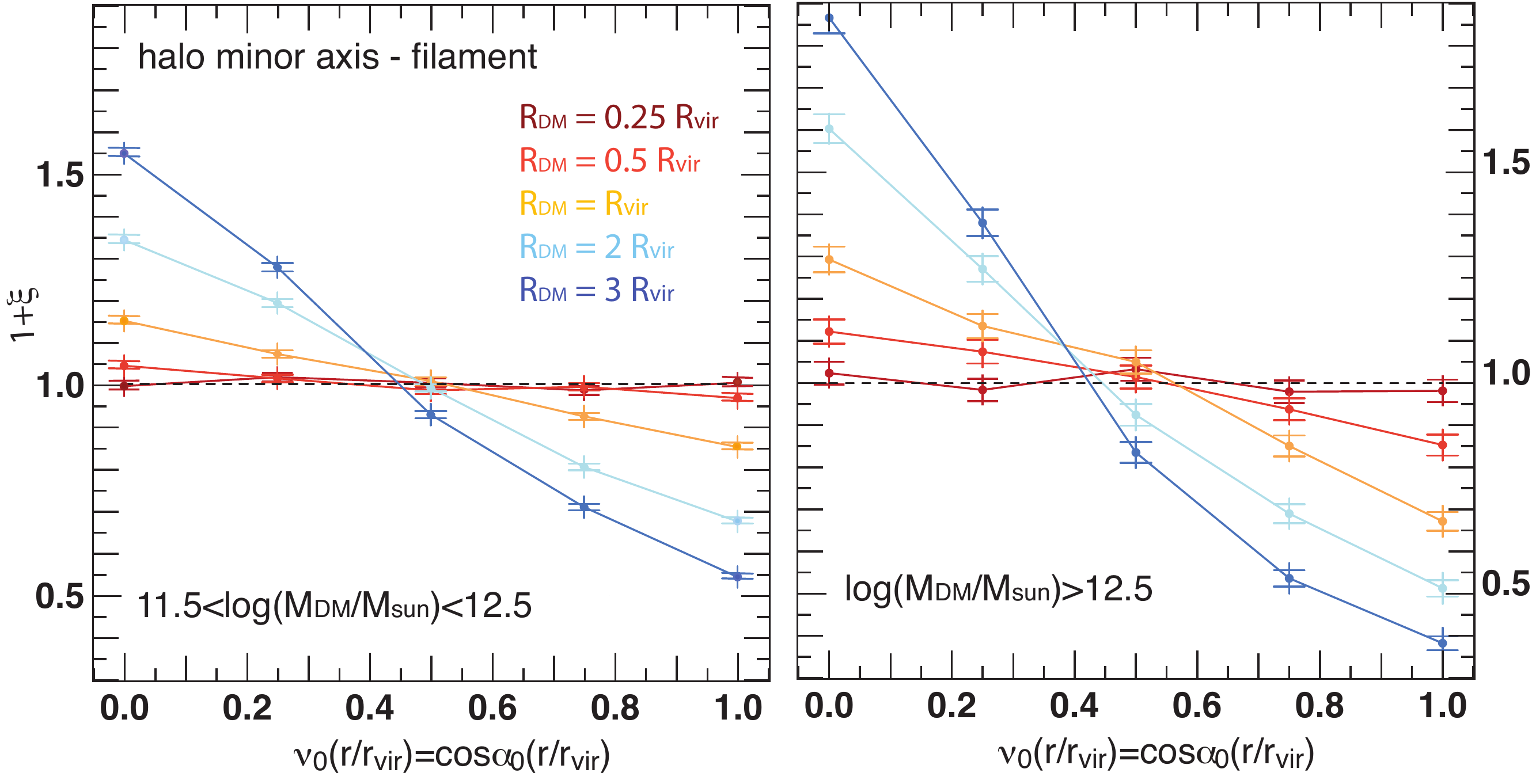}
 \caption{PDF of $\nu_{0}(r/r_{\rm vir})$, the cosine of $\alpha_{0}(r/r_{\rm vir})$, the angle between the minor axis of the DM halo (computed for all DM material within radius $r=R_{\rm DM}$ from its centre of mass) and the nearest filament direction. It is computed in 5  spheres of increasing maximal radius $R_{\rm DM}$ (from red to blue curves) to describe the progressive evolution of the halo shape from the inner core ($0.25\; R_{\rm vir}$) to the outskirts ($3\; R_{\rm vir}$), and for two halo mass bins: $10^{11.5}\, M_{\odot}<M_{\rm0} <10^{12.5}\, M_{\odot}$ (left panel) and $M_{\rm 0}>10^{12.5}\,M_{\odot}$ (right panel). Alignment of haloes shape with the cosmic web is enhanced for more massive haloes.}
\label{fig:fil}
\end{figure*}

Nonetheless,  recall that rather than merely  tracing their host shape, the orientation of satellites in the outer shells of the halo is also reminiscent of their polar infall from the local cosmic filament onto the halo, along gas streams that are possibly more focalised than their their DM counterparts. Paper I thus found that such satellites alignments with their filament extend up to $5\, R_{\rm vir}$ outside the halo with little to no dependence on the central mass \citep[see][for details]{Welker17}.Then their effective alignment with the major axis of the halo is also the result of its likely elongation in the direction of the filament, i.e. the direction of slowest collapse. As satellites start inspiraling deeper into the halo and undergo dynamical friction, this trend progressively fades away and the distribution of satellites is more likely to relax into the shape of its host halo. To confirm this latter dynamical process, let us now analyse the orientation of haloes with respect to the anisotropic features of the nearby cosmic web.

\subsection{connection to the cosmic web.}

Let us now analyse the tendency of haloes to align their inertial axis with the cosmic web. This can be done either by checking the alignment of the major axis with the nearby filament, or conversly by checking the orthogonality of the minor axis (although strictly speaking both angles would be necessary to reconstruct the 3D twist angle). Both estimators were found to give completely consistent results in Horizon-AGN. The main text  focusses on the orthogonality of the minor axis to allow for a straightforward comparison with alignments in the galactic plane, or with kinematic axes (spin axis) in the next sections. Results obtained using the major axis are given in Appendix~\ref{section:major}.

Fig.~\ref{fig:fil} displays the probability density function (PDF) of $\nu_{0}(r/r_{\rm vir})$, the cosine of $\alpha_{0}(r/r_{\rm vir})$, the angle between the minor axis of the DM halo (computed for all DM material within radius $r=R_{\rm DM}$ from its centre of mass) and the nearest filament direction. It is computed in 5  spheres of increasing maximal radius $R_{\rm DM}$ (from red to blue curves) to describe the progressive evolution of the halo shape from the inner core ($0.25\; R_{\rm vir}$) to the outskirts ($3\; R_{\rm vir}$), and for two halo mass bins: $10^{11.5}\, M_{\odot}<M_{\rm 0}<10^{\rm12.5}\, M_{\odot}$ (left panel) and $M_{\rm 0}>10^{12.5}\, M_{\odot}$ (right panel).  

Both mass bins display a similar trend: on $2$ and $3\; R_{\rm vir}$, haloes' minor axis show a strong tendency to be orthogonal to their filament, with an excess probability $\xi=0.35$ and $\xi=0.55$ respectively for $\cos(\alpha_{0})=0$ compared to a uniform distribution (dashed line) in the low mass bin. More massive haloes ($M_{\rm 0}>10^{12.5}\, M_{\odot}$) display a similar trend with an even stronger tendency of orthogonal orientation of the minor axis, with $\xi=0.6$ and $\xi=0.85$ respectively for $\cos(\alpha_{0})=0$ on $2$ and $3\; R_{\rm vir}$  scales respectively. This large scale orientation of the halo minor axis is expected as the anisotropic collapse model predicts that haloes (at least in their outskirts) will be elongated in the direction of their nearby filament, which corresponds to the slowest collapse axis. This results in haloes having their major axis aligned with their nearest filament, hence their minor axis orthogonal to it. 

\begin{figure}
\center \includegraphics[width=0.95\columnwidth]{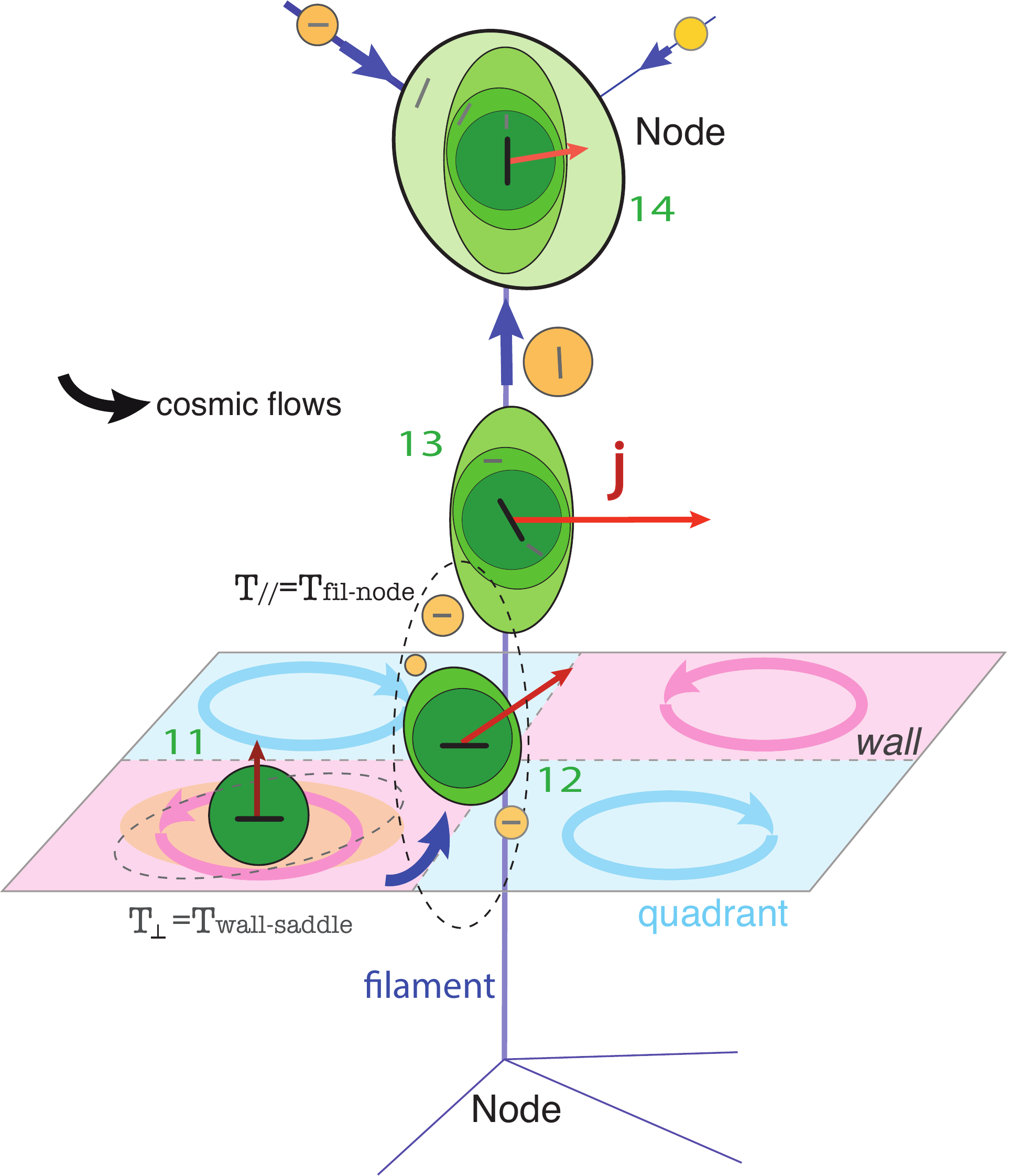}
 \caption{Sketch of the inside-out build-up of a typical halo (with typical masses labeled as ${\rm log}(M_{\rm 0}/M_{\odot})$ and iso-density contours in shades of green) along the cosmic web. Darker shades indicate inner more relaxed parts. Cosmic flows are indicated as dark blue thick arrows and specific angular momentum {\bf j} as thin red arrows.Vorticity quadrants around the saddle-point in the cross-section of the filament are shaded in light blue and pink. The main components of the Lagrangian patch later to be accreted by the halo are shaded in orange (diffuse component or mergers). Rotation planes of hosted galaxies are indicated as short dark lines. At early stages,  the projected components of the tidal shear tensor T are overlaid as dashed ellipses. The misalignments between the tidal tensor and the shape of the Lagrangian patch induce rotation of the patch towards realignment, hence the build-up the angular momentum of the halo. In an Eulerian framework, this transformation occurs through steady diffuse accretion,  followed by mergers. Tidal fields also stretch the halo along the filament. Colder galactic planes lag behind but eventually flip through galaxy mergers. In most massive clusters connected to several contrasted filaments, the connected geometry can be more complex and lead to misalignments between the extreme outskirts and the rest of the halo.}
\label{fig:sketch}
\end{figure}

This fully justifies the alignments of satellites  with the outer shells of the DM halo, with a maximum of the signal above $R_{\rm DM}>3\; R_{\rm vir}$, where the DM  from the filament  dominate  the overall DM budget of the shell. But this trend fades away in the inner halo, while the minor axis orientation becomes compatible with a uniform distribution within $0.25\; R_{\rm vir}$ i.e. in the core of haloes. Even focusing on sub-Virial scales, the shape alignment amplitude within $0.5\; R_{\rm vir}$ is 3 times lower than the one found for the slightly outer $0.5\; R_{\rm vir}<R_{\rm DM}<1\; R_{\rm vir}$ shell. This implies that haloes display a non-zero inertial twist, i.e. a tendency to have their outer shells iso-density contours misaligned with those of their inner parts. This also suggests that the inner parts of the halo do not  have a strong impact on the satellite's distribution. To understand this observation, let us put it in the context of the inside-out build up of haloes as they drift along the cosmic web.

\subsection{Halo inertial twist via anisotropic inside-out build-up.}

The evolution with total halo mass and distance to the centre of mass is consistent with the progressive build-up of halo spins within the metric of the cosmic web as described in \cite{Codis12,Laigle15} and analytically motivated in \cite{Codis15} (via an extension of tidal torque theory conditioned to the vicinity of a filament). This process is developed hereafter, following Fig.~\ref{fig:sketch} which presents a  sketch of the build-up of a typical halo (in green):
\begin{itemize}
\item Small haloes (dark green) are typically born away from the node, slightly offset from the spine of the filament, i.e offset from the region where the anisotropy of the collapse is maximal, and in the vicinity of the filament saddle point (circum-filament medium). In this region, the collapse is impacted both by the filament and the corresponding tidal torques from the nearby wall and the saddle-point ${\rm T}_{\bot}$, which have generated quadrants of coherent angular momentum polarity (as described by the above-mentioned constrained tidal torque theory).
\item From the point of view of Eulerian flows, these quadrants correspond to a multiflow vorticity-rich regions with vorticity aligned with the filament (i.e. a net swirl in a plane orthogonal to the filament), also displayed in alternating quadrants of opposite sign (pink and blue shades). 
\item Consequently, haloes not only undergo limited elongation in the direction of the filament  but also favour coherent, angular momentum rich accretion (orange fluff) within a plane orthogonal to the filament. This  keeps their rotation axis and preferential accretion plane orthogonal, as their spin (red arrow) is aligned with the nearby filament, hence to the slowest collapse direction. This therefore slows the contraction along the faster collapse principal axis: one can expect haloes to remain fairly spheroidal during this phase.
\item These haloes then progressively migrate towards the spine of their closest filament and drift along   towards the cosmic nodes  as they grow in mass  in the process (and overgrow their quadrant) from anisotropic filamentary accretion and mergers. In this phase, they undergo stronger stretching along the filament and accrete from a Lagrangian patch that has experienced enhanced  torques from the filament and the neighbouring node (${\rm T}_{\parallel}$), which tend to flip their spin axis orthogonally to the filament, therefore aligning their minor axis to their rotation axis, and their major axis to their preferential accretion axis. 
\item During that phase, accretion mostly occurs in the form of mergers (in orange) along the filament. They partially transfer their orbital momentum, orthogonal to the filament, to the intrinsic angular momentum of their remnants, therefore flipping their spin orthogonal to the embedding filament.
\end{itemize}

As a result,  more massive haloes are expected to show stronger ellipticity and greater alignment of their major axis to their nearest filament  (hence greater tendency to orthogonality for their minor axis) than their rounder low-mass counterparts. For a given halo, if its evolution is reasonably well-described by an inside-out growth -- i.e. if on average inner shells are created first --   it will therefore  acquire a net inertial twist during its migration in the cosmic web, arising from the aforementioned changes in the favoured orientation of both net torques and net infall. The increased alignment of the major axis in the outskirt of haloes and the increasing degree of misalignment as one probes the innermost parts of the halo is therefore  also a prediction of this scenario \citep[see][for details]{Codis15}. 

\subsection{Signatures of the inside-out build-up of haloes.}

To connect the now well-documented spin flips of haloes at high mass \citep{Codis12,Dubois14,Welker14,Welker17} to their observed inertial twist let us now find signatures of this connection in Horizon-AGN.

\begin{figure}
\center \includegraphics[width=0.85\columnwidth]{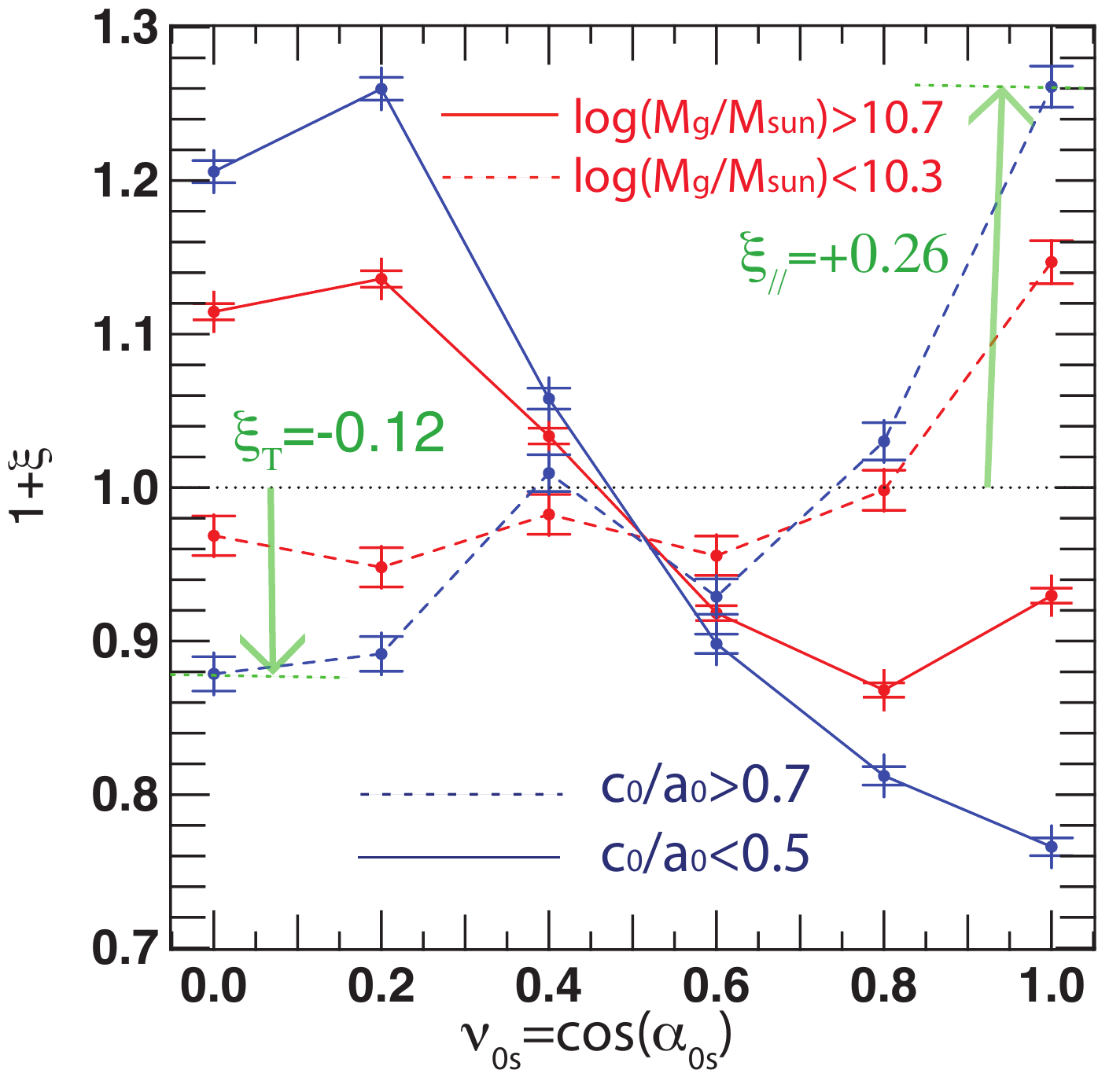}
 \caption{PDF of $ \nu_{0s}=\cos \alpha_{0s}$, the cosine of the angle between the halo's total spin and their nearest filament for two central stellar mass bins in red (dashed and solid lines), and two bins of minor to major halo axis ratio $ c_{\rm 0}/a_{\rm 0}$ in blue (dashed and solid lines).  The definition of $\xi_{T}$ and $\xi_{//}$ is overlaid in green. haloes with higher ellipticity tend to have their spin orthogonal to their nearest filament. }
\label{fig:spinfil0}
\end{figure}

First, the transition of halo orientation with increasing halo mass is easily quantified by studying the orientation of the spin of haloes with respect to the nearby filament. Fig.~\ref{fig:spinfil0} presents the PDF of $ \nu_{0s}=\cos \alpha_{0s}$, the cosine of the angle between the haloes spin and its nearest filament for different central stellar mass bins in red, and different shapes (as traced by the minor-to-major axis ratio of the halo  $ c_{\rm 0}/a_{\rm 0}$ in blue). Here quantities are computed on all the particles bound to the halo identified by the halo finder. Two algebraic quantities are also defined: 
\begin{itemize}
\item $\xi_{\rm T}$, the excess probability at $\nu_{0s}=0$, which characterises the excess probability of having a spin orthogonal to the filament
\item $\xi_{//}$, the excess probability at $\nu_{0s}=1$ which quantifies the excess probability of having a spin aligned with the filament. 
\end{itemize}
This allows us to define:
\begin{eqnarray}
 \Delta \xi = \xi_{//} -\xi_{\rm T},
 \end{eqnarray}
 as a simple measure of how well spins align with  filament in the sample considered. if $\Delta \xi >0$, spins tends to be aligned with the filament, while a sample with $\Delta \xi =0$ has uniformly distributed orientations and $\Delta \xi <0$ shows a  given degree of orthogonality. Note that this definition can be applied to any alignment PDF so long as it is in good approximation monotonous. 

 Let us first focus on the red curves: haloes with massive centrals  ($M_{\rm g} >10^{10.7}\, M_{\odot}$) tend to display a spin orthogonal to their filament, with  $\xi_{\rm T}=0.11$ and $\Delta \xi =-0.18$  while those with less massive centrals are more likely to have a parallel spin ($\Delta \xi =+0.18$). This transition highlights the strong correlation between halo and central masses and traces the known evolution of halo spin with halo mass described at the beginning of the section. As this trend was also found to hold for galaxies in \cite{Dubois14}, our choice of mass bins simply matches the transitional threshold, bracketed between $M_{\rm g} =10^{10.25}\, M_{\odot}$ and $M_{\rm g} =10^{10.75}\, M_{\odot}$ in \cite{Dubois14}.
Focusing now on the blue curves, notice that a similar trend is recovered between strongly elongated haloes with $c_{\rm 0}/a_{\rm 0}<0.5$ (solid blue line) and more spheroidal ones ($c_{\rm 0}/a_{\rm 0}>0.7$) (dashed blue line). The values of the axis ratio thresholds are chosen to match the peak of the distribution, in the high mass ($M_{\rm 0}>10^{12.5}\,M_{\odot}$) and low mass ($M_{\rm 0}<10^{12}\,M_{\odot}$) range respectively.  In fact the amplitude of the signal is enhanced using minor-to-major axis ratio bins rather than central mass bins: more spheroidal haloes tends to have their spin aligned with the neighbouring filament ($\Delta \xi =+0.38$) while elongated haloes are more likely to display a perpendicular orientation ($\Delta \xi =-0.44$). This is fully coherent with the present scenario, in which spheroidal small haloes progressively acquire mass, ellipticity and angular momentum orthogonal to their closest filament as they migrate towards the spine of the cosmic web and towards nodes along filaments.

In order to investigate how this anisotropic mass and spin acquisition in helicity-rich gas flows leads to the observed halo inertial twist, one can perform an analysis similar to that presented in Fig.~\ref{fig:fil}, while replacing the halo minor axis with its spin vector. While the corresponding detailed results can be found in Appendix~\ref{section:haloradialspin},  Fig.~\ref{fig:spinfil01} summarises the main findings of this analysis. It displays the evolution of $\Delta \xi = \xi_{//} -\xi_{\rm T}$  over four increasing mass bins, for halo spins computed on DM material within 5 different radius $R_{\rm DM}$ from the halo centre of mass, from  $R_{\rm DM}=0.25\, R_{\rm vir}$ (dark red dashed line) to  $R_{\rm DM}=3\, R_{\rm vir}$ (navy blue line). Vertical dashed lines indicate the mass bins used and the horizontal dashed line corresponds to uniform random orientations of the spin. The smallest radius bin appears as a dashed line to emphasise the fact that its value in the smallest mass bin ($M_{\rm 0}<10^{11.5}\,M_{\odot}$) is subject to higher uncertainties due to lack of resolution. At first glance, the transition from a parallel orientation of the spin at low mass to a perpendicular orientation at high mass is recovered on all scales. But interesting variations between scales appear:
\begin{itemize}
\item Expectedly, the innermost part of the halo ($R_{\rm DM}=0.25\, R_{\rm vir}$) appears to be the most insensitive to cosmic web. This is consistent with the idea that on that scale the halo material has undergone much more phase mixing and virialization than its outer shells, and was more spheroidal in the first place (this is explicitly tested in the next paragraph) hence more insensitive to cosmic torquing. 
\item For halo masses lower than the transition mass, the inner intermediate parts of the halo ($0.5\, R_{\rm vir}<R_{\rm DM}<1.0\, R_{\rm vir}$) show better spin alignment with the filament than within the outer shells. This is consistent with the fact that the inner parts of haloes form earlier, and consistently acquired angular momentum from the single vorticity quadrant of a given polarity from which they accreted.
\item For halo masses above the transition mass,  the outer parts of the halo ($R_{\rm DM}>1.0\, R_{\rm vir}$) show a stronger tendency to flip their spin orthogonal to the filament than the inner parts, which is consistent with their progressive formation from mergers closer to the spine and nodes of the cosmic web.
\item In the transition mass range, the spin  in the outer parts flips at lower masses than that in inner parts. It is detected from $10^{11.8}\, M_{\odot}$ onwards for $R_{\rm DM}=3\, R_{\rm vir}$, from $10^{12.1}\, M_{\odot}$  for $R_{\rm DM}=2\, R_{\rm vir}$, $10^{12.3}\, M_{\odot}$  for $R_{\rm DM}=1\, R_{\rm vir}$, $10^{12.5}\, M_{\odot}$  for $R_{\rm DM}=0.5\, R_{\rm vir}$
and $10^{12.8}\, M_{\odot}$ for $R_{\rm DM}=0.25\, R_{\rm vir}$. This is consistent with the inside-out build up of haloes as they drift along the cosmic web.
\end{itemize}

\begin{figure}
\center \includegraphics[width=0.95\columnwidth]{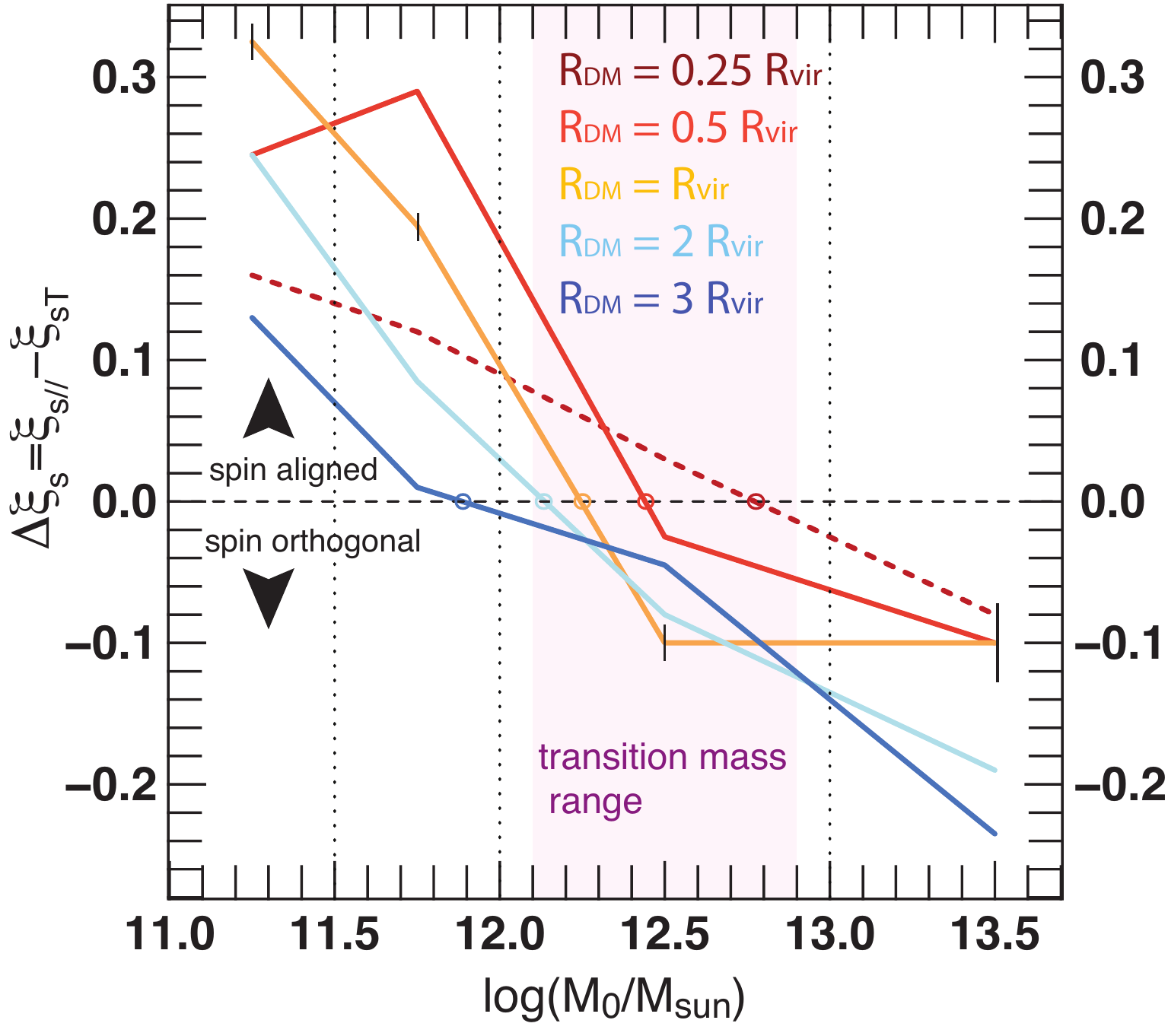}
 \caption{Evolution of $\Delta \xi = \xi_{//} -\xi_{\rm T}$ with halo mass for the PDF of $ \nu_{0s}=\cos \alpha_{0s}$, with halo spins computed within 5 radius $R_{\rm DM}$ from the halo centre, from  $R_{\rm DM}=0.25\, R_{\rm vir}$ (dark red dashed line) to  $R_{\rm DM}=3\, R_{\rm vir}$ (navy blue line). Vertical dashed lines indicate mass bins used. The horizontal dashed line corresponds to  random orientations of the spin. The red dashed line for $M_{\rm 0}<10^{11.5}\,M_{\odot}$ indicates lack of resolution in this bin. The pink shaded region indicates the transition mass bracket predicted in ~\protect\cite{Codis15}. The spin of the innermost part of haloes flips at higher mass than their that of their outskirts.}
\label{fig:spinfil01}
\end{figure}

To confirm the decreasing ellipticity of the diffuse DM component in the inner parts of the halo, let us compute the distributions of the minor-to-major axis ratios $c_{\rm 0}/a_{\rm 0}$ for host haloes. Axis ratios for the inner and outer concentric parts of each halo are produced, defined by their maximal radius $R_{\rm DM}$. 

\begin{figure*}
\center \includegraphics[width=1.9\columnwidth]{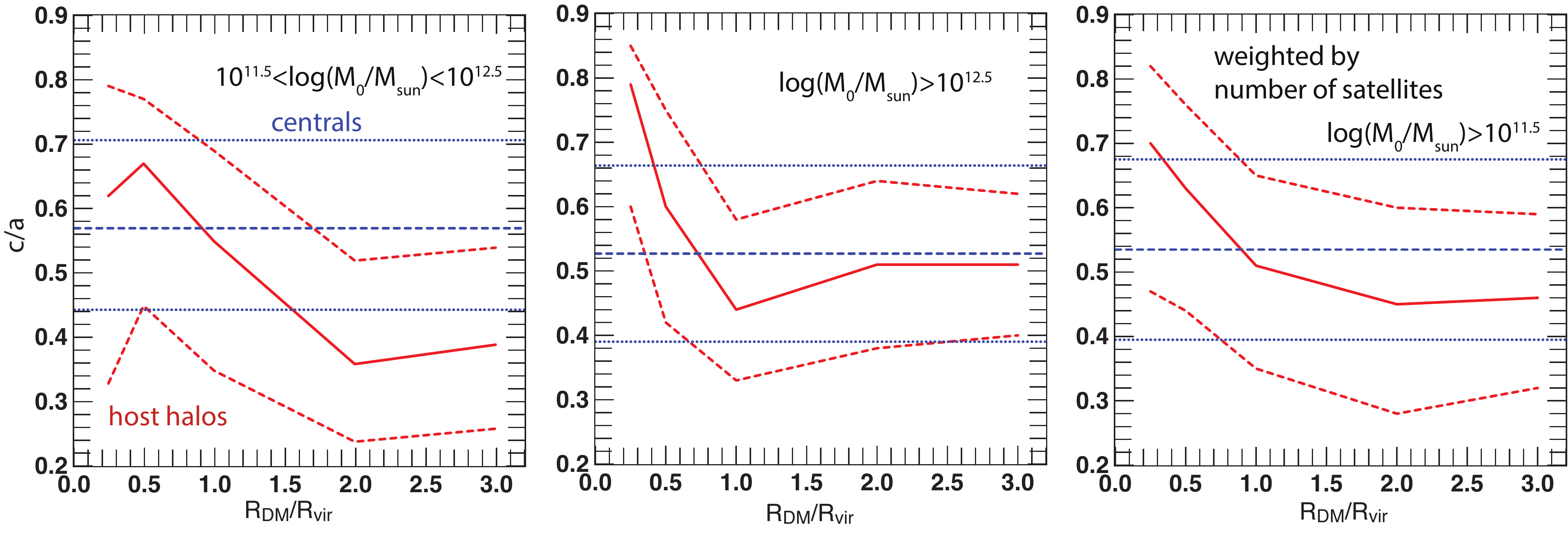}
 \caption{Evolution of $c_{\rm 0}/a_{\rm 0}$ within maximal encompassing radius $R_{\rm DM}$ around the halo centre of mass, for all sufficiently resolved haloes with $10^{11.5} \; M_{\odot}<M_{\rm 0}<10^{12.5} \; M_{\odot}$ ({\it left panel}), all haloes with $M_{\rm 0}>10^{12.5} \; M_{\odot}$ ({\it middle panel}), and all satellites in haloes with $M_{\rm 0}>10^{11.5} \; M_{\odot}$ ({\it right panel}). The red solid line follows the median of the distribution in each radius bin. Red dashed lines show the 16th and the 84th percentiles. The distribution  $c_{\rm g}/a_{\rm g}$ (median, 16th percentile and 84th percentile) for central galaxies in these haloes is overlaid in navy blue. On average halo cores are more spheroidal than their outskirts and the central galaxy they host.}
\label{fig:axisratio}
\end{figure*}

Fig.~\ref{fig:axisratio} displays the evolution of $c_{\rm 0}/a_{\rm 0}$  for parts of the halo of increasing radii  $R_{\rm DM}$ around its centre of mass. The sample contains all haloes with $M_{\rm 0}>10^{11.5} \; M_{\odot}$, for which the central parts are sufficiently resolved.  The left panel focuses on low-mass haloes with $10^{11.5} \; M_{\odot}<M_{\rm 0}<10^{12.5} \; M_{\odot}$, while the middle panel displays the result for more massive haloes with $M_{\rm 0}>10^{12.5} \; M_{\odot}$. Finally, the right panel shows the result for the full sample, weighted by the number of satellites in each halo. The red solid line follows the median of the distribution in each radius bins while red dashed lines account for the 16th and the 84th percentiles. For comparison, the distribution of $c_{\rm g}/a_{\rm g}$ (median, 16th percentile and 84th percentile) for centrals galaxies in these haloes is also overlaid in navy blue. Note that $c_{\rm g}/a_{\rm g}$ values are computed within one half mass radius of the galaxies.

Below the Virial scale, the dark matter component of the halo is indeed significantly more isotropic than at the virial scale and above, with axis ratios $c_{\rm 0}/a_{\rm 0}>0.67$ on average below $R_{\rm DM}<0.5\; R_{\rm vir}$ for $10^{11.5} \; M_{\odot}<M_{\rm 0}<10^{12.5} \; M_{\odot}$ and $c_{\rm 0}/a_{\rm 0}>0.7$ for $M_{\rm 0}>10^{12.5} \; M_{\odot}$, while the median of the distribution is around $c_{\rm 0}/a_{\rm 0}=0.56$  for the lowest mass bin ($c_{\rm 0}/a_{\rm 0}=0.44$ for the highest mass bin) for $R_{\rm DM}=1\; R_{\rm vir}$. Conversely, haloes are found to be highly triaxial above the Virial radius scale, with a median at $c_{\rm 0}/a_{\rm 0} \approx 0.35$ at the $R_{\rm DM}=2\; R_{\rm vir}$ scale ($\langle c_{\rm 0}/a_{\rm 0} \rangle\approx 0.5$ for the highest mass bin), consistently with their strong correlation to the cosmic web.

This result is robust across a wide range of halo masses. Indeed, while low-mass haloes with  $M_{\rm 0}<10^{12.5} \; M_{\odot}$ are generally more spheroidal at the Virial scale ($c_{\rm 0}(1)/a_{\rm 0}(1)\approx 0.56$ on average) than their more massive counterparts with $M_{\rm 0}>10^{12.5} \; M_{\odot}$ ($c_{\rm 0}(1)/a_{\rm 0}(1)\approx 0.44$ on average),  the innermost core of haloes is generally more isotropic than the outskirts across all mass bins. Note however that for most massive haloes the triaxiality above the Virial scale is actually reduced - with $\langle c_{\rm 0}(2)/a_{\rm 0}(2)\rangle \approx 0.51$ for $R_{\rm DM}>2\; R_{\rm vir}$ compared to their low-mass counterparts due to their higher connectivity to the cosmic web: very massive haloes with $M_{\rm 0}>10^{13.5} \; M_{\odot}$ are connected to several highly contrasted filaments of random orientations and undergo strong Virial shocks (as described in the last stage of evolution in Fig.~\ref{fig:sketch}). For these massive groups, even on the Virial scale, the triaxiality is reduced to $\langle c_{\rm 0}(1)/a_{\rm 0}(1)\rangle \approx 0.61$. Once again, the results of Fig.~\ref{fig:axisratio} are left qualitatively unchanged.

In conclusion, haloes are generally strongly triaxial   above the Virial radius scale and clearly elongated in the direction of their nearest filament.
Yet the dark matter  transitions to a much rounder structure within one Virial radius, and the alignment with the filament is essentially lost in the halo's inner core. While the rounder cores of inner haloes might  in part be due to strong relaxation and low asphericity of their early collapse, the net inertial twist of the halo can be explained by its inside-out build up through accretion, and  by synchronous migration from mid-filament vorticity-rich region to the  vicinity of nodes, where tidal torques tend to flip the halo's minor axis and spin orthogonal to its closest filament. 

This first exploration raises naturally the question: to what extent the distribution of satellites is still a reliable tracer of halo shape, and more specifically, on what scales could it be a reliable tracer of halo triaxiality?
 In particular, can the low ellipticity of the inner core of haloes alone still produce anisotropies in the angular distribution of satellites, or are central stellar discs torques dominant  in this region? The next section shows that the impact of the haloes' inner structure on satellite alignments is in fact sub-dominant compared to the influence of the central plane.

\section{Competitive alignments between halo \& centrals}
\label{section:DMvsbaryons}

\subsection{Alignments between halo and central plane}

Let us first compare simultaneously the relative orientations of satellites with respect to their host minor axis and to their central galaxy's minor axis, in order to better understand the fate of satellites as they enter the inner parts of the halo. The left panel of Fig.~\ref{fig:satsphere}  shows the PDF of $\mu_{0}(r/r_{\rm vir})$, following Fig.~\ref{fig:satshell} but this time using all satellite-host pairs identified in the considered $R_{\rm DM}$ {\it sphere} (rather than shell). Results are qualitatively similar but the amplitude of the signal in outer spheres is naturally reduced, which allows us to see more clearly the transition in the inner parts of the halo. 

For comparison,  the right panel  reproduces the PDF of $\mu_{\rm c}(r/r_{\rm vir})$, the cosine of $\theta_{\rm c}(r/r_{\rm vir})$, the angle between the minor axis of the central galaxy and the satellite separation, for all satellite-host pairs identified in the $R_{\rm DM}$ {\it sphere} considered. While the tendency for satellites to align within the galactic plane (i.e. orthogonally to the central's minor axis) is clear on all scales, the alignment within the galactic plane strengthens in the inner parts of the halo, i.e. in the vicinity of the central galaxy. These alignments were studied in details in Paper I, where  a transition was established, from a {\it filamentary} trend in the outskirts of the halo, where satellites mostly align with the filament they are infalling from, to a {\it coplanar} trend in the core of the halo, where satellites undergo dynamical friction and tidal torques from the inner parts of the halo and clearly align in the galactic plane of their central. However Paper I could not distinguish between torques and relaxation in the core of the DM halo and specific alignments with the central galaxy. Indeed our previous study did not compare the relative alignments with the central galaxy plane and with the inner halo shape. However a shared orientation of the central galaxy and its satellites can theoretically be induced by torques from the central galaxy and/or possibly by the past accretion of satellites along the same flows that gave birth to the initial central disc. 

The left panel of Fig.~\ref{fig:satsphere} now makes this distinction clear. In striking contrast with results obtained with the central minor axis, the degree of alignment with the halo shape (as traced by orthogonality to the minor axis) decreases in the vicinity of the halo, and the alignment within the central galaxy plane takes over on scales close to the viral radius ($\Delta \xi =-0.85$ as opposed to $\Delta \xi =-0.62$ for alignments with halo shape for $R_{\rm DM}= R_{\rm vir}$). In the innermost part of the halo, for $R_{\rm DM}= 0.25 \, R_{\rm vir}$, alignments with the halo shape become negligible ($\Delta \xi =-0.15$) while alignments with the central galactic plane are strongly dominant ($\Delta \xi =-2.6$), which represents an increase in the amplitude of the signal by a factor 17. This highlights the importance of the central galaxy and more generally of the baryons (either through mutual torques between the central and its satellite, or through a shared, collimated, initial direction of accretion) on shaping the angular distribution of satellites in the inner parts of the halo. The DM halo triaxiality alone does not account for alignments of satellites in the galactic plane. Note that the dominance of this central correlation is still detectable on scales close to the Virial radius ($R_{\rm DM}= R_{\rm vir}$), where alignments of satellites in the galactic plane still produce a signal 1.4 times greater than alignments with the halo shape. 
 Since galaxies and haloes have a distinct impact on the angular distribution of satellites, an important implication is that both effect should be taken into account in predictions of intrinsic alignments on small scales. In other words, it is crucial to include both the influence of the central and that of the halo in semi-analytic models of alignments used to generate survey mocks.
 
 The baryonic nature of these alignments is further tested in Appendix.~\ref{section:subhaloes}, where we show that the distribution of luminous satellites alone is better aligned with the central galactic plane than the full distribution (including dark satellites) on virtually all scales tested in this paper.

\begin{figure*}
\center \includegraphics[width=1.9\columnwidth]{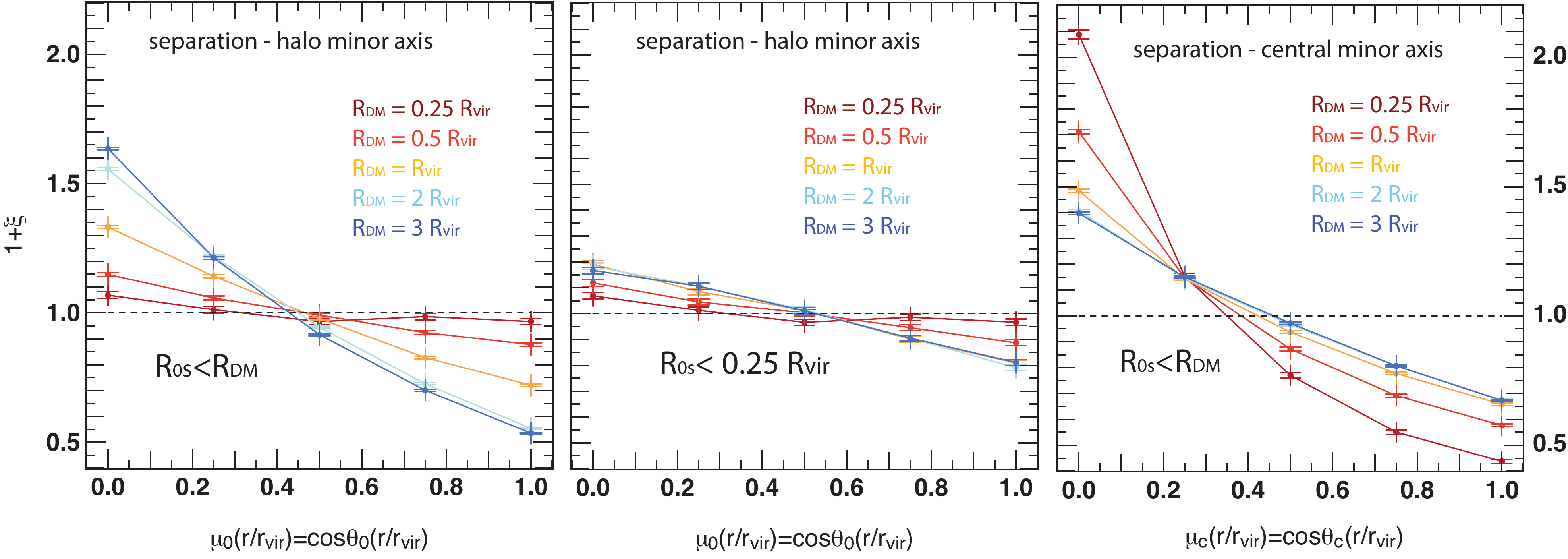}
 \caption{{\it Left panel}: PDF of $\mu_{0}(r/r_{\rm vir})$, similar to Fig.~\ref{fig:satshell} but using all satellite-host pairs identified in the $R_{\rm DM}$ {\it sphere} (rather than shell) considered. The results are qualitatively similar but the amplitude of the signal in the outer spheres is naturally reduced. {\it Middle panel}: PDF of $\mu_{\rm 0}(r/r_{\rm vir})$ restricted to satellites with $R_{\rm DM}<0.25\; R_{\rm vir}$. {\it Right panel}: PDF of $\mu_{\rm c}(r/r_{\rm vir})$  the cosine of $\theta_{\rm c}(r/r_{\rm vir})$,  the angle between the minor axis of the central galaxy and the satellite separation, for all satellite-host pairs identified in eac $R_{\rm DM}$ {\it sphere} considered. The degree of alignment with the halo decreases in its core, while alignments with the central galactic plane increase. }
\label{fig:satsphere}
\end{figure*}

Finally, the middle panel presents the PDF of $\mu_{\rm 0}(r/r_{\rm vir})$, this time focusing only on satellites within the core of the halo, i.e. with $R_{\rm DM}<0.25\; R_{\rm vir}$, and tests their orientation relative to the minor axis of their host halo, computed on the 5 scales used throughout. Interestingly, core satellites align better with the outer halo shape than with the inner halo shape. This confirms that alignment with the halo shape is largely inherited from the cosmic web on large scales and from the subsequent preferential infall of satellites along the most contrasted filament that feeds the halo. As detailed in \cite{Welker17}, this also explains why the alignment of satellites with the galactic plane, although fainter, is also detected at large separations ($\Delta \xi =-0.7$ for $R_{\rm 0s} < 3\, R_{\rm vir}$): Centrals above $M_{\rm c}>10^{10.5} \; M_{\odot}$ tend to have their spin and minor axis orthogonal to the filament, hence their galactic plane aligned with it, and consecutively also aligned with both the preferred satellite infall direction and the  major axis of their host halo. 

While clear evidence was found that baryon-driven torques are dominant when shaping the satellite angular distribution on sub-virial scales, one needs now to understand how such torques also take over the more massive core halo torques. A logical explanation could be that the core might be too spheroidal to generate significant torques, therefore allowing central disks anisotropic accretion and torques to become the dominant shaping processes. This assertion is tested in the next subsection.

\subsection{Relative shapes of the halo and its central galaxy.}

For this analysis, let us go back to Fig.~\ref{fig:axisratio}, which compares the distributions of $c_{\rm 0}/a_{\rm 0}$ for centrals galaxies and host haloes simultaneously. Recall that  Fig.~\ref{fig:axisratio} displays the evolution of   $c_{\rm 0}/a_{\rm 0}$  for parts of the halo of increasing radius  $R_{\rm DM}$ for all haloes with $M_{\rm 0}>10^{11.5} \; M_{\odot}$, for which the central parts are sufficiently resolved. The red solid line follows the peak of the distribution in each radius bins while red dashed lines account for the 16th and the 84th percentiles. The distribution for centrals (peak, 16th percentile and 84th percentile) is overlaid in navy blue.

It is striking that below the Virial scale, the dark matter component of the halo is indeed significantly more isotropic than the stellar component of the central galaxy, with axis ratios $c_{\rm 0}/a_{\rm 0}>0.7$ on average below $R_{\rm DM}<0.5\; R_{\rm vir}$ while the median of the distribution for central galaxies is around $c_{\rm 0}/a_{\rm 0}=0.53-0.57$ depending on mass scale. Interestingly, at fixed mass, little evidence was found for a correlation between the halo shape on $R_{\rm DM}=0.25\; R_{\rm vir}$ scale and the central shape.   Less than $3\%$ variation is found on average for  $c_{\rm g}/a_{\rm g}$ between haloes with $c_{\rm 0}(0.25)/a_{\rm 0}(0.25)<0.6$ and and those with $c_{\rm 0}(0.25)/a_{\rm 0}(0.25)>0.75$ ($\approx 25\%$ variation between bins) at fixed mass range. This is consistent with \cite{Chisari17} who found that the shape of the central is not correlated with the shape of its host halo.

As mentioned before, haloes are in fact highly triaxial above the Virial radius scale, with a median at $c_{\rm 0}/a_{\rm 0} \approx 0.4$ on $R_{\rm DM}=2\; R_{\rm vir}$ scale, where they correlate more strongly with the cosmic web, although the outskirt triaxiality on the $R_{\rm DM}>2\; R_{\rm vir}$ scale is actually reduced to $c_{\rm 0}(3)/a_{\rm 0}(3)\approx 0.54$ for most massive haloes ($M_{\rm 0}>10^{13} \; M_{\odot}$) due to their higher connectivity. 

 This result is robust across a wide range of halo masses. Indeed,  while  low-mass haloes with  $M_{\rm 0}<10^{12.5} \; M_{\odot}$ are generally more spheroidal on Virial scales than their more massive counterparts with $M_{\rm 0}>10^{12.5} \; M_{\odot}$, the innermost core of haloes is generally more isotropic than the stellar material of the central across all mass bins and that the outskirts of haloes are always on average more triaxial than their central galaxy.

\begin{figure*}
\center \includegraphics[width=1.9\columnwidth]{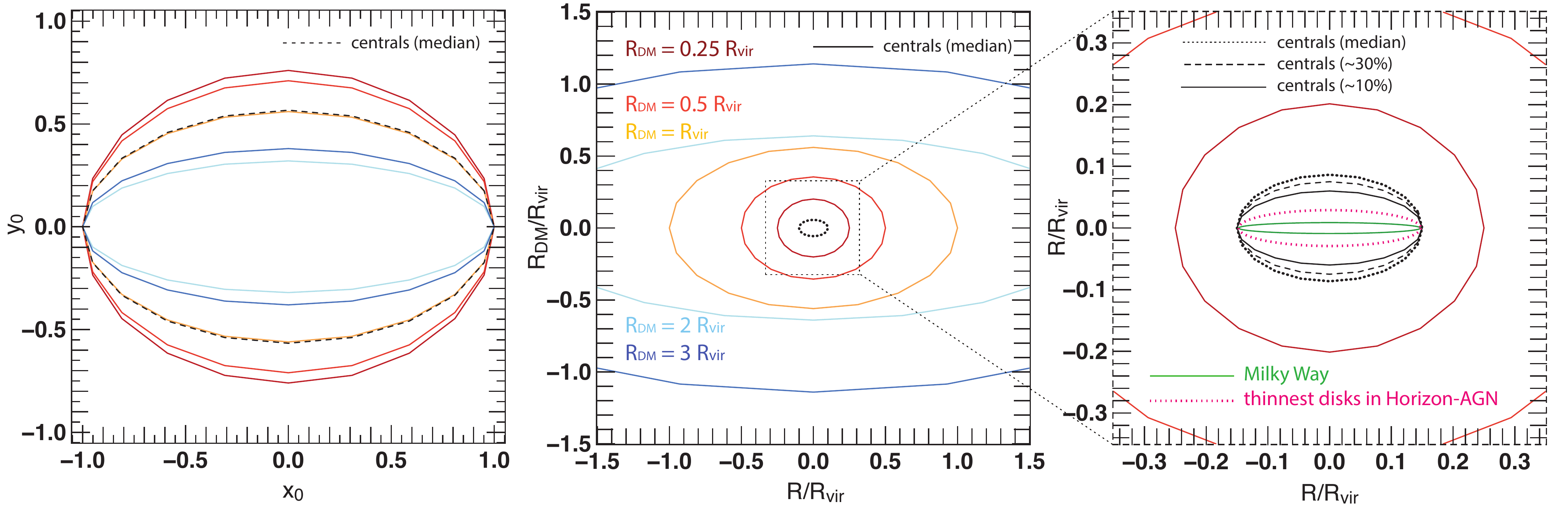}
 \caption{{\it Left panel}: Median projected ellipsoids derived from the inertia tensor of the halo computed within spheres of increasing radius (dark red to navy blue). Major axes are normalised to $a=1$. The dashed black line indicates the median for central galaxies. {\it Middle panel}: Projected ellipsoids on scale with the central galaxy median ellipsoid as a dotted black line. {\it Right panel}: zoom of previous panel on sub-Virial scales. The 10th percentile (solid black line) and 30th percentile (dashed black line) ellipsoids drawn from centrals in Horizon-AGN are overlaid, as well as the thinnest disk centrals identified at z=0 in Horizon-AGN (dashed pink line) and a typical Milky-Way galaxy (green). Central galaxy ellipsoids are 3 to 5 times bigger than their actual size.The inner halo-central shape contrast in Horizon-AGN is only a lower limit of that of real galaxies at $z=0$. }
\label{fig:ellipses}
\end{figure*}

To illustrate these findings, Fig.~\ref{fig:ellipses} displays the median ellipsoid (weighted by the number of satellites) -- projected along the intermediate axis - derived from the inertia tensor of the DM halo material computed within concentric spheres of increasing radius (ellipses from dark red to navy blue). On the left panel, the ellipses are plotted with a major axis normalised to $a=1$ to help compare their relative ellipticity. The dashed black line indicates the median for central galaxies in Horizon-AGN. The middle panel shows the same projected ellipsoids on scale with a focus on the Virial radius scale, and the right panel display a zoom on sub-Virial scales. The central galaxy's median inertia ellipsoid is displayed as a dotted black line. The right panel also displays the 10th percentile (solid black line) and 30th percentile (dashed black line) ellipsoids drawn from the distribution of axis ratios for centrals in Horizon-AGN is overlaid, as well as the thinnest disk centrals identified at z=0 in Horizon-AGN (dashed pink line). To guide the eye, the central galaxy ellipsoids are not on scale but 3 to 5 times bigger than their actual size. Note also that the central inertia tensor does not reflect the fact that most of the galaxies within the 30th percentile are in fact bulge+disk systems, so that the inertia tensor largely underestimates the flatness of the real stellar disk alone. 

Finally, keep in mind that galaxies in Horizon-AGN tend to show puffed-up disks compared to real spiral galaxies with thin disks in the Local Universe due to its spatial resolution limited to 1 kpc  affecting both star formation and feedback. As a comparison, a Milky Way-like galaxy typically has $c_{\rm g}/a_{\rm g}=0.06$ (solid green line) as opposed to $c_{\rm g}/a_{\rm g} \approx 0.2$ at best for diskiest systems in Horizon-AGN at $z=0$ (dashed pink line). Hence the striking contrast between the shape of the inner halo and that of the central galaxy presented on  Fig.~\ref{fig:ellipses} and  Fig.~\ref{fig:axisratio} is actually only a lower limit of the actual contrast, expected to be stronger in real systems.

\subsection{Interpretation and discussion}

This section showed that satellites above $1\,R_{\rm vir}$ align with the major axis of their halo, all the more so that the halo is elongated in the direction of the filament from which it accretes satellites (see Paper I for satellite alignments with the cosmic filaments). This alignment is however lost in the rounder core of the halo, where alignments with the much more anisotropic stellar disk dominate. This highlights the major impact of baryons - and the corresponding thin disks in particular - in producing coherent planes of satellites. 

One possible explanation is that tidal torques form the anisotropic disk dominate in this region. Another, and possibly concomitant explanation is that the distribution of satellites within the inner region region of the halo traces their past preferential direction of accretion, which is also the direction of the anisotropic inflows that gave rise to the initial central disc at higher redshift. In other words, the satellites and the central galactic plane could share a common direction because they were fed into the halo along the same flows that built the central plane into its current orientation. The existence of cold collimated gas flows at high redshift could in particular justify the strength and detectability of such alignments at low redshift. Exactly in which regime and for which type of galaxies one scenario becomes dominant over the other is beyond the scope of this paper and will require further analysis of relative timescales, including in higher resolution simulations with similar physics (Dubois et al., in prep). But these alignments confirm that tracing the triaxiality of haloes through the distribution of its satellites within $1-1.5 \, R_{\rm vir}$, assuming the relaxation of such systems of satellites, will generally lead to poor estimates of the triaxiality of the halo if the orientation and mass of the central disc is not controlled.

\begin{figure}
\center \includegraphics[width=0.85\columnwidth]{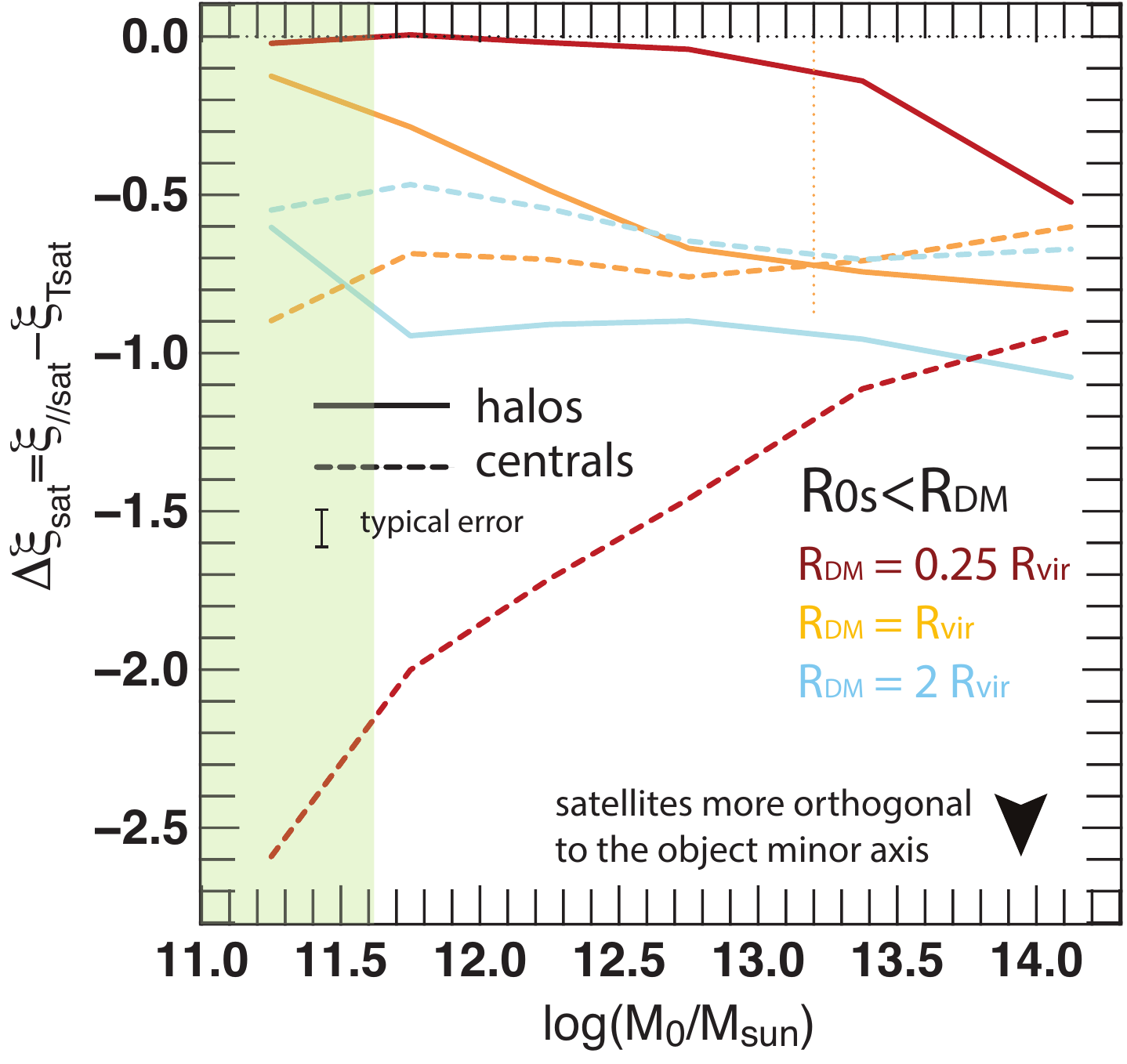}
 \caption{$\Delta \xi_{\rm sat} = \xi_{\rm // sat} -\xi_{\rm T sat}$ as a function of halo mass, similar to $\Delta \xi$ but derived from the PDF of $\mu_{\rm c}(r/r_{\rm vir})$ (satellite vs. central minor axis, dashed lines) and $\mu_{\rm 0}(r/r_{\rm vir})$ (satellite-halo minor axis, solid lines), for satellites within increasing radii from the halo centre of mass (dark red to light blue). The green shaded are indicates low resolution bins. The vertical yellow dotted line indicates the intersection between the signals derived from $\mu_{\rm c}$ and $\mu_{\rm 0}$. Alignments around centrals dominate over alignments with the halo shape.}
\label{fig:dxi_sat0}
\end{figure}

Moreover, this effect seems to persist for high halo masses, including for large groups and clusters with $M_{\rm 0}>10^{13.5} \; M_{\odot}$. This can be seen on Fig.~\ref{fig:dxi_sat0}, which plot $\Delta \xi_{\rm sat} = \xi_{\rm // sat} -\xi_{\rm T sat}$ as a function of halo mass, with a definition similar to $\Delta \xi_{\rm} = \xi_{//} -\xi_{\rm T}$ but this time derived from the PDF of $\mu_{\rm c}(r/r_{\rm vir})$ (i.e. satellite-central minor axis orthogonality signal, dashed lines) and $\mu_{\rm 0}(r/r_{\rm vir})$ (i.e. satellite-halo minor axis orthogonality signal, solid lines), for satellites (and DM material in the case of $\mu_{\rm 0}$) within increasing distances from the halo centre of mass. Within $R_{\rm DM}=0.25\; R_{\rm vir}$ (red curve) alignments within the galactic plane dominate at all halo masses: $\Delta \xi_{\rm sat}$ is more strongly negative for anti-alignments with the central minor axis than with the DM core minor axis. Although the impact of the DM core increases with halo mass as its ellipticity rises, the signal is still about twice stronger around centrals than around DM cores for $M_{\rm 0}>10^{13.75} \; M_{\odot}$. Even at the Virial scale, setting $R_{\rm DM}=\; R_{\rm vir}$ (yellow curves), the halo shape alignment signal (solid line) only takes over for haloes above $M_{\rm 0}>10^{13.2} \; M_{\odot}$ where centrals are more often spheroidal, and only dominates by less than $25\%$. This seems insufficient to consider satellite distribution to be an unbiased tracer of the halo shape, even in that mass range.

Paper I  established  that when the galactic plane is aligned with the nearest filament, alignments of satellites within the central plane are strengthened as the {\it filamentary} alignments on large scales now coherently add up to the  tidally induced {\it coplanarity} on smaller scales, both trends bending satellites in the same plane. A similar signal enhancement is observed around haloes which align best with the cosmic web down to the Virial radius, as presented in Appendix~\ref{section:haloeshape}. \cite{Dubois14} found that this case is slightly favoured statistically for $M_{\rm 0}>10^{10.5} \; M_{\odot}$  and all the more so that the central mass is higher. Indeed, such galaxies above that transition mass have generally undergone several mergers along the filament throughout their cosmic history. This tend to flip their spin orthogonal to the filament, through the transfer of orbital momentum of the pair into the intrinsic angular momentum of the remnant. This is the scenario established for haloes by \cite{Codis12,Codis15}   described in Fig.~\ref{fig:sketch}, which statistically leads to  orthogonal minor axis for galaxies, as their minor axis is  approximatively aligned with their spin axis. 

This also suggests that in order to accurately estimate the degree of satellite coplanarity around synthetic central galaxies, it is crucial that a simulation should not only model the baryon physics, but also resolve simultaneously and with sufficient precision:
\begin{itemize}
\item the cosmic web along which the central galaxy drifts and along which satellites are accreted across the cosmic history of the galaxy (on a few Mpcs scale). This should in particular include the gaseous cosmic web, along which galaxies form and whose filaments are more concentrated and have smaller cross-sections than the corresponding dark matter filaments (Ramsoy et al., in prep) due to their dissipative nature \citep[see][for details]{Pichon11} but are the most relevant to galaxy formation and satellite infall.
\item the gas and stellar discs so as to allow the formation of thin discs leading to robust bulge to disc decomposition. This implies to reach a spatial resolution $< 100$ pc. Additionally, resolving dwarf galaxies down to $10^6 \, M_{\odot}$ appears necessary to quantify the alignments around local galaxies such as the Milky Way. Indeed, the strength of satellite alignments around the central are strongly dependent on the satellite-to-central mass ratio, while alignments with the inner halo are not, as detailed in Appendix~\ref{section:satmass1}.
\end{itemize}
\subsection{Comment on stacked signals.}
\begin{figure}
\center \includegraphics[width=0.9\columnwidth]{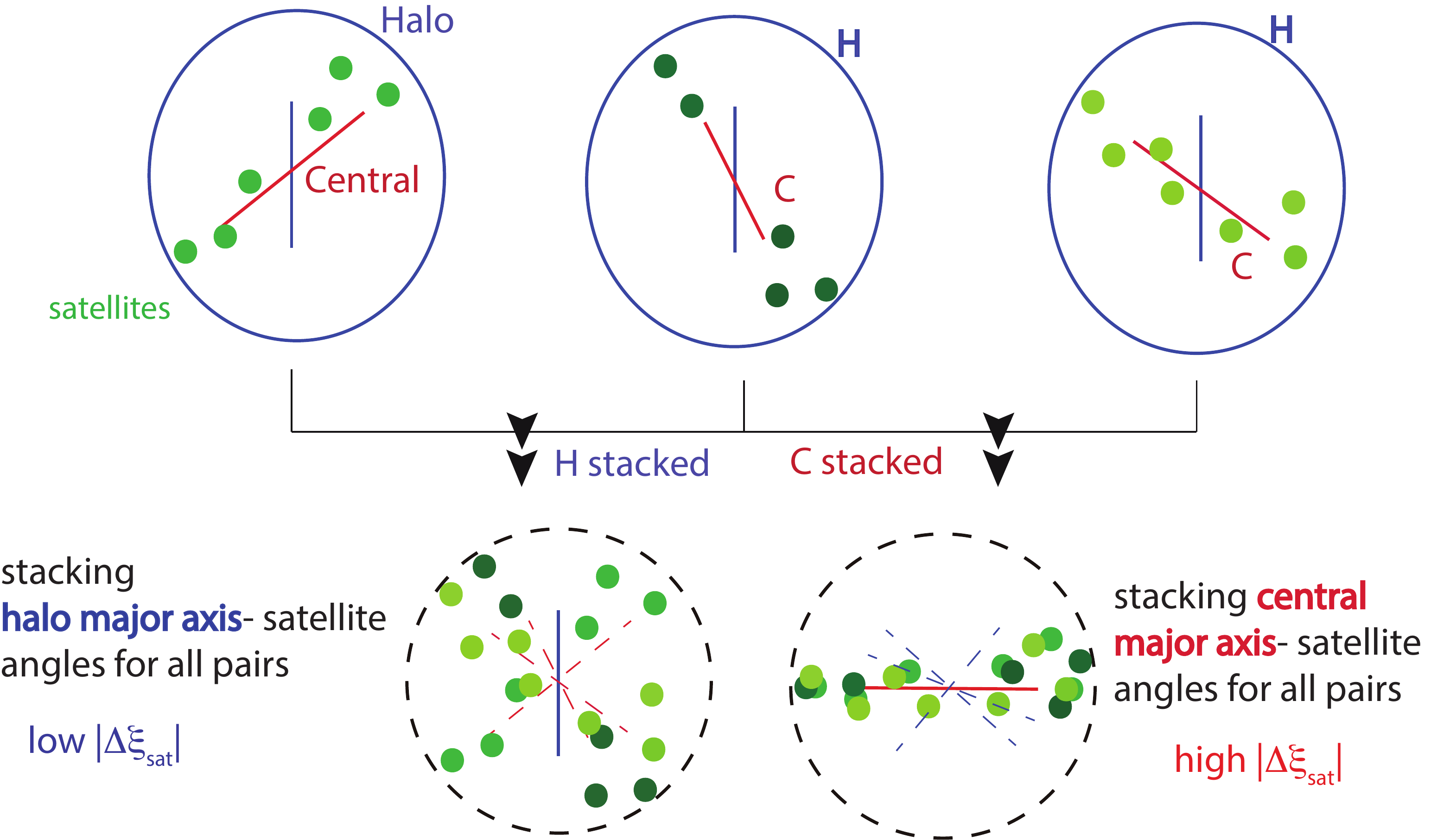}
 \caption{Sketch of the equivalent synthetic distributions obtained through the stacking procedure used to produce PDFs $\mu_{\rm c}$ (in red) and $\mu_{\rm 0}$ (in blue). Here, three individual populations of satellites (green dots) in distinct host haloes are stacked aligning either their halo's or their central's major axis. In this case, satellites align better with their central's galactic plane, which clearly shows on the stacked signal.}
\label{fig:stack}
\end{figure}

Alignment PDFs so far were obtained by taking into account all satellite-central galaxy pairs  for all haloes in a given mass range (or within a given normalised separation range). These are therefore {\it stacked} PDFs. Indeed, computing the PDFs of $\mu_{\rm c}$ and $\mu_{\rm 0}$\footnote{Recall that the difference between  $\mu_{\rm c}$ and $\mu_{\rm 0}$ is the axis of reference along which all the distribution of satellites is stacked, the minor axis of the central galaxy and the minor axis of the DM halo respectively.}   for all pairs for all haloes in a given mass or distance range is equivalent to considering a synthetic halo and satellite distribution corresponding to the stacked distributions in a preferential frame of satellites of all individual haloes. This procedure is illustrated in Fig.~\ref{fig:stack} using major axis stacking rather than minor axis for better visibility. The method is nonetheless identical. As it becomes apparent, this has the advantage of clearly highlighting which axis is the most relevant to describe alignments: the halo minor axis, the central minor axis or the cosmic web. Indeed, since the stacked signal is strong around the central minor axis ($\mu_{\rm c}$) for satellites within the Virial radius, this suggests that the galactic plane is indeed the plane in which satellites bend. Conversely, stacking satellites along the halo minor axis returns a much lower signal on that scale, pointing towards the fact that  halo shape is not the relevant feature to describe the alignments within dark matter haloes. However, this does not mean that anisotropic distributions of satellites are not observed in haloes, only that the direction and origin of such alignments are either better predicted or strengthened by the impact of the central galaxy.

Quantifying how much alignments in the halo shape are lost, enhanced or more generally modified by the coupling between these features implies to get a closer look at the relative orientation of such features. Tidal torques from centrals and/or past baryon accretion play an important part in bending their satellites in specific planes, hence this may lead to misestimate the host halo's ellipticity (i.e. the eigenvalues of the halo inertia tensor) from the distribution of satellites. So far the focus was only on the comparative amplitude of the two stacked signals ($\mu_{\rm c}$ and $\mu_{\rm 0}$), i.e. determining which one is dominant in as a function of mass or distance to the centre. However, finding a dominant signal in a given range of mass or distance does not ensure that the other signal is sufficiently negligible not to impact at all the perceived alignments. For instance, the fact that the stacked signal around the halo minor axis becomes slightly dominant at the Virial scale for haloes with $M_{\rm 0}>10^{13.2}\, M_{\odot}$ does not imply that the central has no impact on the satellite distribution on these scales. Since signals remain comparable in this range, it may well be that the central and the halo shapes are very strongly aligned, as suggested in \cite{Chisari17} which found strongest halo-central alignment for the high-mass end. In this case, satellite correlations with the central might still artificially enhance the perceived alignments of satellites along the major axis of the halo.

 More generally, one should ask whether or not the influence from centrals leads to poor estimates of the directions of the halo principal axis (i.e. the eigenvectors of its inertia tensor). If centrals galaxies show strong misalignments with their inner halo, estimates of such directions from the distribution of satellites will likely be very poor if radial shells in which satellites are chosen are not carefully picked. However, since respective effects on satellites from the central and from the halo are not degenerate in this case, one can possibly expect to find such shells in which the impact of the halo is detectably dominant. On the other hand, if central galaxies are sufficiently aligned with their inner halo, their impact does not affect the measurement of the direction in which the halo is elongated and can in fact strengthen the signal. This also means that the error on the ellipticity will be the strongest, and since effects from the halo and its central are in this case degenerate, it may be very hard to separate the two even above the Virial scale. 
 
 A careful analysis of alignments between haloes and centrals is therefore necessary to draw a  definite conclusion. Previous studies in Massive-Black II, EAGLE and Horizon-AGN \citep{Tenneti14,Velliscig15,Chisari17} found that such misalignments between the central galaxy and the dark matter component of the halo decrease with halo mass and show only a weak dependence on redshift. The following section investigates further the radial dependence of  these misalignments and relate them to the scenario  detailed in Fig.~\ref{fig:sketch}.

\section{Correlations of halo shape and central  plane.}
\label{section:halogal}

This section evaluates in some details the correlations between the orientation of the galactic plane and the halo shape.

\subsection{Alignments between the halo and the galactic plane}

\begin{figure*}
\center \includegraphics[width=1.9\columnwidth]{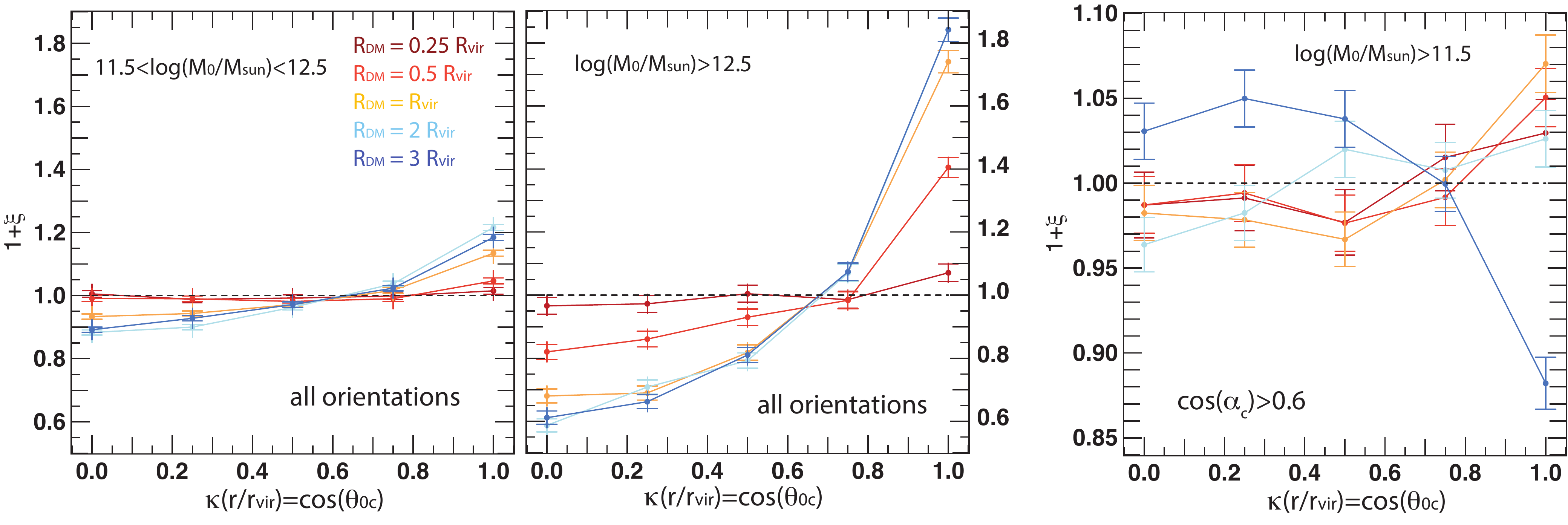}
 \caption{PDF of  $\kappa(r/r_{\rm vir})$, the cosine of $\theta_{0c}$, the angle between the minor axis of the central galaxy and that of the halo, computed in spheres of increasing radius (dark red to navy blue curves), for two different mass ranges: $10^{11.5}\, M_{\odot}< M_{\rm 0}\!< 10^{12.5}\, M_{\odot}$ ({\it left panel}) and $ M_{\rm 0}> 10^{12.5}\, M_{\odot}$ {\it middle panel}, and for a sample restricted to centrals with a spin aligned with their nearby filament ($\nu_{\rm c}=\cos\alpha_{\rm c}>0.6$ i.e. $\alpha_{\rm c}<50^{\rm o}$ ) ({\it right panel}).}
\label{fig:halogal}
\end{figure*}

 Fig.~\ref{fig:halogal}  displays the PDF of  $\kappa(r/r_{\rm vir})$, the cosine of $\theta_{0}^{C}$, the angle between the minor axis of the halo and that of the central galaxy. The minor axis of the halo is computed in spheres of increasing radius from its centre of mass, similarly to what was done in previous sections (dark red to navy blue curves).   The corresponding PDF is computed for haloes within two different mass ranges: $10^{11.5}\, M_{\odot}< M_{\rm 0}\!< 10^{12.5}\, M_{\odot}$ on the left panel and $ M_{\rm 0}> 10^{12.5}\, M_{\odot}$ on the middle panel.

Notice that the alignment between the two minor axis (positive excess probability $\xi$ in $\kappa(r/r_{\rm vir})=1$) is  strongest for massive galaxies, for which the signal is four times stronger than for their low-mass counterparts. This mass dependence was already found in \cite{Chisari17}. It is expected, as for massive haloes that have statistically more massive centrals, both minor axis and spins are expected to be orthogonal to the cosmic web and this effect should strengthen with cosmic time as the mass intake due to the steady infall from the filament -- which is also the direction of slowest collapse -- increases. In other words, at high mass, all processes that can possibly impact the relative orientation of the halo and the central add up in the same direction, leading to increased alignments. Conversely, at the  low-mass end, even though both halo and central spins are statistically expected to be aligned with their nearby cosmic filament, this correlation is limited at low-redshift and does not easily transfer to mutual alignment. This is especially true for minor axis alignment as the galactic rotation plane is orthogonal to the direction of slowest collapse in that case. However, strikingly the alignment is  strongest in the outer shells of the halo, while it decreases towards the core. The limited correlation within $0.5 \, R_{\rm vir}$ is expected since the halo core shape is essentially spheroidal compared to its central galaxy, all the more for low-mass haloes. However the alignments on large scales are  directly related to the strong influence of the cosmic web in shaping haloes and galaxies.
decrement

Indeed, haloes are primarily elongated in the direction of their nearest filament, especially at high masses. \cite{Dubois14}, \cite{Welker14} and \cite{Welker17} also detailed how galaxies progressively flip their spin orthogonal to their host filament through mergers along the spine of the filament, resulting in massive galaxies having a spin (and a minor axis) orthogonal to their nearby filament. Although this follows the same scenario as the halo spin flips detailed in \cite{Codis12}, the timescales and specific merging processes vary between the two. Indeed, on the one hand, the diffuse DM component of a subhalo - infalling from a filament into a more massive halo - progressively disolves within its host through dynamical friction as it orbits and inspirals down its potential well, resulting in little to no impact on the inner core of the host halo. On the other hand, the stellar component of a satellite galaxy hosted in such a subhalo is more likely to be preserved down to the core of its host, where it merges at last with the central galaxy and directly impacts its structure. Through this process, the preferential direction of infall determined by the cosmic web steadily impacts the shape of the central and globally that of the halo, but not the core of the halo, the shape of which is likely more determined by its state of relaxation, earlier infall and possibly by AGN feedback from the central. This possibly causes the galactic spin flips to lag behind the corresponding halo flips (see Fig.~\ref{fig:sketch}).


To test that the influence of the cosmic web is indeed dominant in shaping the aforementioned alignments, let us now focus on the right panel of Fig.~\ref{fig:halogal}. In this panel, the sample is restricted to haloes with centrals that have a spin axis broadly aligned with their nearby filament ($\nu_{\rm c}=\cos\alpha_{\rm c}>0.6$ i.e. $\alpha_{\rm c}<50^{\rm o}$ ), hence a galactic plane orthogonal to it, hence to the preferential direction of infall. This time the signals are very different: the outermost shell  of the halo ($R_{\rm DM}> 2\; R_{\rm vir}$) is statistically not anymore  aligned with the central shape. In fact there are even hints of a trend for this shell's minor axis to be orthogonal to the galaxy minor axis, which emphasises once again the dominance of the filament DM budget in the halo on such scales. In this case, the inner shells ($R_{\rm DM}<\; R_{\rm vir}$) of the halo are on average better aligned with the central galaxy than the outer shells. Although this signal is faint ($\Delta \xi \approx 0.06-0.09$) and the PDF remains close to a random distribution, this suggests a partial alignment trend between the inner halo and centrals which is not related to their respective alignments with the cosmic web. As galaxies with a spin aligned with their nearest filament are predominantly low-mass galaxies that have undergone few mergers and therefore retain the spin orientation they were created with (in circum-filament vorticity rich region), it could either be that the inner halo also retain some orientation from its earlier build-up, or a hint of the central's tidal influence on the inner halo.
bending the halo spin in a direction orthogonal to the filament
Finally, it should be noted that even in the case of maximal alignment probability between the central galaxy and its host halo's minor axis --  as was found for the outermost sphere $R_{\rm DM}>2\, R_{\rm vir}$ on the middle panel of Fig.~\ref{fig:halogal}, there is still a fairly high proportion of centrals misaligned from their host halo. Indeed, around $50\%$ of centrals have a minor axis misaligned with their host minor axis by more than $40^{\rm o}$ in that case, hence only half of the central-halo pairs are roughly aligned with one another. Although this is a significantly higher fraction than the $25\%$ expected for uniformly distributed orientations, this does not allow us   to assume centrals and haloes minor axis are aligned simply based on their respective masses.

\subsection{Spin correlations between haloes and central galaxies.}

\begin{figure*}
\center \includegraphics[width=1.9\columnwidth]{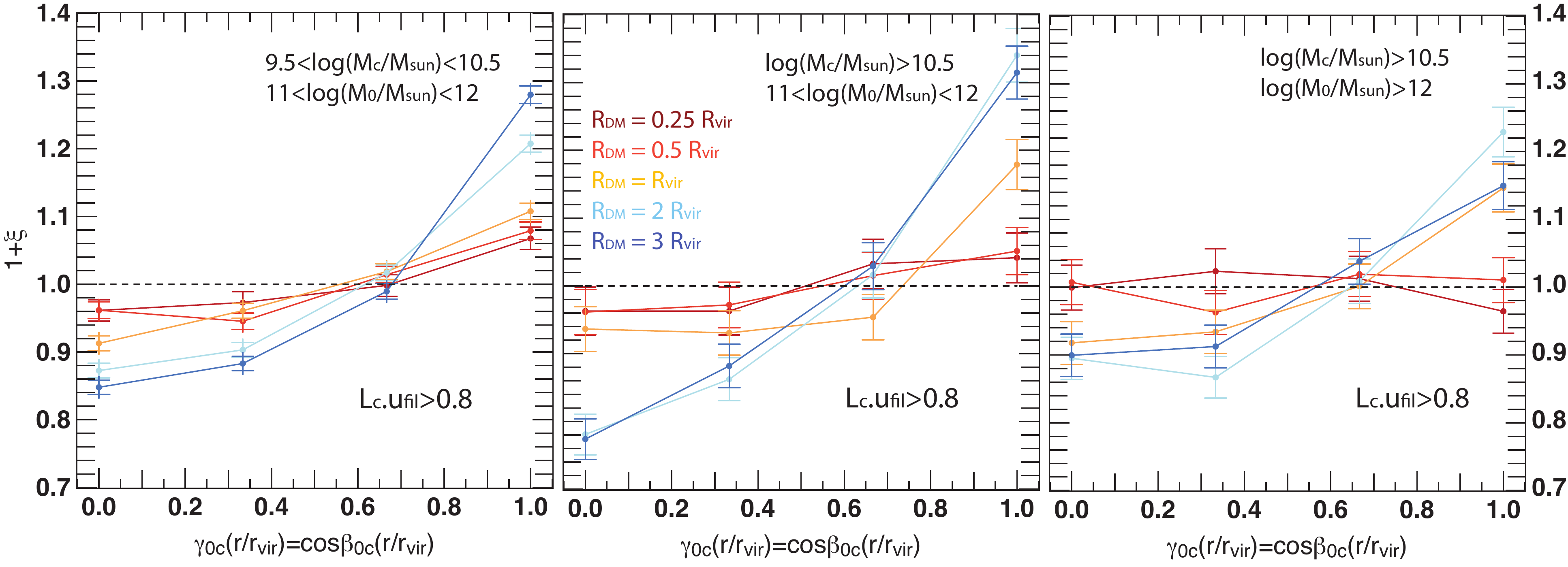}
\center \includegraphics[width=1.9\columnwidth]{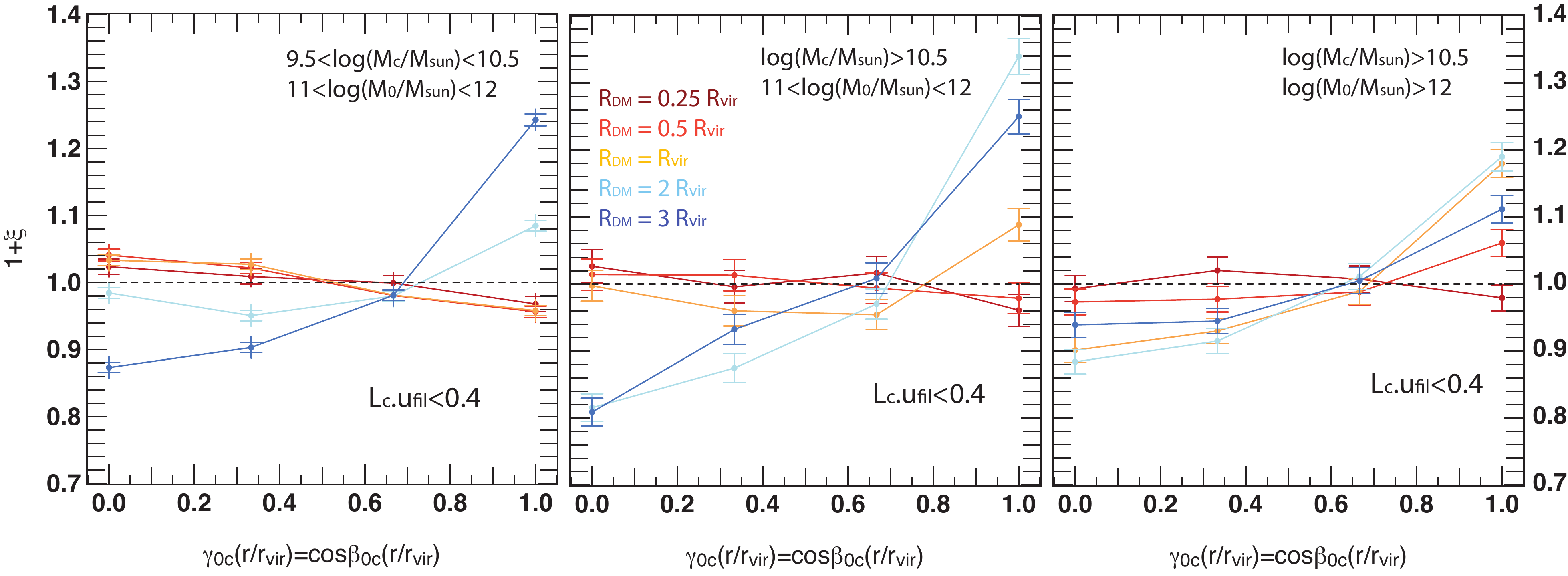}
 \caption{PDF of $\gamma_{0c}(r/r_{\rm vir})$, the cosine of $\beta_{0}^{C}$, the angle between the spin of central galaxies and that of their host halo, computed in spheres of increasing radius (dark red to navy blue curves). {\it Upper panels} : centrals with a spin aligned with their closest filament ($\alpha_{\rm c}<37^{\rm o}$), for three halo and central mass bins, from left to right: low-mass centrals $(10^{9.5}\, M_{\odot}< M_{\rm c}< 10^{10.5}\, M_{\odot})$ in low-mass haloes $(10^{11.5}\, M_{\odot}< M_{\rm 0}\!< 10^{12.5}\, M_{\odot})$, high mass centrals $(M_{\rm c}>10^{10.5}\, M_{\odot})$ in low-mass haloes and high mass centrals in high mass haloes $(M_{\rm 0}>10^{12.5}\, M_{\odot})$. {\it Lower panels}: Same centrals with a spin clearly misaligned with their closest filament ($\alpha_{\rm c}>67^{\rm o}$).
}
\label{fig:spinhg}
\end{figure*}

Let us now investigate the correlations between the orientation of the spin of the host halo and that of the central galaxy. Fig.~\ref{fig:spinhg} displays the PDF of  $\gamma_{0c}(r/r_{\rm vir})$, the cosine of $\beta_{0}^{C}$, the angle between the spin of the halo and that of its central galaxy. The spin of the halo is computed in spheres of increasing radius from its centre of mass,  as was  measured in the previous sections (dark red to navy blue curves). The upper panels focus on the sample of haloes that have centrals with a spin well-aligned with their closest filament ($\alpha_{\rm c}<37^{\rm o}$), and displays the PDF of  $\gamma_{0c}(r/r_{\rm vir})$ for three different halo and central mass bins, from left to right: low-mass centrals $(10^{9.5}\, M_{\odot}< M_{\rm c}< 10^{10.5}\, M_{\odot})$ in low-mass haloes ($10^{11.5}\, M_{\odot}< M_{\rm 0}\!< 10^{12.5}\, M_{\odot}$), high mass centrals ($M_{\rm c}>10^{10.5}\, M_{\odot}$) in low-mass haloes and high mass centrals in high mass haloes ($M_{\rm 0}>10^{12.5}\, M_{\odot}$)\footnote{Results for low-mass centrals in high mass haloes are not displayed due to the poor statistics of this sample.}. The lower panels show the PDF for similar mass bins, this time focusing on centrals with a spin clearly misaligned with their closest filament ($\alpha_{\rm c}>67^{\rm o}$).

Focusing on centrals with their {\it spin aligned } to their nearest filament (upper panels) one finds that:
\begin{itemize}
\item Low-mass centrals in low-mass haloes show spins alignment,  with however a decrement by a factor $\approx 3-4$ from the outskirt to the innermost parts of the halo. This is consistent with the build-up of young haloes and galaxies from vorticity rich filaments that align their spin with the cosmic web, and with the low degree of relaxation of the halo's outskirts which better trace the cosmic environment surrounding the halo.
\item Massive centrals in low mass haloes show a slightly stronger tendency for spin alignment on the Virial scale and above but little variation compared to their low-mass counterparts on sub-Virial scales. The tidal impact of the central is therefore strengthened since the central to halo mass ratio is increased. 
\item Massive centrals in massive haloes show a decrease of the tendency of spins to align on all scales. Indeed, in such cases, although the central spin alignment suggests a low merger rate -- hence a halo still far away from nodes, the higher mass of the halo suggests that it has already migrated closer to the spine of the filament. The tidal torques from the halo are therefore progressively changing direction and start bending the halo's spin more orthogonal to its host's filament.
\end{itemize}
%
%
Observing centrals with their spin orthogonal to their nearest filament (lower panels) yields the following conclusions:
\begin{itemize}
\item Low-mass centrals in low-mass haloes show a faint yet distinct tendency to present a spin orthogonal to that of the halo on all scales within $R_{\rm vir}$. This is consistent with previous findings that the core of such haloes tend to display a spin reminiscent of their earlier build-up in vorticity quadrants, hence parallel to the nearest filament. However, the spin of low-mass centrals is on the contrary well-aligned with the outskirts of their host halo. Once again, this is consistent with their progressive drift along the cosmic web and consecutive preferential accretion along the filament. 
\item Massive centrals in low mass haloes start to display alignments with their halo at the Virial scale. Alignments with the halo's outskirts are amplified. This reflects their evolution through cosmic drift: such massive centrals have usually started undergoing significant mergers.
\item Massive centrals in massive haloes tend to have their spin aligned with their DM host on virtually all scales except in the innermost, most relaxed core of the halo, where the PDF is compatible with random relative orientations of the galaxy and its host' spins. Note that in this later case, the alignment signal in the outermost parts of the halo is actually decreased as this population of massive hosts contains more haloes connected to multiple contrasted filaments, all of which significantly feed the halo with satellites in distinct directions.
\end{itemize}
These trends reflect an incremental dynamical build-up for haloes and their central galaxy as they change location within the cosmic web, hence modify the geometry of their accretion. In this scenario, while the overall outcome is similar for both haloes and their central galaxy (spin flips orthogonal to the filament at high mass), misalignments arise from distinct evolutionary timescales due to the specific physical processes at stake when haloes or galaxies merge. While the slow dilution of sub-haloes in their host through dynamical friction and collision-less relaxation generates a shell-like evolution of the spin in haloes, the preservation of satellite stellar discs down to the core of haloes, merging with the central through violent relaxation and dissipation through gas shocks (even during minor mergers) generate sharper spin flips for central galaxies, which bend their spin orthogonally to their closest filament later than their host as a whole, but before its core is significantly impacted\footnote{Note that in this latter case, AGN feedback might play a role in freezing the spin of the central's stellar disc as has been previously suggested by \cite{Dubois14}. This will require further investigation, including comparison with Horizon-noAGN a simulation with similar characteristics than Horizon-AGN but with AGN feedback switched off.
}. %
These results highlight the fact that misalignments between the core of host haloes and their central galaxy are common place using either the minor axis or the spin as a tracer of the structure's shape. In fact, even in configurations where a trend for alignment is detected, it is much fainter than the corresponding trend derived from the outskirts of the halo. Central-halo shape alignments on sub-Virial scales are poor in every configuration (in terms of cosmic web environment and mass ratio) compared to the strength of the signal obtained for the alignment of satellites in the central galactic plane on similar scales ($\Delta \xi <0.15$ at best compared to $\Delta \xi_{\rm sat} \approx 1$). This suggests that torques from the central galaxy and relaxation within the halo have at least partially competitive effects on the angular distribution of satellites, including in the inner core of the halo.

Figs.~\ref{fig:satsphere} and~\ref{fig:dxi_sat0}  showed that on large scales satellites lay orthogonal to their host's minor axis (and actually align with their host's major axis), but Paper I suggested this is mostly due to its  corresponding alignment with the embedding filament. In this region the distribution is unrelaxed and cannot be used to derive the shape of the halo. But the impact of the cosmic web fades off, as satellites  enter deeper layers of their host and relax. However, in the core of the halo, torques from the central galaxy are actually dominant in shaping the satellites' distribution. Since on such scales the central and the inner halo display strong misalignments, one expects a transition from a halo-impacted satellite distribution to a central-impacted distribution at around the Virial scale. This is however not always the case as alignments between centrals and their host's halo strengthen with halo mass. In massive groups and clusters, the net effect might then be enhanced alignments due to the cumulative effects of the halo and the central. This should limit one's ability to reliably measure a cluster's ellipticity solely from the distribution of its satellites. This point is developed in the next section.

\section{Coupling  alignments around central  and  halo. }
\label{section:virialshell}

This section  first  investigates how satellite alignment  -- around both the central minor axis (i.e using $\mu_{\rm c}$ to compute $\Delta \xi_{\rm sat}$) and the host halo minor axis (i.e using $\mu_{\rm 0}$ to compute $\Delta \xi_{\rm sat}$) -- are impacted by the central galaxy's shape and orientation up to $1\, R_{\rm vir}$.
It  then focuses on the Virial shell
to identify a range (in terms of halo mass and central shape) where the angular distribution of satellites likely traces better the halo shape than the galactic plane.  Finally it tests whether the satellite distribution in this range can be reliably used to estimate the host halo's ellipticity, and quantifies the loss of precision on this measurement due to secondary alignments with the central.

\subsection{Competitive alignments: Impact of  shape and orientation}

\begin{figure*}
\center \includegraphics[width=1.4\columnwidth]{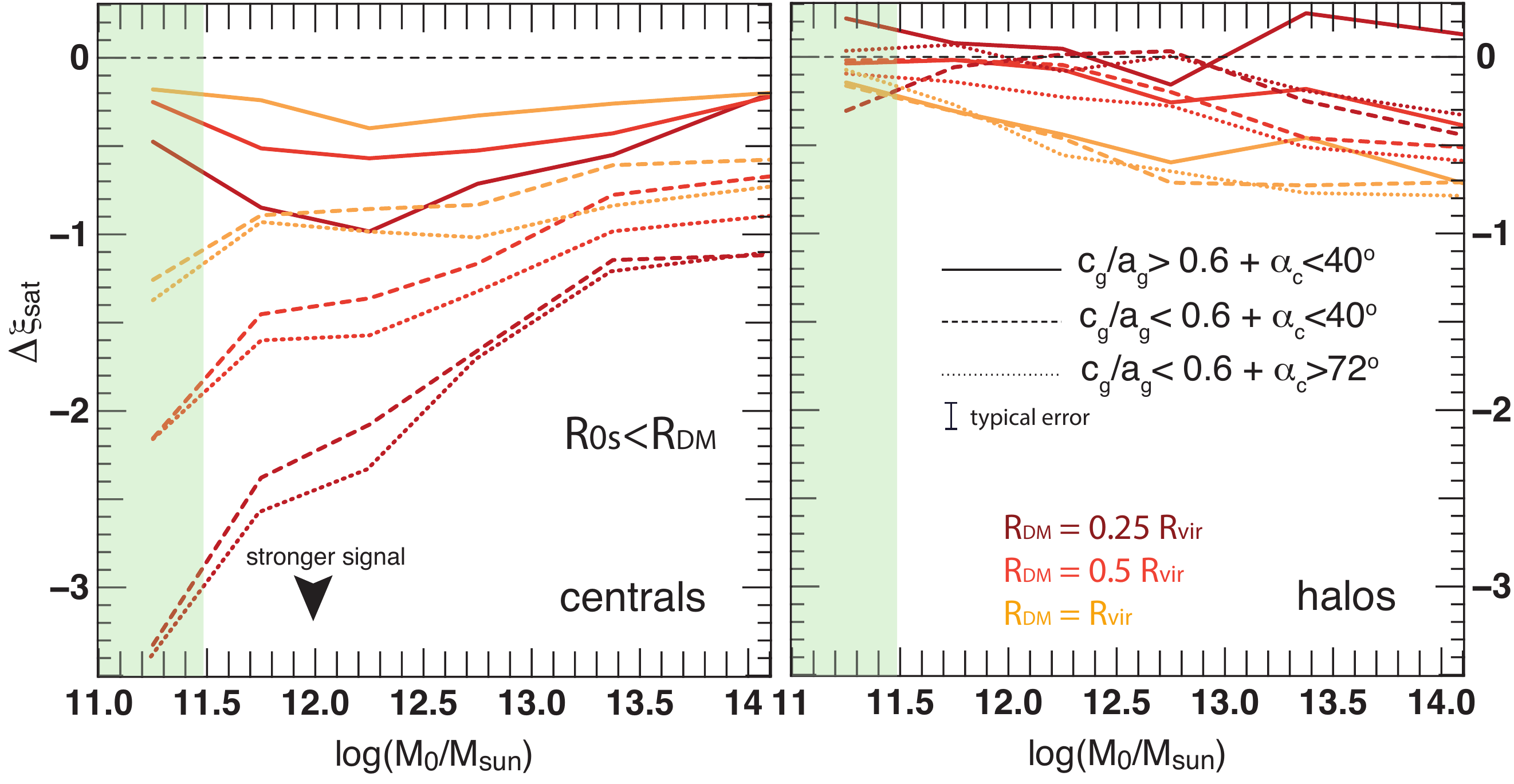}
 \caption{Evolution of $\Delta \xi_{\rm sat}$, the magnitude of the alignment of satellites with the minor axis of either the central galaxy (left panel) or the halo (right panel), as a function halo mass. This measurement is performed for satellites and DM material within three spheres of increasing radius from $0.25\, R_{\rm vir}$ (dark red curve) to $1\, R_{\rm vir}$ (orange curve), and three cuts in terms of central shape and orientation: centrals with a minor axis aligned with their nearest filament ($\alpha_{\rm c}<40^{\rm o}$) and a spheroidal shape ($c_{\rm g}/a_{\rm g}> 0.6$) (solid lines), centrals with their minor axis aligned and a discier shape ($c_{\rm g}/a_{\rm g}< 0.6$) (dashed lines), and centrals with their minor axis more orthogonal to their nearest filament $\alpha_{\rm c}>72^{\rm o}$ and a discier shape (dotted lines). Discier centrals trigger stronger alignments around both their galactic plane and the halo shape but this effect fades away in the cluster regime. The effect of the cosmic web is fainter in comparison.}
\label{fig:dxall}
\end{figure*}

Fig.~\ref{fig:dxall} displays the evolution of $\Delta \xi_{\rm sat}$ computed from $\mu_{\rm c}$ -- ``around centrals'' -- (left panel) and from $\mu_{\rm 0}$ --  ``around haloes'' -- (right panel) as a function of halo mass. It is computed for satellites and DM material within three spheres of increasing radii, from $0.25\, R_{\rm vir}$ (dark red curve) to $1\, R_{\rm vir}$ (orange curve), and for three different cuts in terms of central's shape and orientation: centrals that have a minor axis aligned with their nearest filament ($\alpha_{\rm c}<40^{\rm o}$) and a spheroidal shape ($c_{\rm g}/a_{\rm g}> 0.6$) as solid lines, centrals with their minor axis aligned with the filament and a diskier shape ($c_{\rm g}/a_{\rm g}< 0.6$) as dashed lines, and centrals with their minor axis more orthogonal to their nearest filament $\alpha_{\rm c}>72^{\rm o}$ and a diskier shape as dotted lines. 

 Let us first focus on the solid and dashed lines on the left and right panels. Central galaxies in these samples have their minor axis aligned with their nearest filament, hence display a galactic plane orthogonal to it. This configuration preferentially selects galaxies for which the cosmic infall of satellites is the least aligned with the central plane of the galaxy. The effect of the central shape is nonetheless clear: the tendency of satellites to bend within the central galactic plane is much stronger around the diskiest centrals. Even on the Virial scale ($R_{\rm DM}=R_{\rm vir}$) and across a wide range of halo masses, the signal is on average twice stronger around centrals with a low minor-to-major axis ratio ($c_{\rm g}/a_{\rm g}< 0.6$) than around their counterparts with axis ratios closer to unity ($c_{\rm g}/a_{\rm g}>0.6$).  
On the other hand, no significant variation in the satellite alignment with halo minor axis is measured between haloes hosting spheroidal or disc galaxies.
This provides yet another piece of evidence that either tidal torques or past collimated (baryonic) anisotropic accretion bend satellites and their centrals in a common plane more efficiently than the inner halo. This could either be that tidal torques are stronger for anisotropic discs or that the morphology of the central was itself imprinted by the anisotropy and strength of past inflows that fed its disc and brought the satellites in.
It seems however that the impact of the galactic plane on satellites decreases with halo mass, most likely because the two central shape bins are chosen very broad for the sake of statistics, hence cannot not capture well the lower amplitude evolution of central shape with halo mass within one bin, in particular the progressive disappearance of central discs with large central-to-halo mass ratios.
    
The filamentary infall from the nearby filament has a detectable yet much fainter impact on alignments below the Virial radius: comparing the signal around ``disc'' centrals with their closest filament reasonably aligned with their galactic plane ($\alpha_{\rm c}>72^{\rm o}$) to those with their galactic plane more orthogonal to it  ($\alpha_{\rm c}<40^{\rm o}$),  notice a systematic enhancement of the signal around centrals, by 10 to $20\%$ in the former case. The effect is strongly reduced  around haloes, which generally show much better alignment of their major axis with the nearest filament, irrespective of the orientation of their central galaxy.
 
Notice that below the Virial scale - thus for all three measurements - the signal around haloes is systematically fainter than that around all kinds of central galaxies. Even at the Virial scale (orange lines), only satellite alignments around haloes which host the most spheroidal centrals (solid orange line) clearly take over the alignment  around their central (twice stronger  for $M_{\rm c}> 10^{12.5}\, M_{\odot}$). However, focusing on all orange lines, let us observe that on the Virial scale, the amplitude of the alignment  around haloes and their central becomes close to comparable, even for central discs (dashed and dotted lines).  Given that the signal around centrals decreases with distance while the signal around haloes increases, one can expect to find an intermediate area around the Virial sphere where the halo signal is distorted neither by the central influence nor by the filamentary infall, at least for some range of central morphologies.

\subsection{Competitive alignments in the Virial shell}

\begin{figure*}
\center \includegraphics[width=1.4\columnwidth]{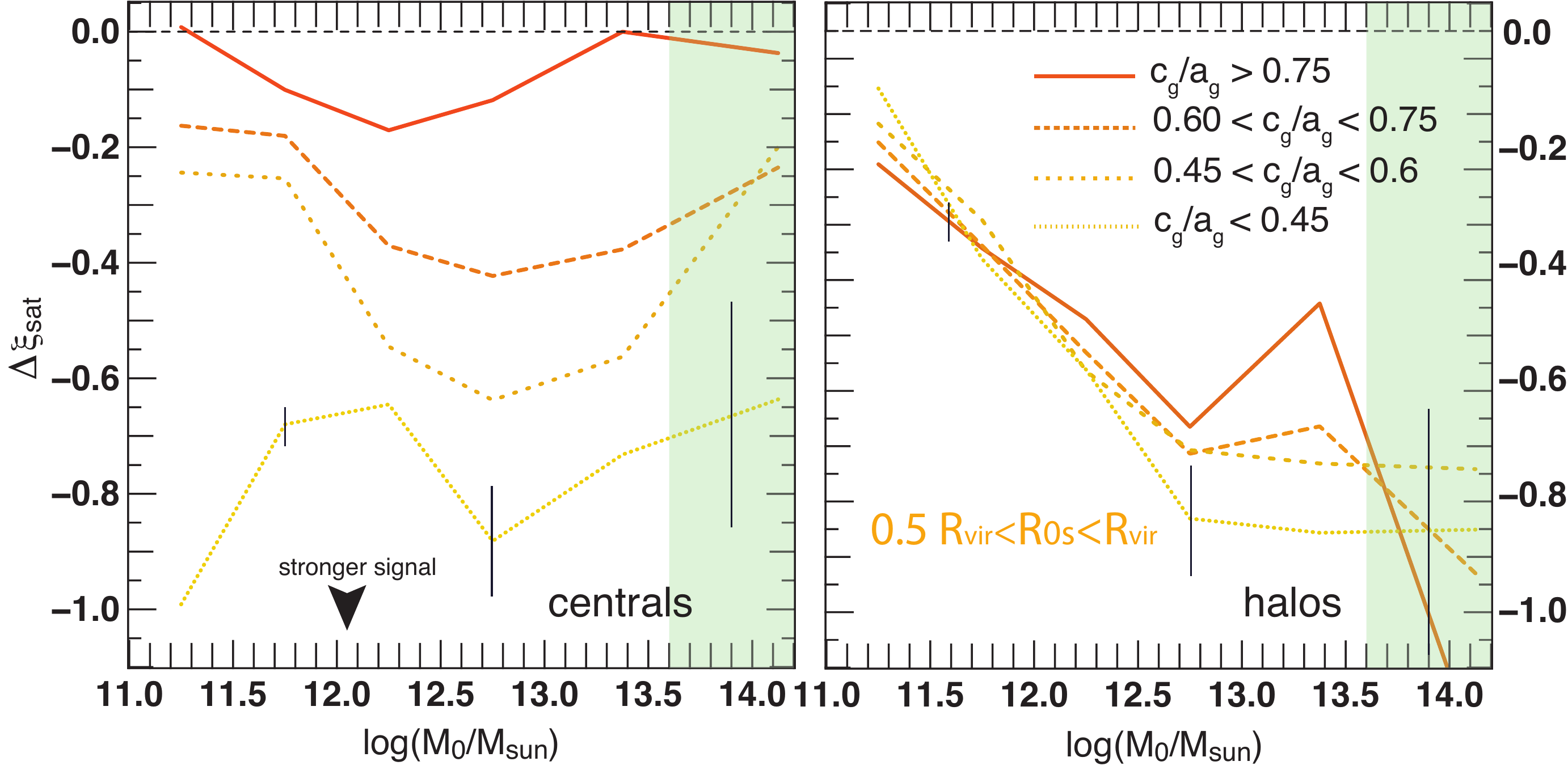}
 \caption{Evolution of $\Delta \xi_{\rm sat}$, the magnitude of the alignment of satellites with the minor axis of either the central galaxy (left panel) or the halo (right panel), as a function halo mass. Contrary to Fig.~\ref{fig:dxall}, the satellites are taken within a ``Virial shell'' (material within $0.5\, R_{\rm vir}< r < R_{\rm vir}$), for four bins of central minor-to-major axis ratios from $c_{\rm g}/a_{\rm g}<0.45$ (light yellow dotted lines) to $c_{\rm g}/a_{\rm g}>0.75$ (dark orange solid lines), respectively from disc-like to spheroid-like shapes.}
\label{fig:dxshell}
\end{figure*}

Let us focus now on the previously defined {\it Virial shell} ($0.5\, R_{\rm vir}< r < R_{\rm vir}$), and investigate further the effect of the central morphology on the alignment signal within that shell. Fig.~\ref{fig:dxshell} displays the evolution of $\Delta \xi_{\rm sat}$ computed around centrals (left panel) and haloes (right panel) with halo mass, following Fig.~\ref{fig:dxall}, but this time focusing solely on satellites in the Virial shell, for four refined bins of increasing central minor-to-major axis ratios, from $c_{\rm g}/a_{\rm g}<0.45$ (light yellow dotted lines) to $c_{\rm g}/a_{\rm g}>0.75$ (dark orange solid lines).
Note that the minor axis of the halo is still computed on all DM material within $ R_{\rm vir}$ so that the ellipsoid approximation remains valid to derive the axis ratios.
 
Even within the Virial shell, the impact of the central ellipticity is striking: $\Delta \xi_{\rm sat}$ around centrals increases by factor 7-8 from spheroidal centrals ($c_{\rm g}/a_{\rm g}>0.75$) to clear stellar disks ($c_{\rm g}/a_{\rm g}<0.45$), while one does not observe such sharp variations around haloes. Moreover, this effect is not strongly mass dependent, apart for a small drop around the spin transition mass $ \approx 10^{12.5}\, M_{\odot}$, which might be a trace of the asynchronous flips between centrals and haloes.
One of its effects on the halo's signal is however still visible, at least within the $10^{12}\, M_{\odot}< M_{\rm c}< 10^{13.5}\, M_{\odot}$ mass range where it is sufficiently resolved: it is up to $35\%$ stronger around haloes hosting disky centrals than around those hosting spheroids. Expectedly, this effect tends to increase with the mass of the central. 
However, for most massive groups and clusters with $ M_{\rm 0}>\,10^{13.2}\, M_{\odot}$, the alignment  with the halo's shape becomes clearly dominant for sufficiently spheroidal centrals with $c_{\rm g}/a_{\rm g}>0.45$, and the alignment  in the galactic plane even becomes negligible for $c_{\rm g}/a_{\rm g}>0.75$. 

\subsection{Discussion}
\label{section:discussion}

Can one straightforwardly conclude that the overall effect of the satellite correlations with the central is a simple enhancement of the alignment with the halo's shape, compared to the case without a central galaxy?  It is not  straightforward, as significant degrees of misalignment between centrals and inner haloes were found in this study. The Virial shell may indeed well be a zone of transition from alignment around the central to alignment with the halo. In which case, the integrated effect in such a zone might well be a decrement in alignment  around the host halo compared to the central-free case, resulting in an underestimation of the halo's ellipticity. Conversely, if centrals and haloes are well aligned, the central may indeed amplify the measured alignments with haloes' shape, even within the range where the halo's contribution  dominates. This is a particular concern for massive groups and clusters.  Indeed, in this mass range where surrounding cosmic filaments are highly contrasted and where central galaxies are more evolved and strongly merger-dominated, host haloes and centrals show much stronger shape alignment  at the Virial scale. This is related to the mutual strong alignment of their major axis/galactic plane with their nearest cosmic filament. In this range, direction of infall, gravitational torques from haloes, centrals and cosmic structures are not expected to compete but add up to one another, therefore enhancing the perceived alignment of satellites orthogonally to their host minor axis. In order to explicitly test for the effect of central alignments on the estimation of a halo's ellipticity, one needs to relate the ellipticity (or minor-to-major axis ratio in this study) computed directly from dark matter particles to that estimated from the system of satellites alone.

\begin{figure}
\center \includegraphics[width=1\columnwidth]{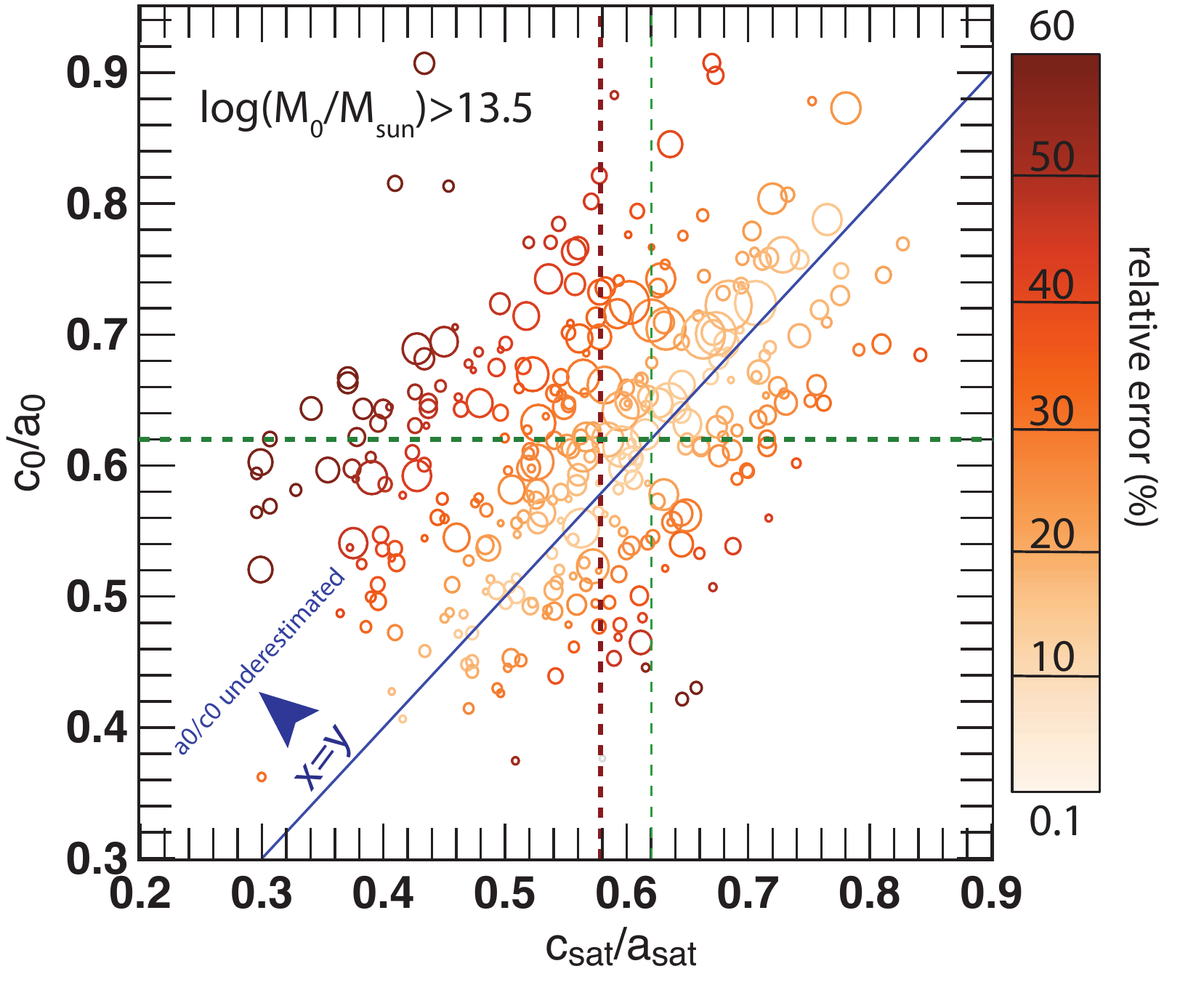}
 \caption{Comparison of minor-to-major axis ratios computed from the DM particles of haloes within $R_{\rm vir}$, $c_{\rm 0}/a_{\rm 0}$, and from their system of satellites within $R_{\rm vir}$ only, $c_{\rm sat}/a_{\rm sat}$. Only the most massive haloes with $M_{\rm 0}>10^{13.5}\, M_{\odot}$ are considered here. The color scale encodes the absolute value of the relative error when equating the axis ratio of satellites to that of the halo. Green dashed lines indicate the median axis ratio of the DM haloes and the dark red dashed line the median axis ratio of their systems of satellites. The size of circles varies linearly with the mass of haloes from $10^{13.5}\, M_{\odot}$ to $10^{14.6}\, M_{\odot}$. Differences in mass of satellites are not taken into account. Using the axis ratio of the system of satellites to infer that of the halo leads to an underestimation in $66\%$ of cases, with an average error of $7\%$, reaching more than $10\%$ in $43\%$ of cases, and more than $20\%$ in $27\%$ of cases.
}
\label{fig:sataxis}
\end{figure}

Fig.~\ref{fig:sataxis} shows the direct comparison of minor-to-major axis ratios computed from the DM particles of haloes within $R_{\rm vir}$, $c_{\rm 0}/a_{\rm 0}$, and from their system of satellites within $R_{\rm vir}$ only, $c_{\rm sat}/a_{\rm sat}$. Only the most massive haloes with $M_{\rm 0}>10^{13.5}\, M_{\odot}$ are considered here. The low number of clusters with  $M_{\rm 0}>10^{14}\, M_{\odot}$ limits the number of satellites per halo, and consequently the precision of $c_{\rm sat}/a_{\rm sat}$ computed from the inertia tensor of satellites. The inertia tensor is in particular more sensitive to the (less populated) outskirts of the system than $\Delta \xi_{\rm sat}$. The variation in mass of satellites is therefore discarded in the calculation- as is often the case in observations - to avoid overweighting a given satellite, and all satellites within $R_{\rm vir}$ are considered. This also limits the overestimation of the halo's ellipticity as most massive satellites are usually more strongly aligned with their central galaxy. The color scale encodes the absolute value of the relative error when equating the axis ratio of satellites to that of the halo. The green dashed lines indicate the median axis ratio of the DM haloes and the dark red dashed line the median axis ratio of their systems of satellites. The size of circles varies linearly with the mass of haloes from $10^{13.5}\, M_{\odot}$ to $10^{14.6}\, M_{\odot}$. The median values  are $\langle c_{\rm sat}/a_{\rm sat}\rangle =0.58\pm 0.12$ and $\langle c_{\rm 0}/a_{\rm 0}\rangle =0.62\pm 0.1$. One can notice that using the axis ratio of the system of satellites to infer that of the halo generally leads to an underestimation of the axis ratio, hence an overestimation of its ellipticity. This is indeed the case for $66\%$ of massive haloes, with an average error of $7\%$. The error is however reaching more than $10\%$ in $43\%$ of the cases, more than $20\%$ in $27\%$ of the cases, and more than $40\%$ in $5\%$ of  the cases.

The statistics to perform this comparison in Horizon-AGN are arguably limited given the rather low number of massive groups and the sensitivity of the inertia tensor to outliers given their low number of satellites. To confirm this analysis,  Appendix~\ref{section:ellipticity}  produces large samples of synthetic haloes with constrained minor-to-major axis ratios (radially varying or not) which are populated with satellites under the assumption that they simply follow the local shape of their host halo.   This construction is repeated for various radial concentrations marginalising over the number of satellites following the distribution found in Horizon-AGN for haloes with $M_{\rm 0}> 10^{13.5}\, M_{\odot}$.  The quantity $\Delta \xi_{\rm sat}$ is  then computed for each of this halo (less sensitive to outliers than the inertia tensor), computed with respect to the halo's minor axis. This allows us to establish a theoretical relation between measured $\Delta \xi_{\rm sat}$ and underlying values of $c_{\rm 0}/a_{\rm 0}$ if satellites simply trace the shape of their host, and to test them against the values of $\Delta \xi_{\rm sat}$  found in Horizon-AGN. Independently of the radial concentrations considered, this confirms that $\Delta \xi_{\rm sat}$ values measured around massive groups in Horizon-AGN are higher than expected due to additional satellite alignments with the central galaxy, hence lead to overestimations of $c_{\rm 0}/a_{\rm 0}$ by around $10\%$ on average with errors for individual haloes frequently reaching $40\%$.

\section{Implication for local planes of satellites}
\label{section:MW}

It has long been observed in local surveys that satellites of nearby galaxies tend to orbit their hosts in thin, extended and possibly co-rotating planes. Polar parallel planes of satellites have been described around both M31 \citep{Ibata13,Shaya13} and Centaurus A \citep{Tully15}, and a slightly bent plane was also found around the Milky-Way. It has also been (surprizingly) argued that such distributions were incompatible with $\Lambda$CDM, but recent studies suggest that taking into account both the more focalised infall of baryons and cosmic large scale structure may reconcile  observations with theory \citep{Cautun15}. The following section makes use of available data to assess these alignments with respect to the scenario developed in the present study, through the analysis of hydrodynamical simulations.

\begin{figure}
\center \includegraphics[width=0.9\columnwidth]{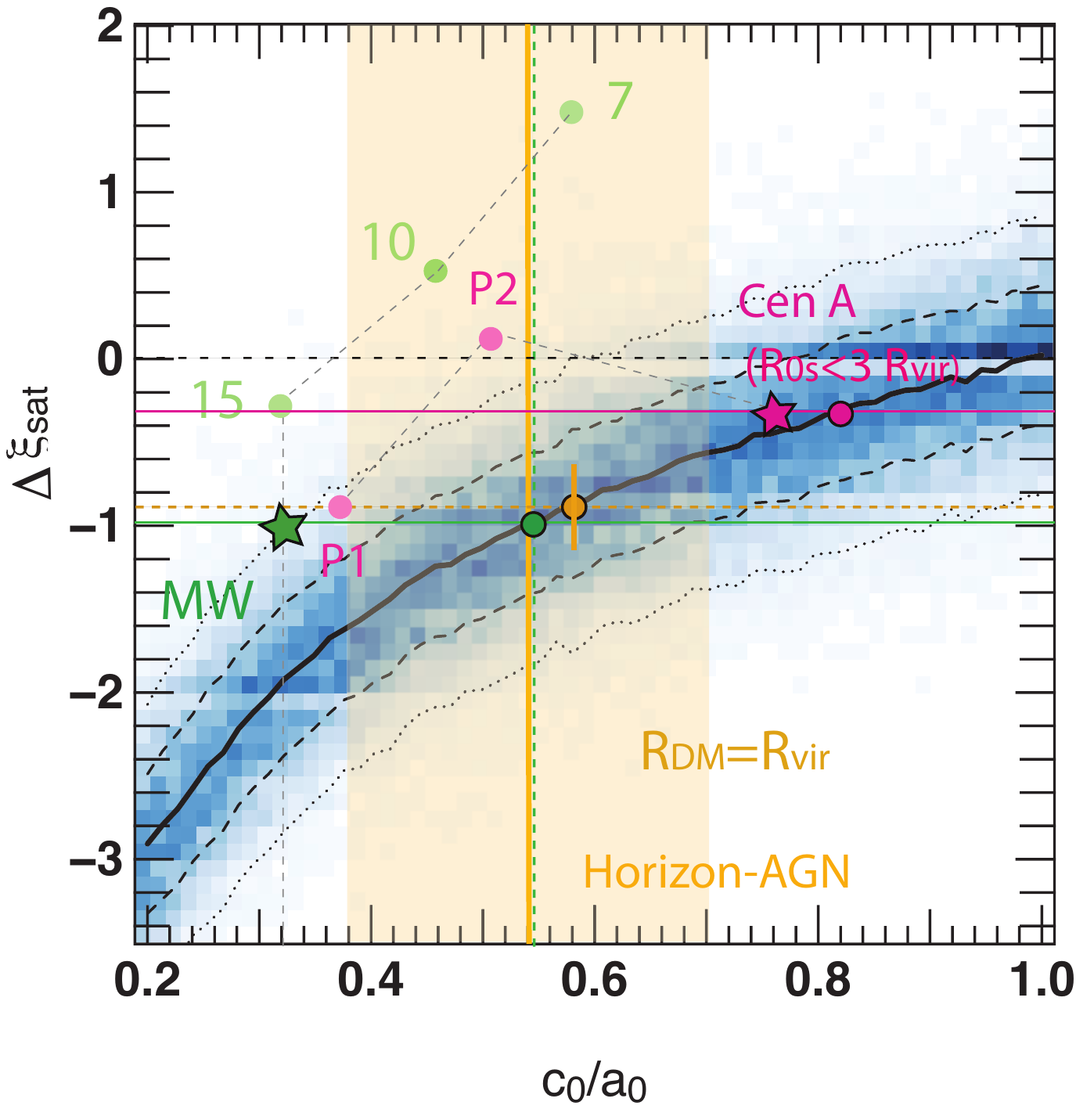}
 \caption{Average evolution of $\Delta \xi_{\rm sat}$ in the Virial radius with $c_{\rm 0} /a_{\rm 0}$ (solid line), and $1-\sigma$ and $2-\sigma$ contours (dashed and dotted lines respectively) drawn from 1000 realisations of satellite distributions per axis ratio bin, for 100 such bins, sampling 20 to 60 satellites per halo to mimick typical number of objects used to trace alignments with the MW. The solid horizontal green line indicates  $\Delta \xi_{\rm sat}^{\rm MW}$  for the Milky Way  equating the minor axis of the halo to that of the total system of satellites. The green star indicates the minor-to-major axis ratio of the system of satellites, the green dot the most likely halo minor-to-major axis ratio corresponding to $\Delta \xi_{\rm sat}^{\rm MW}$. The vertical orange line shows the average axis ratio for haloes with $10^{11.8}\, M_{\odot} <M_{\rm 0}<10^{12.3} \, M_{\odot}$ in Horizon-AGN, with the 16th to 84th percentile region shaded in orange. Light green dots indicate $\Delta \xi_{\rm sat}^{\rm MW}$ for the N innermost satellites only, with N from 7. The dashed horizontal orange line (with poissonnian error bars on the mean) shows the average stacked $\Delta \xi_{\rm sat}$ for the $50\%$ most elongated haloes with $10^{11.8}\, M_{\odot} <M_{\rm 0}<10^{12.3} \, M_{\odot}$ and with misalignment with their cosmic filament limited to $37^{\rm o}$.}
\label{fig:sampleMW}
\end{figure}

\begin{itemize}
\item {\bf Orientation:} Planes of satellites around Andromeda and the Milky Way galaxies are found to be fairly parallel to one another and orthogonal to the galactic plane of the central they orbit. In this scenario, the mass range in which haloes of such galaxies are found ($10^{11.8}\, M_{\odot} <M_{\rm 0}<10^{12.3} \, M_{\odot}$) -- i.e. just below the spin flip transition mass -- combined with rather high stellar masses around $10^{10.5}\, M_{\odot}$ points towards galactic planes expected to lie orthogonally to their nearest cosmic filament (MW) or starting to bend towards it (M31) but close to its saddle point and spine, which, therefore, provides a preferential direction of halo stretching, accretion and satellite infall (2nd to 3rd stage in Fig.~\ref{fig:sketch}). Recent cosmic flows study by \cite{Libeskind15} suggests that it is indeed the case, clearly showing that the flow towards the Virgo Cluster, as well as the local filament that bridges the Local Group with the Virgo Cluster, exhibits clear features in both the density and velocity fields, along which planes of satellites align. MW and M31 are found lying in the saddle point region, less than 500 kpc away from the filament spine. Such planes of satellites are therefore easily interpreted as arising from steady infall from the filament and little to no orbit orientation induced by the halo shape, as it is most likely aligned with the filament. 
\item {\bf Strength}: Fig.~\ref{fig:sampleMW} shows the expected average evolution of $\Delta \xi_{\rm sat}$ with $c_{\rm 0}/a_{\rm 0}$ within $R_{\rm vir}$ (solid line), and $1-\sigma$ and $2-\sigma$ contours (dashed and dotted lines respectively) if satellites trace the local shape of their host. Such synthetic distribution is drawn from 1000 realisations of satellite distributions per axis ratio bin, for 100 such bins (sampling 20 to 60 satellites per halo to mimic the typical number of satellite objects used to analyse alignments around Local galaxies). These  are obtained for uniformly populated ellipsoids. Results tailored to the radial distribution of satellites in the MW can be found in Appendix~\ref{subsection:MWrad}. The solid horizontal green line indicates the value $\Delta \xi_{\rm sat}^{\rm MW}$ found for the Milky Way using the 41 satellites identified in \cite{Pawlowski15} and equating the minor axis of the halo to that of the total system of satellites. The green star indicates the minor-to-major axis ratio of the system of satellites and the black circled green dot the most likely halo minor-to-major axis ratio corresponding to $\Delta \xi_{\rm sat}^{\rm MW}$. The vertical orange line shows the average axis ratio for haloes with $10^{11.8}\, M_{\odot} <M_{\rm 0}<10^{12.3} \, M_{\odot}$ in Horizon-AGN, with the 16th to 84th percentile region shaded in orange. One can see that  while the system of satellites might seem elongated, the most likely halo shape is still compatible with the average among haloes of similar mass in Horizon-AGN. The tension with  $\Delta \xi_{\rm sat}$ stacked over the $50\%$ most elongated haloes of similar mass, with misalignment with their cosmic filament limited to $37^{\rm o}$ (dashed horizontal orange line with Poissonnian error bars on the mean, see also Appendix~\ref{section:haloeshape}) is limited to $\Delta \xi \approx 0.05$. Taking into account the fact that the MW signal  is compared with a stacked signal, equating the minor axis of the MW system of satellites to that of the halo, this value seems compatible with cosmic web enhancement inducing strong alignments between satellites and both the large scale structure and the elongated halo, with little impact of the central. 
\item {\bf Effect of the central}: At first sight, the impact of the central seems low compared to some of the transitions observed in Horizon-AGN. While this is consistent with  the preservation of strong orthogonal orientation  w.r.t. the galactic plane, it may seem unexpected. But recall  that one is not observing a signal stacked over centrals in this case, but the distribution around a {\sl single} central. This sharply decreases the statistics for satellites and it becomes difficult to resolve the innermost parts of the system. However, looking at the innermost satellites of the MW, hints of a transition towards alignment with the galactic plane can be found. Light green dots on Fig.~\ref{fig:sampleMW} indicate values of $\Delta \xi_{\rm sat}^{\rm MW}$ around the minor axis of the full system obtained for the N innermost satellites only, with N from 7  to 15. These satellites seem to align with the overall system minor axis, hence to bend in the galactic plane, since the planes of satellites of the MW lie roughly orthogonally to the galactic plane of the MW. Correlations with the central orientation seem however undetectable above 15 satellites ($\approx 60$ kpc from MW). This may be related to the type of  satellites. Note that most of the MW system consists of dwarves with $M_{\rm s}<10^{8}\, M_{\odot}$. Hence these satellites are not resolved in Horizon-AGN, as they correspond to satellite-to-central mass ratios lower than 0.002. While this regime  cannot be probed directly in this  simulation, a broader study of the impact of the satellite-to-central mass ratio on alignments is given in Appendix~\ref{section:satmass1}. It shows that while this ratio has little impact on alignments with the halo at the Virial radius, the alignments around the central on the same scale are strongly damped for the lighter satellites. The actual position of the MW close to the stretching spine of its filament, and the orthogonal orientation of its galactic plane  -- therefore minimising the coupling between halo/filament and central alignments,  may  explain  the limited impact of the MW on its population of satellites.
\item {\bf Comparison with Centaurus A}: Using data from \cite{Tully15},   $\Delta \xi_{\rm sat} - c_{\rm 0}/a_{\rm 0}$ was computed matching for the the full system of satellites around Centaurus A (pink star) and for the two significant planes identified among them, P1 the ``inner'' plane ($\langle R_{\rm 0s} \rangle \approx 300$ kpc) and P2 the ``outer'' plane ($\langle R_{\rm 0s} \rangle \approx 420$ kpc $\approx R_{\rm vir}$). In each case,  $\Delta \xi_{\rm sat}$ is computed around the minor axis of the full system of satellites. Results are expectedly different from the ones found around the MW. Cen A is a much more massive galaxy (up to $10^{12}\, M_{\odot}$) in a massive halo ($10^{12.8}-10^{13}\, M_{\odot}$). Here, such a halo is found away from the mid-filament region and the saddle point, closer to a more massive multi-connected node, most likely dynamically connected to more than one filament itself, and has undergone a significant merger history. The disturbed elliptical morphology of Cen A suggests it is indeed the case. Its location in the cosmic flows identified in \cite{Libeskind15} reveals that it is indeed closer to the Virgo cluster and further away from the saddle-point than the MW. Interestingly, it appears to be offset from the main cosmic filament spine, in high-helicity flows impacted not only by the MW filament but also by a more remote filament branching into the Virgo cluster as well. This suggests that Cen A is indeed in the outskirt of a multi-connected region. One can therefore expect it to have undergone variations in the preferential accretion direction and spin flips, and to exhibit a strong inertial twist (as defined in Section~\ref{section:shapes}). This might explain the perceived layout of satellites in two distinct planes, all the more that Cen A satellites are less concentrated and found significantly further away from the central than MW satellites, with about $65\%$ above $0.5\,R_{\rm vir}$ and more than $30\%$ above $R_{\rm vir}$, while the MW has around $75\%$ within $0.5\,R_{\rm vir}$ and virtually no one above $R_{\rm vir}$. While the position of the overall system of  satellites show in Fig.~\ref{fig:sampleMW} (pink star) points towards a round outer halo ($c_{\rm 0}/a_{\rm 0}\approx 0.76$), the ``inner plane'' P1 distribution of satellites is fully compatible with a Virial halo stretched by the cosmic web similar to that of the MW, and the ``outer'' plane P2 with a perturbed direction of alignment which could trace a shifted/secondary direction of infall and/or a previous orientation of the galactic plane.
\end{itemize}

Hence, a nuanced analysis implies that planes of satellites around local galaxies do not particularly stand out in our model. It is even remarkable that these distributions of satellites correlate so well with their expected location in the cosmic web on the basis of their central mass and morphology, as has now been observationally confirmed \citep{Libeskind15}. Notwithstanding, this preliminary study also highlights the requirement to combine cosmic volume, high spatial resolution and baryonic physics to reach definite conclusions on this matter. 


\section{Conclusion}
\label{section:conclusion}
In a series of two papers, we investigated the distributions of satellites around centrals in the cosmological simulation \hagn\!\!.

Paper I showed that the distribution of satellites in a halo transitions radially from an alignment with the nearest cosmic filament they are infalling from at large scales, to an alignment in the plane of the central galaxy within $R_{\rm vir}$. It did not however fully explain the origin of these alignments. In particular, it did not discriminate between friction and torquing from the innermost parts of the halo on the one hand, and  the specific effect of baryonic processes  on the other hand (amongst which the formation and buildup of the central within cosmic flows -- along with the inner satellites population, and possible re-torquing from the central disc onto the satellite's population). 

The  results presented in the present paper underline this distinction between dark matter only processes and baryons and highlight a radial variation of satellite alignments w.r.t. the DM halo's minor axis that differs strikingly  from satellite alignments w.r.t. the central galactic plane.
In order to  quantify the tendency of satellites to trace their host's structure, we first investigated this structure in details and followed the evolution of dark matter halo shapes,  from their innermost core to their extreme outskirts (where they dilute in their cosmic environment). We  then related the radial evolution of their ellipticity and net inertial twist (progressive bend of their major axis) to their inside-out cosmic assembly history, within the anisotropic metric of the cosmic web that defines their environment.  This allowed us to understand alignments (and misalignments) of satellite population with respect to their host halo's shape at different separations, and to compare them directly to  their alignment  with respect to  their central galactic plane. Our main findings are summarised hereafter (see also Fig.~\ref{fig:sketch}):
\begin{enumerate}
\item  In the outskirts of  haloes,  satellites are distributed  orthogonally to their halo's minor axis (and  aligned with its major axis); this is consistent with infall along the cosmic filament's direction -- also the direction in which the halo is elongated. It is also consistent with the fact that the halo is more elongated in its outer parts.
\item 
This alignment  fades  as one probes the inner parts of the halo: the stacked angular distributions of satellites around  their haloes' minor axis become nearly uniform in its  core.
 This is consistent with the fact that
the inner parts of haloes are typically rounder and less aligned with cosmic filaments than their outskirts.
\item The radial evolution of the geometric (shape) and dynamical (spin) properties of haloes is consistent with a cosmic evolution within the anisotropic cosmic web, involving halo spin flips (displaying an  transition, from parallel to the filament at low-mass to perpendicular  at high mass).
\item  The spin of outer shells  flips at smaller halo mass than that of the inner ones, which flip last. It reflects the earlier formation time of the halo core, offset from the spine of the cosmic filament, in region where consistent vorticity~\citep{laigle2015} build up the spin of young haloes parallel to the axis of their nearby filament (therefore favouring a  plane of accretion orthogonal to the filament), and where the anisotropy of the collapse (and stretching along the filament) is  reduced (compared to when it  reaches the spine of the filament). 
\item The trend for satellites to settle in the galactic plane strengthens as one probes deeper into the halo and dominates within $0.5 \,R_{\rm vir}$, which is the exact opposite of the satellite-DM alignement trend (point 2).
Since i) satellites do not accurately follows the shape on the DM inner core but rather the shape of stars of the central, ii) the ellipticity of the central is higher than that of the DM, and iii) the alignment of satellites is stronger around centrals with higher ellipticity (aka discs rather than ``ellipticals''), thus, it highlights the importance of baryon driven torques, either tidal torques from the central galaxy or shared direction of collimated accretion at earlier cosmic time, possibly through cold flows.
\item Alignment of  central galactic planes with their closest filament further enhances the signal  around centrals by up to $20\%$. While this effect is subdominant to the effect of the central's shape, satellite population around a disc-like central with a minor axis orthogonal to   the cosmic filament and aligned with the halo's minor axis are  more  strongly aligned to their host's filament, compared to populations around central discs oriented differently w.r.t. to filaments or haloes: this signal is three times stronger within $R_{\rm vir}$ for $10^{12.5}\, M_{\odot}<M_{\rm 0}<10^{13.5}\, M_{\odot}$. Note that this galactic plane-filament-halo alignment configuration is also  favoured statistically in this mass range, as it is above the transition mass for spin flips for haloes and galaxies likewise \citep{Codis12,Dubois14}, yet still in the range where haloes lie on the spine of one dominant filament rather than at the centre of a multi-connected node.
\item   The misalignments between the central galactic plane and the shape of the innermost part of the halo ($R<0.5 \,R_{\rm vir}$) are common at all mass scales, and lead to a competition between satellite alignments with the central on the one hand, and alignments with the inner halo's shape within this region on the other hand. In particular, the shape of the central galaxy has a significant impact on  alignments around haloes for halo masses between $10^{12}\,M_{\odot}$ and $10^{13.5}\, M_{\odot}$. Thus,  satellites in big galaxy groups do not  trace their host's shape well enough to allow for its precise estimation solely from their distribution. Our low statistics above that latter mass scale ($\approx 350$ objects above $10^{13.5}\, M_{\odot}$, and 24 objects above $10^{14}\, M_{\odot}$ {\it after stacking over redshift}) do not  yield a  good estimate for the impact of the central's shape on massive clusters (it yields important uncertainties, with average relative error of $10\%$, rising up to $30-40\%$ for individual haloes). This suggests that the distribution of satellites in clusters cannot  be used to derive the halo's minor-to-major axis ratio.
\item Finally the location of the Milky Way near the saddle point of its cosmic filament \citep{Libeskind15} together with its estimated halo mass (slightly below the spin transition mass)  explain the  orientation of its known satellites in a plane along the spine of the local cosmic web. While further studies at higher resolution   (resolving dwarf galaxies) will be necessary to draw definite conclusions, no significant discrepancy is found at this stage between the current data on the one hand, and Milky-Way mass systems of similar cosmic location in Horizon-AGN on the other hand.
\end{enumerate}

Overall, this  paper (together with paper I) quantifies the distribution of satellites in haloes as the ever-changing competition between dynamical processes with distinct evolution timescales, feeding on one another. 
The main driver for evolution is  anisotropic accretion imposed by the large scale structures,  continuously varying as haloes drift along the cosmic web. Initially, while centrals and inner older satellites retain the orientation inherited from their past accretion, the younger population of satellites follows the dynamics dictated by its more recent environment (the filament, then the halo). However, as they approach the central galaxy,   the stellar disc may apply  torques  onto the satellites distribution, which corresponds to  a retroaction/dynamical feedback from the disc's past accretion history. On even longer timescales (and higher mass scales), major mergers may eventually ``update'' the orientation of their central to its more recent environment, leaving satellites in its vicinity to seemingly relax into the halo, until they adapt again to their centrals's new orientation. This intricate interplay  is in stark contrast with the (naive) assumption that satellites simply trace adiabatically the shape of their halo host. 
 
Some of the findings presented in this paper will  require further investigations using high resolution runs that  resolve dwarf galaxies down to stellar masses below $10^{7}\, M_{\odot}$ (Dubois et al., in prep), in order to estimate how  alignments around Milky-Way like systems are impacted, and allow for a fair comparison to observations. This will  also allow us to quantify explicitely the relative torque from each component (halo, central, filament).
 Notwithstanding, the present analysis emphasises the importance of resolving cosmic structures well beyond $3\, R_{\rm vir}$ e.g. via zoom simulations, so as to reconstruct filamentary infall and derive reliable predictions on the anisotropic distribution of satellites.
As the ability of galaxies to ``freeze or evolve'' in these distinct phases is key to the dynamics of satellites, one may wonder  if internal physical  processes such as stellar and AGN feedback-- which have been argued to freeze the orientation of host galaxies~\citep[see][]{Dubois14}, have a significant impact on defining these trends.

\begin{acknowledgements}
Parts of this research were conducted via the Australian Research Council Centre of Excellence for All-sky Astrophysics (CAASTRO), through project number CE110001020.
This work was supported by the Flagship Allocation Scheme of the NCI National Facility at the ANU.
This work was also granted access to the HPC resources of CINES (Jade and Occigen) under the allocation 2013047012, c2014047012 and c2015047012 made by GENCI.
This research is part of the Spin(e) (ANR-13-BS05-0005, \url{http://cosmicorigin.org}) and Horizon-UK projects. 
Let us thank D.~Munro for freely distributing his {\sc \small  Yorick} programming language and opengl interface (available at \url{http://yorick.sourceforge.net/}). 
We warmly thank Claudia Lagos and Pascal Elahi for valuable discussions during the preparation of this manuscript and Elisa Chisari for extensive comments. We also thank
S.~Rouberol for running  the {\tt Horizon} cluster on which the simulation was  post-processed.
\end{acknowledgements}
\vspace{-0.5cm}

\bibliographystyle{aa}
\bibliography{author}

\appendix

%

\section{Alignments of major axis \& filament.}
\label{section:major}

To analyse the tendency of haloes to align their axis with the cosmic web in Fig.~\ref{fig:fil}, the focus so far was put on the orthogonality of the minor axis, so as to allow for a straightforward comparison with alignment trends in the galactic plane as well as along the kinematic  (spin) axis of haloes and galaxies. But a more direct measurement can be done checking directly the alignment of the major axis with the nearby filament. To check for consistency between both signals, this measurement is presented hereafter.
 
\begin{figure*}
\center \includegraphics[width=1.6\columnwidth]{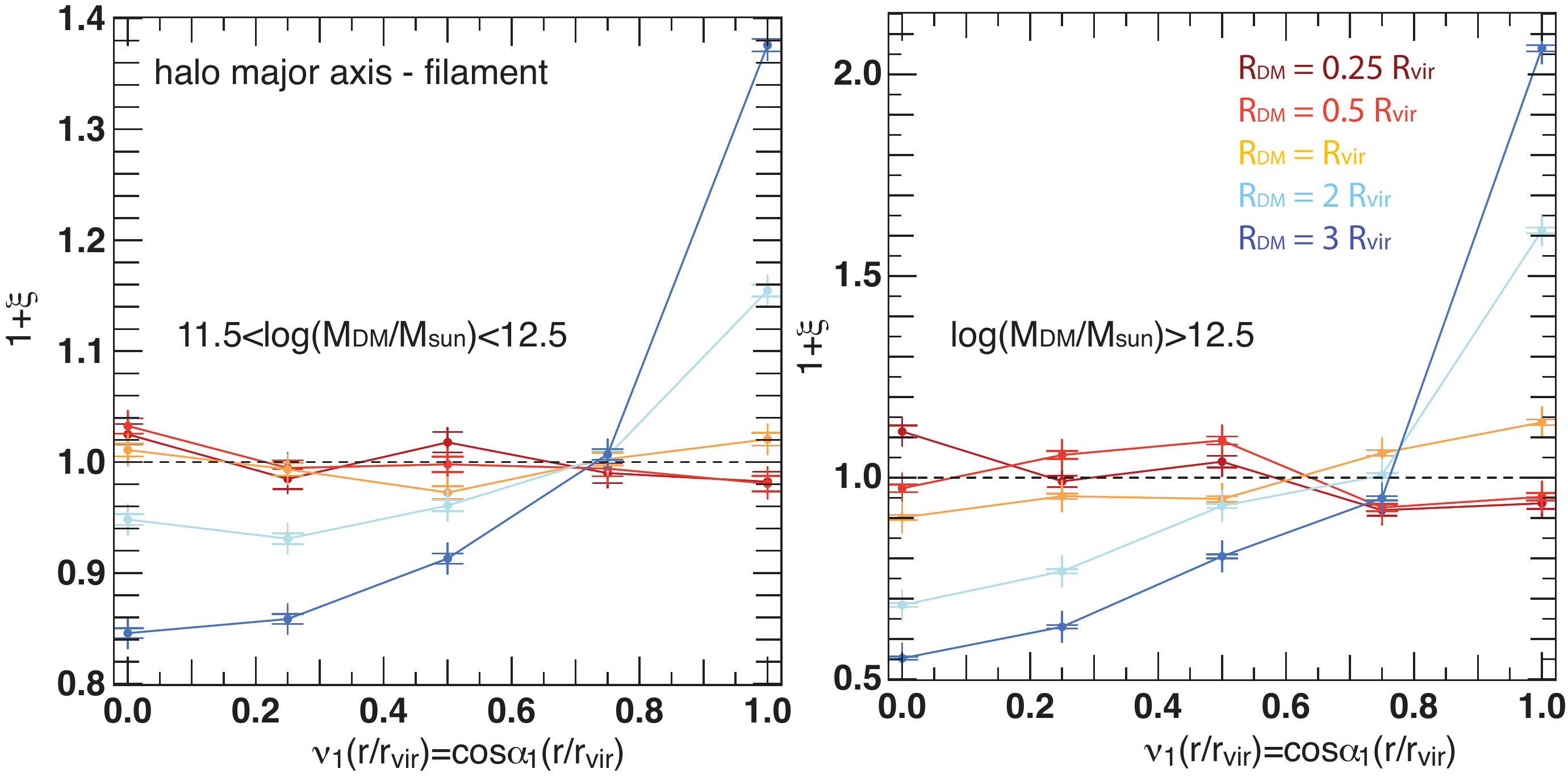}
 \caption{PDF of $\nu_{\rm 1}(r/r_{\rm vir})$, the cosine of $\alpha_{\rm 1}(r/r_{\rm vir})$, the angle between the major axis of the DM halo (computed for all DM material within radius $r=R_{\rm DM}$ from its centre of mass) and the nearest filament direction. It is computed in 5  spheres of increasing maximal radius $R_{\rm DM}$ (from red to blue curves) to describe the progressive evolution of the halo shape from the inner core ($0.25\; R_{\rm vir}$) to the outskirts ($3\; R_{\rm vir}$), and for two halo mass bins: $10^{11.5}\, M_{\odot}<M_{\rm0} <10^{12.5}\, M_{\odot}$ (left panel) and $M_{\rm 0}>10^{12.5}\,M_{\odot}$ (right panel). Alignment of haloes shape with the cosmic web is enhanced for more massive haloes.}
\label{fig:maj}
\end{figure*}

Fig.~\ref{fig:maj} displays the probability density function (PDF) of $\nu_{1}(r/r_{\rm vir})$, the cosine of $\alpha_{1}(r/r_{\rm vir})$, the angle between the major axis of the DM halo (computed for all DM material within radius $r=R_{\rm DM}$ from its centre of mass) and the nearest filament direction. Just as for the minor axis, it is computed in 5 spheres of increasing maximal radius $R_{\rm DM}$ (from red to blue curves) to describe the progressive evolution of the halo shape from the inner core ($0.25\; R_{\rm vir}$) to the outskirts ($3\; R_{\rm vir}$), and for two halo mass bins: $10^{11.5}\, M_{\odot}<M_{\rm 0}<10^{\rm12.5}\, M_{\odot}$ (left panel) and $M_{\rm 0}>10^{12.5}\, M_{\odot}$ (right panel).  

Both mass bins display a similar trend: at $2$ and $3\; R_{\rm vir}$, haloes'~major axis  are strongly  aligned with their filament, with an excess probability of $\xi=0.16$ and $\xi=0.38$ respectively for $\cos(\alpha_{\rm 1})=0$ compared to the uniform distribution (dashed line) in the low mass bin. More massive haloes ($M_{\rm 0}>10^{12.5}\, M_{\odot}$) display a similar trend with an even stronger tendency to align with the halo major axis, with $\xi=0.6$ and $\xi=1.05$ respectively for $cos(\alpha_{\rm 1})=0$ at the $2$ and $3\; R_{\rm vir}$  scales respectively. This large scale orientation of the halo major axis is expected, as the anisotropic collapse model predicts that haloes will be elongated towards the direction of their nearby filament, which corresponds to the slowest collapse axis. This results in haloes having their major axis aligned with their nearest filament. 

Note that this trend decreases sharply within the Virial radius. Within $0.25\, R_{\rm vir}$, the trend, although very faint, is reversed: haloes show a slight tendency to orient their major axis orthogonally to their nearest filament. This may be the hint of the impact of the central galaxy on the innermost parts of the halo.

\section{Radial evolution of  spin with the cosmic web.}
\label{section:haloradialspin}

Let us explore the radial evolution of the trend of for haloes to align their spin with their nearby cosmic filament.
While this mass-dependent trend has already been shown in numerous studies, let us focus here in understanding how this trend evolves within different concentric shells of the halo.

\begin{figure*}
\center \includegraphics[width=1.4\columnwidth]{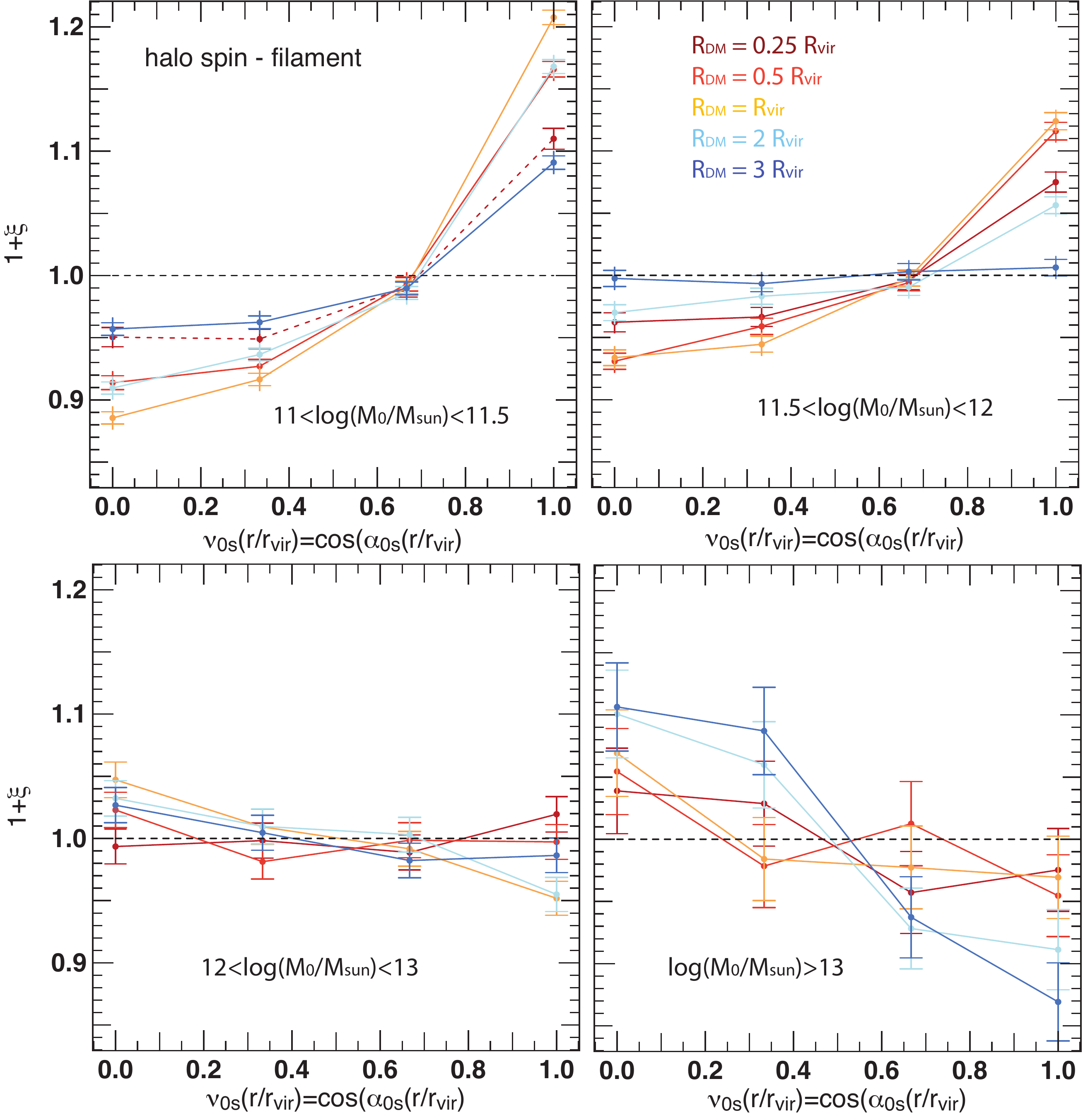}
 \caption{PDF of $\nu_{\rm 0s}$, the cosine of $\alpha_{\rm 0s}(r/r_{\rm vir})$, angle between the spin of the halo DM material contained within radius $r/r_{\rm vir}$ and its nearest filament, for five radial bins from $r/r_{\rm vir}< R_{\rm DM}/r_{\rm vir} = 0.25$ (dark red) to  $r/r_{\rm vir}<3$ (navy blue), and for four halo mass bins:
$10^{11}\, M_{\odot}<M_{\rm 0}<10^{11.5}\, M_{\odot}$ {\it upper left panel}, $10^{11.5}\, M_{\odot}<M_{\rm 0}<10^{12}\, M_{\odot}$ {\it upper right panel}, $10^{12}\, M_{\odot}<M_{\rm 0}<10^{13}\, M_{\odot}$ {\it lower left panel} and $M_{\rm 0}>10^{13}\, M_{\odot}$ {\it lower right panel}. Dashed lines are used when the particle resolution is low. Outskirts of haloes flip at lower mass than their inner core.}
\label{fig:spinhf}
\end{figure*}

Fig.~\ref{fig:spinhf} displays the PDF of $\nu_{\rm 0s}$, the cosine of $\alpha_{\rm 0s}(r/r_{\rm vir})$ the angle between the spin of the halo DM material contained within radius $r/r_{\rm vir}$ and its nearest filament, for five different radial bins from $r/r_{\rm vir}< R_{\rm DM}/r_{\rm vir} = 0.25$ (in dark red) to  $r/r_{\rm vir}<3$ (in navy blue). Additionally, the PDFs are represented for haloes in four different mass bins:
$10^{11}\, M_{\odot}<M_{\rm 0}<10^{11.5}\, M_{\odot}$ on the upper left panel, $10^{11.5}\, M_{\odot}<M_{\rm 0}<10^{12}\, M_{\odot}$ on the upper right panel, $10^{12}\, M_{\odot}<M_{\rm 0}<10^{13}\, M_{\odot}$ on the lower left panel and $M_{\rm 0}>10^{13}\, M_{\odot}$ on the lower right panel. Note that dashed lines are used for bins where the particle resolution is low, therefore making the estimation of the spin less reliable.
%
%
First, one can notice that the well-known ``spin swings'' from low-mass to high-mass haloes is recovered: while haloes with $M_{\rm 0}<10^{12}\, M_{\odot}$ show a tendency to align their spin with their nearest filament ($\xi(\nu_{\rm 0s}=1)>0$) regardless of the radius in which it's calculated), more massive haloes more likely display the opposite trend, their spin being orthogonal to their nearest filament ($\xi(\nu_{\rm 0s}=0)>0$). The transition mass if confidently bracketed between $M_{\rm 0}=10^{12}\, M_{\odot}$ and $M_{\rm 0}=10^{13}\, M_{\odot}$, coherently with previous studies \citep[see][for details]{Codis12, Codis15}.

Focusing now on how the spin of inner and outer parts of the halo evolve, let us observe that:
\begin{itemize}
\item In the lowest mass bins ($M_{\rm 0}<10^{12}\, M_{\odot}$), the inner parts (orange lines) of the halo show stronger alignment with the nearby filament than the outer parts (blue lines). 
\item Notice that the alignment  for the total spin -- including material in the outermost part of the halo ($r/r_{\rm vir}>2$),  is lost for masses as low as $10^{11.5}\, M_{\odot}<M_{\rm 0}<10^{12}\, M_{\odot}$ , while in the inner halo ($r/r_{\rm vir}<0.25$) it is detectable up to $M_{\rm 0}=10^{13}\, M_{\odot}$.
\item In the highest mass bins, the spin becomes orthogonal to its filament on all scales, but the orientation is strongest when the outermost parts of the halo are included in the calculation of the spin. The spin orthogonality signal remains poor in the inner core ($r/r_{\rm vir}<0.25$) of haloes, even for $M_{\rm 0}>10^{13}\, M_{\odot}$ ($\Delta\xi=-0.06$, compared to $\Delta\xi=-0.24$ for $r/r_{\rm vir}<3$).
\end{itemize}

This evolution must  follow from the inside-out build up of haloes in the cosmic web, from their birth in vorticity-rich regions in the vicinity of the filament where they grow a parallel spin due to coherent accretion of vorticity rich material, to their drift along the spin of the filament after outgrowing their coherent vorticity quadrant. In this second phase, they accrete material and other haloes along the filament, inducing  a partial transfer of the orbital momentum of the pair into intrinsic angular momentum of the remnant. This  re-orientants   the remnant's spin orthogonal to the filament and explains the net inertial twist of the halo and its evolution with halo mass.


\section{Comparison of central alignment trends between luminous and dark satellites.}
\label{section:subhaloes}

In this section, we compare the evolution of alignment trends around central galaxies for luminous satellites on the one hand and for all DM sub-haloes (hence, including dark satellites) on the other hand. We remind the reader that a luminous satellite is defined here as a galaxy orbiting a halo with $M_{\rm*}>1.7\times10^{8}\,M_{\odot}$ (excluding central galaxies), hence resolved with at least 50 star particles. Sub-haloes are identified directly from the dark matter particles as sub-structures identified by the halo finder.

\begin{figure}
\center \includegraphics[width=0.85\columnwidth]{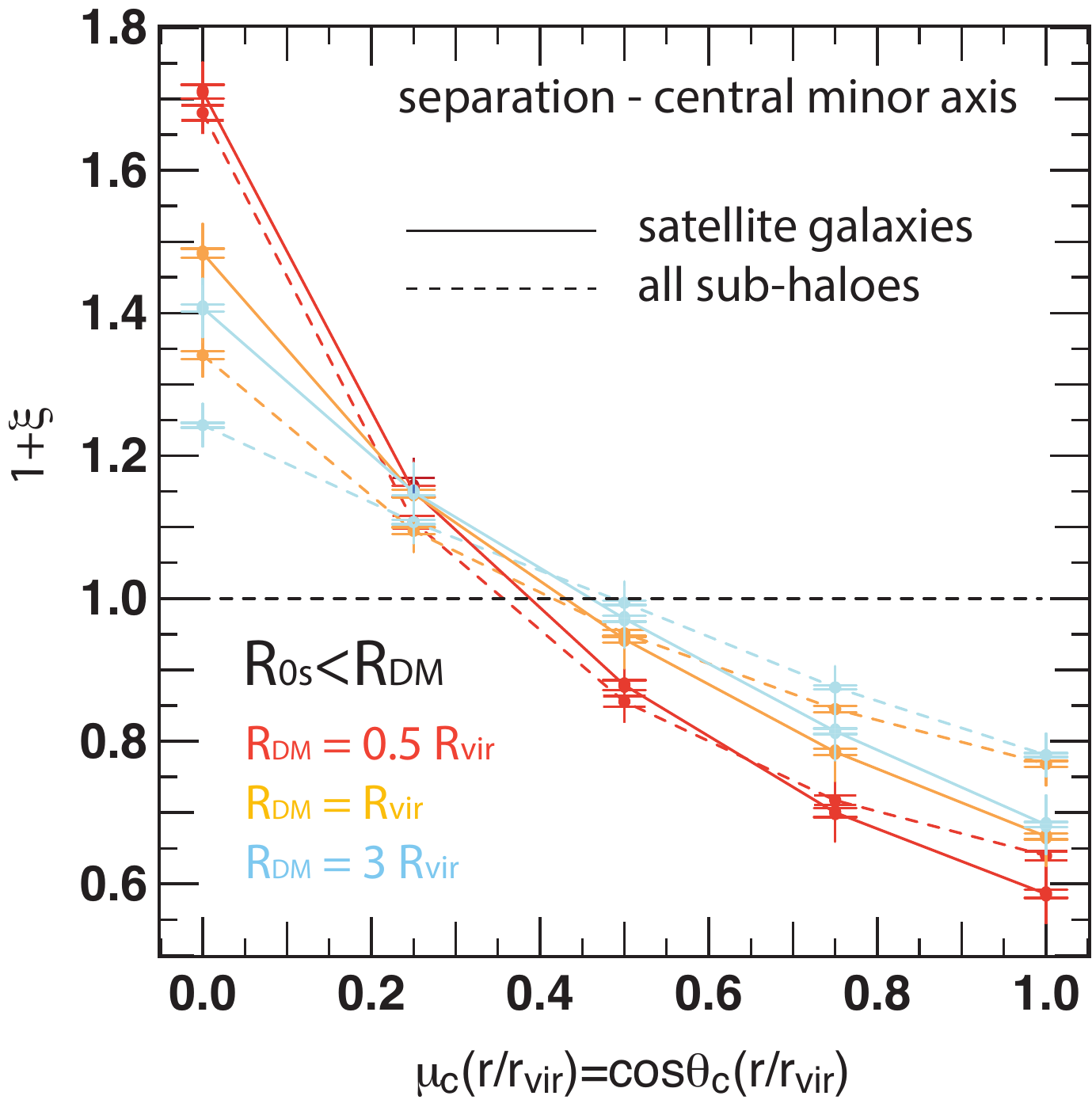}
 \caption{PDF of $\mu_{\rm c} $, the cosine of $\theta_{\rm c} (r/r_{\rm vir})$, angle between the minor axis of central galaxy and the satellite separation vector for luminous satellites alone (solid line) and for all dark matter sub-haloes (dashed lines), within  three radial bins from $r/r_{\rm vir}< R_{\rm DM}/r_{\rm vir} = 0.5$ (dark red) to  $r/r_{\rm vir}<3$ (navy blue). The signal is stronger for luminous satellites alone on all scales. }
\label{fig:sub}
\end{figure}

Fig.~\ref{fig:sub} displays the PDF of $\mu_{\rm c}$, the cosine of $\theta_{\rm c}(r/r_{\rm vir})$ the angle between the minor axis of the central galaxy and the satellite separation vector for luminous satellites alone (solid line) and for all dark matter subhaloes (dashed lines), within three different radial bins from $r/r_{\rm vir}< R_{\rm DM}/r_{\rm vir} = 0.5$ (in red) to  $r/r_{\rm vir}<3$ (in blue). 

The results highlight the increased degree of alignment of luminous satellites compared to their dark counterparts. On all scales probed, the signal obtained for the full population of sub-haloes is lower ($30\%$ lower within $r_{\rm vir}$) than the one obtained for luminous satellites alone.
Note that as you probe deeper parts of the halo, the amount of well-defined dark sub-haloes with no galaxy sharply decreases, hence the fraction of dark satellites in the full sample becomes negligible on scales lower than $0.5 \, R_{\rm vir}$. This explains why the discrepancy between both signals is reduced in the inner core of the halo.


\section{Evolution of alignment  with  mass ratio}
\label{section:satmass1}

\begin{figure*}
\center \includegraphics[width=1.9\columnwidth]{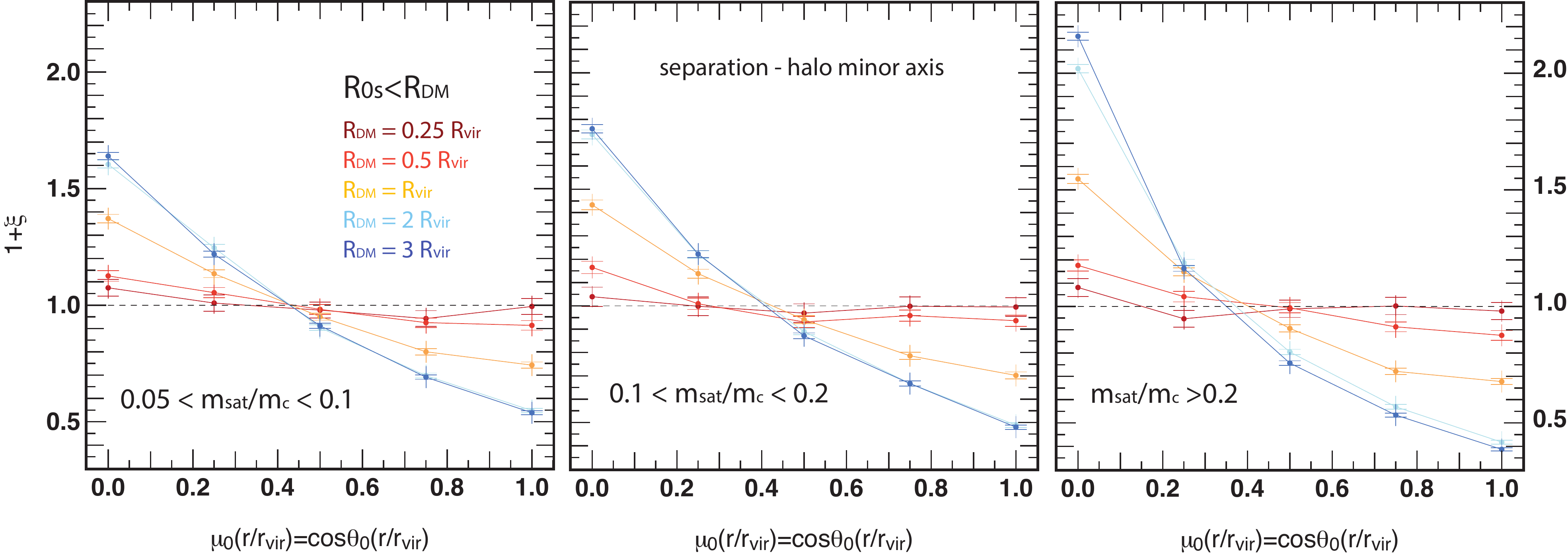}
 \caption{PDF of $\mu_{\rm 0}$, the cosine of $\theta_{\rm 0}(r/r_{\rm vir})$, angle between the minor axis of the halo DM material contained within radius $r/r_{\rm vir}$ and the satellite separation vector for five radial bins from $r/r_{\rm vir}< R_{\rm DM}/r_{\rm vir} = 0.25$ (dark red) to  $r/r_{\rm vir}<3$ (navy blue), and for three bins of satellite to central mass ratios: $0.05<M_{\rm sat}/M_{\rm c}<0.1$ ({\it left panel}), $0.1<M_{\rm sat}/M_{\rm c}<0.2$ ({\it middle panel}) and $M_{\rm sat}/M_{\rm c}>0.2$ ({\it right panel}). The signal strengthens in the outskirts of the halo for high satellite-to-central mass ratios.}
\label{fig:mrh}
\end{figure*}

Let us now focus on the evolution of the satellite alignment trends with the satellite-to-central mass ratio $M_{\rm sat}/M_{\rm c}$.
Fig.~\ref{fig:mrh} displays the PDF of $\mu_{\rm 0}$, the cosine of $\theta_{\rm 0}(r/r_{\rm vir})$ the angle between the minor axis of the halo DM material contained within radius $r/r_{\rm vir}$ and the satellite separation vector for five different radial bins from $r/r_{\rm vir}< R_{\rm DM}/r_{\rm vir} = 0.25$ (in dark red) to  $r/r_{\rm vir}<3$ (in navy blue). The results are shown for three different bins of satellite to central mass ratios: $0.05<M_{\rm sat}/M_{\rm c}<0.1$ on the left panel, $0.1<M_{\rm sat}/M_{\rm c}<0.2$ on the middle panel and $M_{\rm sat}/M_{\rm c}>0.2$ on the right panel. 

While the evolution of the alignment trend is similar to what was described in previous sections in all bins (satellites tend to lie orthogonally to their host minor axis and this effect strengthens in the outskirts of the halo), observe that the strength of the signal varies also with the satellite-to-central mass ratio. While  little to no variation is found in the inner core of the halo (red and orange solid lines), the most massive satellites ($M_{\rm sat}/M_{\rm c}>0.2$) show a stronger tendency to align with their host minor axis in the outskirts of the halo ($r/r_{\rm vir}< 3 $) with a signal up to $60\%$ stronger than what is observed for their less massive counterparts, with $M_{\rm sat}/M_{\rm c}<0.1$. Even on the Virial scale ($r/r_{\rm vir}< 1$ the signal is still $35\%$ stronger for most massive satellites. This  is consistent with the fact that more massive haloes are more likely to lie near the spine of the cosmic web (hence trace the filament) than less massive ones which display a much more isotropic distribution and are more frequently found away from the spine of the filament.

\begin{figure*}
\center \includegraphics[width=1.9\columnwidth]{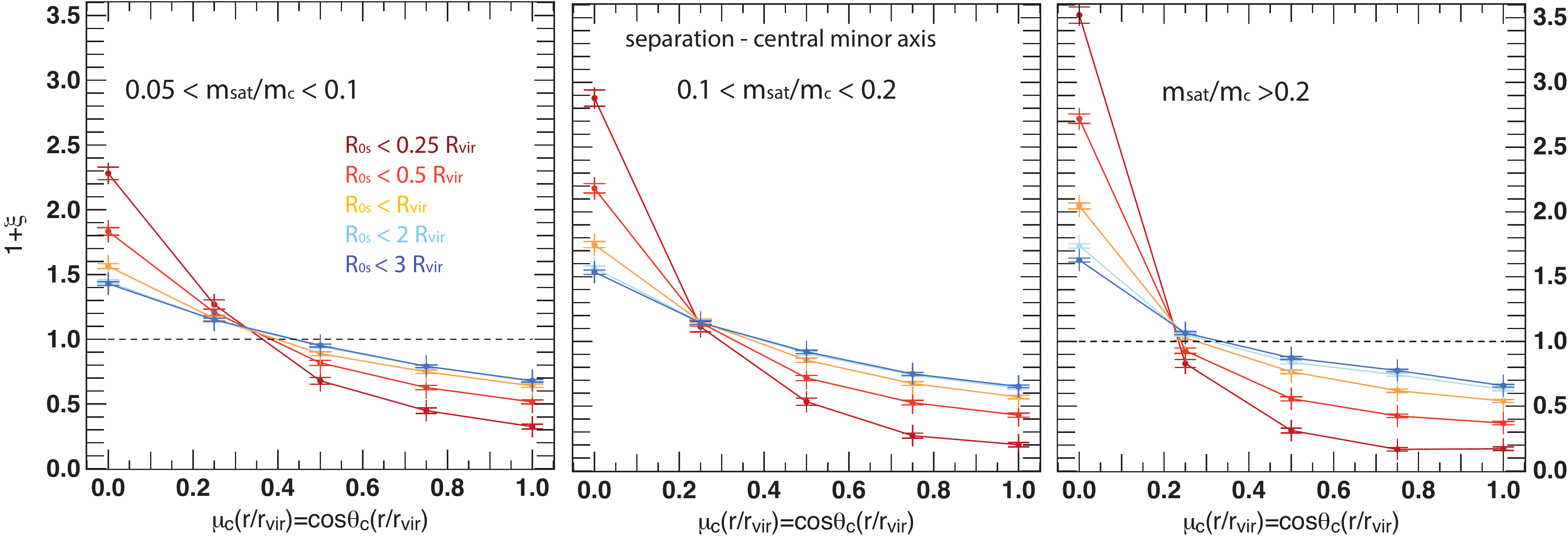}
 \caption{PDF of $\mu_{\rm c} $, the cosine of $\theta_{\rm c} (r/r_{\rm vir})$, angle between the minor axis of central galaxy and the satellite separation vector for five radial bins from $r/r_{\rm vir}< R_{\rm DM}/r_{\rm vir} = 0.25$ (dark red) to  $r/r_{\rm vir}<3$ (navy blue) and for three bins of satellite-to-central mass ratios:$0.05<M_{\rm sat}/M_{\rm c}<0.1$ ({\it left panel}), $0.1<M_{\rm sat}/M_{\rm c}<0.2$ ({\it middle panel}) and $M_{\rm sat}/M_{\rm c}>0.2$ ({\it right panel}). The signal strengthens in the inner core of the halo for high satellite-to-central mass ratios. }
\label{fig:mrc}
\end{figure*}

Let us now focus on alignments of satellites in the central galactic plane. Fig.~\ref{fig:mrc} displays the PDF of $\mu_{\rm c} $, the cosine of $\theta_{\rm c} (r/r_{\rm vir})$ the angle between the minor axis of central galaxy and the satellite separation vector for satellites in five different radial bins from $r/r_{\rm vir}< R_{\rm DM}/r_{\rm vir} = 0.25$ (in dark red) to  $r/r_{\rm vir}<3$ (in navy blue). The results are displayed  for the same three bins of satellite to central mass ratios: $0.05<M_{\rm sat}/M_{\rm c}<0.1$ on the left panel, $0.1<M_{\rm sat}/M_{\rm c}<0.2$ on the middle panel and $M_{\rm sat}/M_{\rm c}>0.2$ on the right panel.

Once again, the trend for satellites to bend in the galactic plane of the central (i.e. orthogonally to its minor axis) is observed in all three bins, but unlike the signal around haloes it strengthens in the inner parts of the halo, where it reaches amplitudes 1.5 to twice stronger than the strongest alignment signal observed for the shape of the halo. This is consistent with the idea that torques from central discs efficiently bend satellites in the galactic plane in the centre of haloes.

Focusing now on the impact of the satellite-to-central mass ratio,  more massive satellites tend to align more strongly in the galactic plane than their less massive counterparts in the inner parts of the halo. Below the Virial scale, the signal is indeed close to being $65\%$ stronger for satellites with $M_{\rm sat}/M_{\rm c}>0.2$ than the one for satellites with $0.05<M_{\rm sat}/M_{\rm c}<0.1$ (dark red lines on the left and right panels), while on the outskirt the boost in the signal is limited to $20\%$ for most massive satellites. This evolution is expected, since for more massive satellites, both the satellite and the central exert significant torques on one another, resulting in a stronger alignment  of the satellite  in the central galactic plane.

\section{Effect of central alignments on halo axis ratio measurements}
\label{section:ellipticity}

\subsection{Radially varying distributions of satellites.}
\label{subsection:einasto-nfw}

In order to explicitly test for the effect of central torques on the estimation of halo ellipticity, one needs to relate the quantity $\Delta \xi_{\rm sat}$, computed with respect to minor axis of halo, to the estimated ellipticity (or minor-to-major axis ratio in this case) of the corresponding distribution of satellites. To do so, 
the following experiment is implemented: satellites are distributed following a NFW and Einasto density distribution within ellipsoids of increasing  $c_{\rm 0}/a_{\rm 0}$ (100 bins between 0.2 and 1) and reproducing both the number of satellites in Horizon-AGN measured from the $M_{\rm 0}> 10^{13.5}\, M_{\odot}$ mass range of Horizon-AGN (80 to 400 satellites per halo) and the intermediate-to-major axis ratio following the distribution found in Horizon-AGN for the high mass range. 
More specifically, the following assumption is made for the radial evolution of the halo's shape:
the minor-to-major axis ratio of each synthetic ellipsoid is assumed to vary linearly between $0.5\; R_{\rm vir}$ and $1\;R_{\rm vir}$, with the constraint that:
\begin{eqnarray}
 c_{\rm 0}/a_{\rm 0}(0.5)=c_{\rm 0}/a_{\rm 0}(1)+0.1+\epsilon,
 \end{eqnarray}
 $\epsilon$ being randomly drawn from a gaussian distribution centred on 0.1 with standard deviation $\sigma=0.05$ for each ellipsoid sampled, following the average evolution in this mass range in Horizon-AGN

To distribute satellites in these ellipsoids, they are assumed to follow a given radial density profile, and their circularised radius is randomly drawn from the corresponding distribution. This in turn determines the local shape of the ellipsoid on which their position is randomly drawn (uniformly). Results are presented for two different radial density profiles, resp. 
an Einasto profile and a NFW profile:
\begin{equation}
 \rho(r) \propto \exp\left[-\frac{2}{\rm \alpha}\left(\frac{r}{c R_{\rm vir}}\right)^{\rm \alpha}\right],\,\,
  \rho(r) \propto \frac{4}{\left(\displaystyle\frac{r}{c R_{\rm vir}}\right)\left(\displaystyle 1+\frac{r}{c R_{\rm vir}}\right)^{2}}.
  \end{equation}
  
\begin{figure*}
\center \includegraphics[width=1.65\columnwidth]{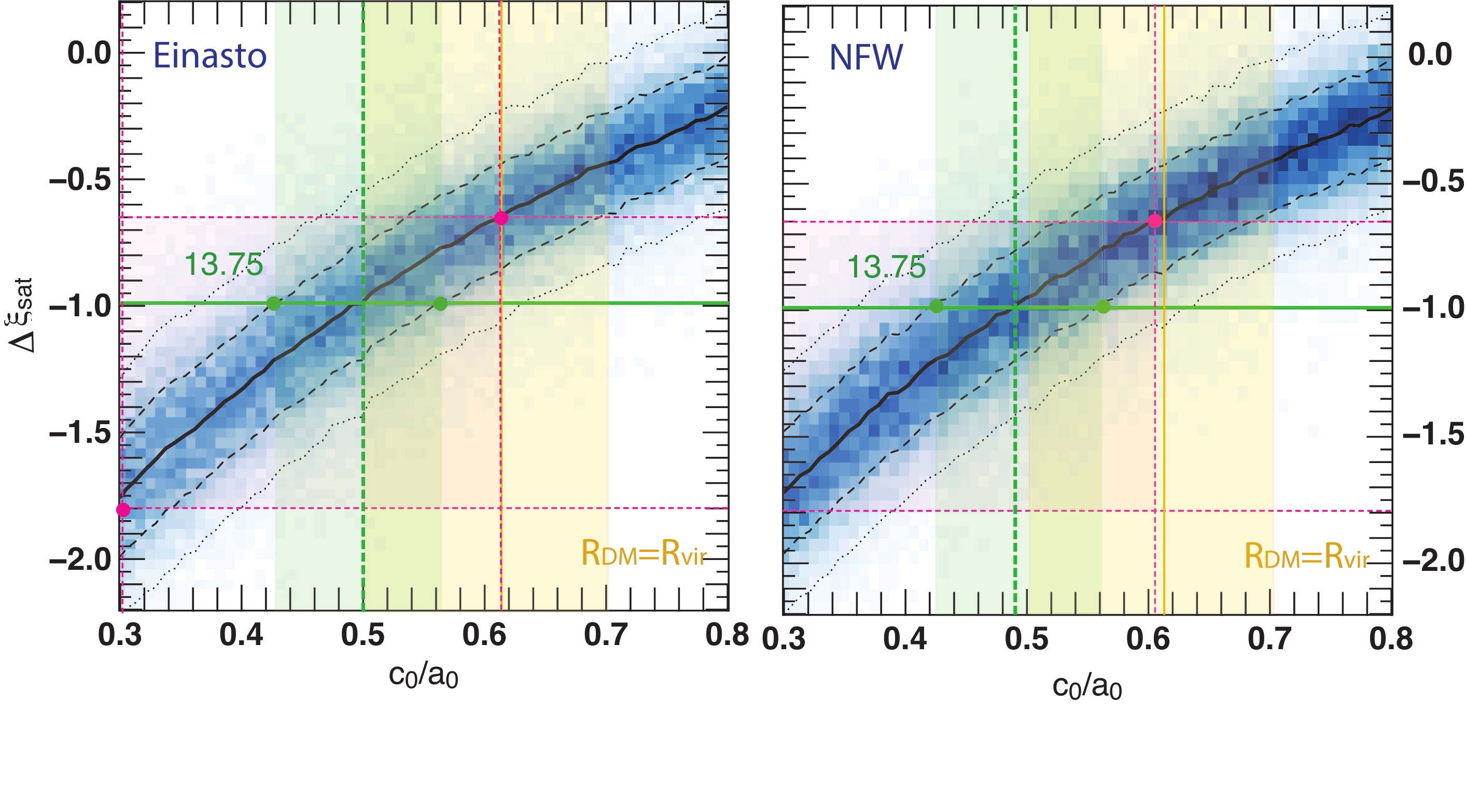}
 \caption{Average evolution of $\Delta \xi_{\rm sat}$ in the Virial shell with $c_{\rm 0}/a_{\rm 0}$ (solid line), and $1-\sigma$ and $2-\sigma$ contours (dashed and dotted lines respectively) drawn from 1000 realisations of satellite distributions per axis ratio bin, for 100 such bins, sampling 80 to 400 satellites per halo. $\Delta \xi_{\rm sat}\, - \, c_{\rm 0}/a_{\rm 0}$ matching for $M_{\rm 0}> 10^{13.5}\, M_{\odot}$ is overlaid  in green with $1-\sigma$ region shaded in green. The average value of $c_{\rm 0}/a_{\rm 0}$ for this mass bin within $R_{\rm vir}$ is overlaid in orange with the $1-\sigma$ region shaded in light orange. The 16th to 84th percentile region for $\Delta \xi_{\rm sat}$ found for $M_{\rm 0}> 10^{13.5}\, M_{\odot}$ is shaded in pink, with extremal $c_{\rm 0}/a_{\rm 0}$ matching as dashed pink lines. In the highest mass range, axis ratios derived from the distribution of satellites are reasonably matched to real ones but the central still accounts for a systematic underestimation. {\it Left panel:}  $\Delta \xi_{\rm sat}$ is computed assuming a radially decreasing axis ratio and an underlying Einasto profile for the radial satellite distribution. {\it Right panel}: Same with a NFW profile.
}
\label{fig:nfw1}
\end{figure*}

This choice is motivated by the assumption that satellites trace the radial density of their host halo, and supported by observations in the SDSS \citep{Tal12,Wang14}. The results are presented for typical values $\alpha=0.25$ and $c=6$. Note that several observations find lower concentration parameters for satellite distributions than what is expected for their dark matter haloes, on average $c=2-4$ for groups and clusters \citep{Carlberg97,Muzzin07,Budzynski12}. Nonetheless, it was tested that these parameters can vary significantly within physically motivated ranges described in \cite{Ludlow16} ($0.15-0.35$ and $2-10$ respectively below $z=1$) without any significant effect on the results presented hereafter. This is mostly due to the fact that   only satellites in the Virial shell are considered here, while it is mainly in the inner core of the halo that is affected by these parameters. Note that  the model dependence on these assumptions was tested by doing the same analysis considering uniformly distributed satellites in constant shape ellipsoids in Appendix~\ref{subsection:uniform}. This does not affect the qualitative results of this analysis.
We provide a sample of 100\,000 mock-generated haloes (1000 per bin of $c_{\rm 0}/a_{\rm 0}$). For each of these,  the corresponding $\Delta \xi_{\rm sat}$ is shown on Fig.~\ref{fig:nfw1}. Results for the Einasto profile are presented on the left panel, those for the NFW profile on the right panel. In this case, $\Delta \xi_{\rm sat}$ is measured within the Virial shell to limit as much as possible the impact of the central. It shows the corresponding distribution of these 100\,000 data points in light blue, as well as the the average of $\Delta \xi_{\rm sat}$ as a function $c_{\rm 0}/a_{\rm 0}$ as a black solid line. The matching $\Delta \xi_{\rm sat}\, - \, c_{\rm 0}/a_{\rm 0}$  is overlaid for the highest mass bin $M_{\rm 0}> 10^{13.5}\, M_{\odot}$ in green (with the $1-\sigma$ region shaded in green). For comparison, the average value of $c_{\rm 0}/a_{\rm 0}$ derived from the inertia tensor of all DM material within $R_{\rm vir}$ for that mass range is also overlaid in orange. The $1-\sigma$ region is shaded in light orange.

The focus is now  on the highest mass bin $M_{\rm 0}> 10^{13.5}\, M_{\odot}$ -- as only haloes in this bin host hundreds of satellites or more, and allow for a halo-per-halo analysis without requiring stacking. In this range, the alignment of the central with its inner halo is also increased. Using the average $\Delta \xi_{\rm sat}$ in this mass bin (thick green dashed line),  an average halo minor-to-major axis ratio, and  an average halo minor-to-major axis ratio  $\langle c_{\rm 0}/a_{\rm 0}\rangle =0.5\pm 0.06$ is found for the Einasto model and $\langle c_{\rm 0}/a_{\rm 0}\rangle =0.49 \pm 0.06$ for the NFW model. Although slightly decreased compared to the uniform model ($7$ to $9\%$ variation: $\langle c_{\rm 0}/a_{\rm 0}\rangle =0.54\pm 0.06$), it remains compatible with recent observations in the SDSS by \cite{Shin17} and in RCS-2 and SpARCS by \cite{Just16}. Notice that, although estimating the average $c_{\rm 0}/a_{\rm 0}$ from stacked $\Delta \xi_{\rm sat}$  computed in the Virial shell (thick green dashed line) leads to an underestimation compared to the average value obtained from the inertia tensor of the dark matter material (i.e. overestimation of the ellipticity), both values remain compatible within $1\sigma$, with a relative error around $20\%$ ($10\%$ in the uniform case, see Appendix~\ref{subsection:uniform}).
Hence, in massive enough haloes which contain hundreds of satellites, the influence of the central on the stacked signal for satellites {\it in the Virial shell } might be low enough to use the distribution of satellites to derive a halo's ellipticity, but with an important relative error.  However, for individual haloes, the error soars quickly. The pink shaded area highlights the 16th to 84th percentile region for $\Delta \xi_{\rm sat}$ values computed for each individual halo with $M_{\rm 0}> 10^{13.5}\, M_{\odot}$, with $c_{\rm 0}/a_{\rm 0}$ matching as dashed pink lines. The inferred distribution of $c_{\rm 0}/a_{\rm 0}$ (pink area) is significantly shifted and skewed to lower values of $c_{\rm 0}/a_{\rm 0}$ compared to the true distribution $c_{\rm 0}/a_{\rm 0}$ (orange area): the estimated ellipsoids are, thus, {\it systematically} flatter than the true ellipsoids of the DM inertia tensor, and haloes appear more triaxial through their distribution of satellites than they actually are. This now suggests that the relative error on the axis ratio measurement can reach more than $50\%$ for individual haloes.

This is expected as in the cluster mass range where surrounding cosmic filaments are highly contrasted and where central galaxies are more evolved and strongly merger-dominated, host haloes and centrals show much stronger shape alignment  at the Virial scale. This is related to the mutual strong alignment of their major axis/galactic plane with their nearest cosmic filament. In this range, gravitational torques from haloes, centrals and cosmic structures are not expected to compete but add up to one another, therefore enhancing the perceived alignment of satellites orthogonally to their host minor axis.

\subsection{Uniform distributions of satellites.}
\label{subsection:uniform}

\begin{figure*}
\center \includegraphics[width=1.7\columnwidth]{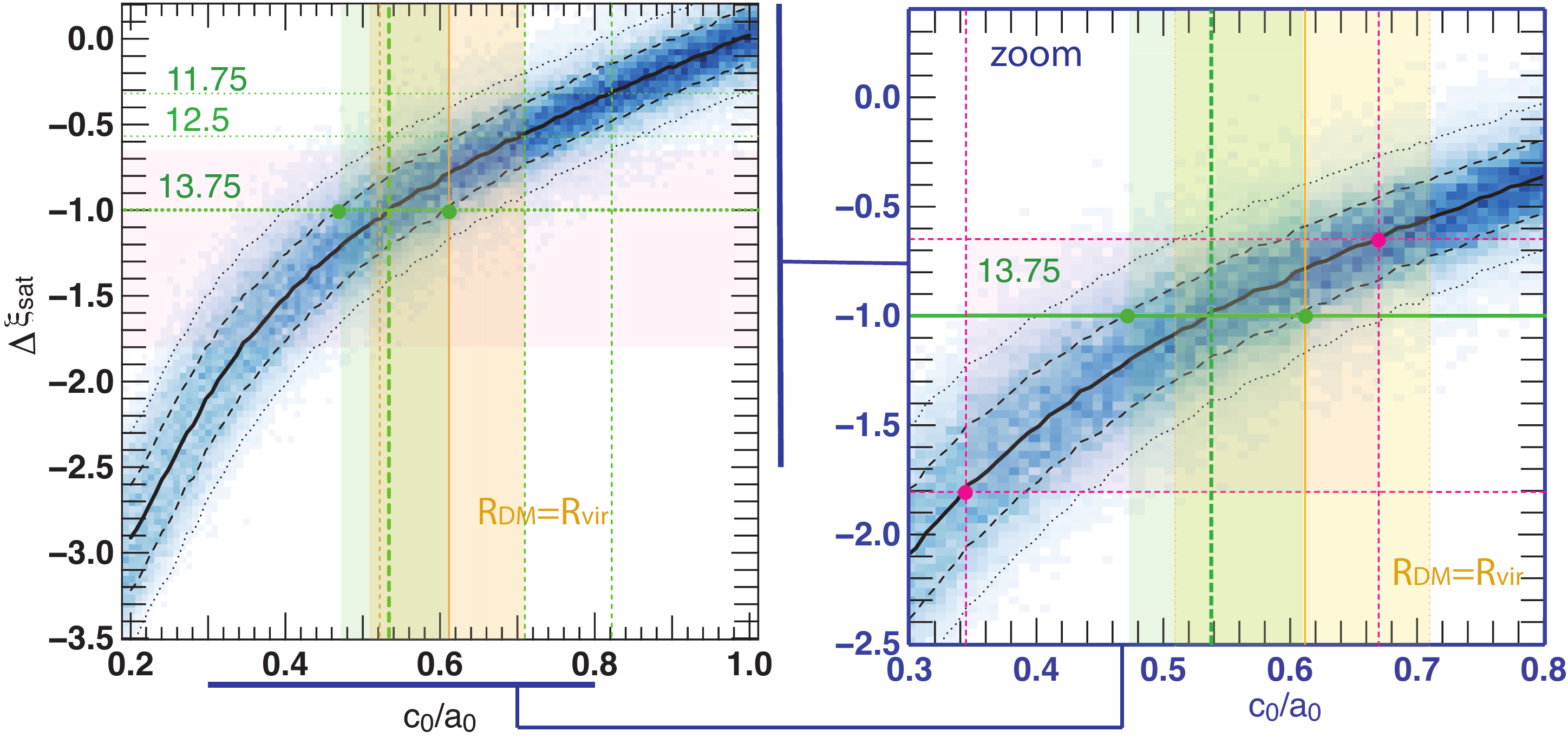}
 \caption{Average evolution of $\Delta \xi_{\rm sat}$ in the Virial shell with $c_{\rm 0}/a_{\rm 0}$ (solid line), and $1-\sigma$ and $2-\sigma$ contours (dashed and dotted lines respectively) drawn from 1000 realisations of satellite distributions per axis ratio bin, for 100 such bins, sampling 80 to 400 satellites per halo (blue shades of the distribution). {\it Left panel}: $\Delta \xi_{\rm sat}\, - \, c_{\rm 0}/a_{\rm 0}$ matching for three halo mass bins is overlaid in green (with $1-\sigma$ region shaded in green for the highest mass bin). The average value of $c_{\rm 0}/a_{\rm 0}$ within $R_{\rm vir}$ is overlaid in orange: dashed line for the full halo population and solid line for the average in the highest halo mass bin $M_{\rm 0}> 10^{13.5}\, M_{\odot}$, with the $1-\sigma$ region shaded in light orange. {\it Right panel}: zoom around the highest mass bin region. Orange/green lines and shades are defined as previously. The 16th to 84th percentile region for $\Delta \xi_{\rm sat}$ found for $M_{\rm 0}> 10^{13.5}\, M_{\odot}$ is shaded in pink, with extremal $c_{\rm 0}/a_{\rm 0}$ matching as dashed pink lines. In the highest mass range, axis ratios derived from the distribution of satellites are reasonably matched to real ones but the central still accounts for a systematic underestimation.}
\label{fig:samples}
\end{figure*}

In order to relate the quantity $\Delta \xi_{\rm sat}$ to the estimated ellipticity (or minor-to-major axis ratio here) of the corresponding underlying halo,  sets of satellites were distributed within ellipsoids of radially decreasing  $c_{\rm 0}/a_{\rm 0}$ (100 bins between 0.2 and 1) in Section~\ref{section:ellipticity}, after marginalising over both the number of satellites (80 to 400, following statistics in the high mass range, $M_{\rm 0}> 10^{13.5}\, M_{\odot}$ in Horizon-AGN) and the intermediate-to-major axis ratio following the distribution found in Horizon-AGN.  It aimed to mimic the radially varying shapes, from more isotropic cores to more elongated outskirts, found in Horizon-AGN.  To estimate how model-dependent the results derived from this assumption are, Fig.~\ref{fig:samples} reproduces the same analysis with the assumption that satellites are uniformly distributed in haloes. Note that the choice of populating the ellipsoids with uniformly distributed satellites implies that the axis ratio $c_{\rm 0}/a_{\rm 0}$ is assumed to be constant throughout the radial range explored for satellites (i.e. the Virial shell), and equal to its inertial value within $R_{\rm vir}$. Since the inertial tensor, although sensitive to the outskirts, takes better into account the whole distribution than iso-density contours, and since only satellites within the Virial shell are taken into account, this remains a reasonable approximation. 

For each $c_{\rm 0}/a_{\rm 0}$ bin, the corresponding distribution of $\Delta \xi_{\rm sat}$ is shown in blue (with the average of $\Delta \xi_{\rm sat}$ as a function $c_{\rm 0}/a_{\rm 0}$ as a black solid line) in the left panel of Fig.~\ref{fig:samples}. Recall that in this case, $\Delta \xi_{\rm sat}$ is measured within the Virial shell to limit as much as possible the impact of the central.  $\Delta \xi_{\rm sat}\, - \, c_{\rm 0}/a_{\rm 0}$ is overlaid matching for three different halo masses derived from Fig.~\ref{fig:dxshell} in green (with the $1-\sigma$ region shaded in green for the highest mass). For comparison, the average value of $c_{\rm 0}/a_{\rm 0}$ derived from the inertia tensor of all DM material within $R_{\rm vir}$ is also overlaid in orange. The dashed orange line corresponds to the full halo population and the solid line is the average in the highest halo mass bin $M_{\rm 0}> 10^{13.5}\, M_{\odot}$. In this latter case, the $1-\sigma$ region is also shaded in light orange.

It is easily observed that below $10^{13.5}\, M_{\odot}$, stacking satellites distributions around centrals does not allow us to recover the average minor-to-major axis ratio of dark matter haloes in this mass range (dashed orange line). Axis ratios derived from $\Delta \xi_{\rm sat}$ computed in the Virial shell are largely overestimated: the stacked distribution of satellites tends to be more isotropic than its average host halo. This is due to the fact that in that mass range, as dwarf galaxies are not resolved in Horizon-AGN, haloes in this mass range host few satellites, biased towards the highest mass range of satellites. These still mainly align with their central galactic plane. The orientation of this plane is also not well correlated to that of the halo within this mass bin. As a result, stacking angles for all haloes with near-random central galaxy orientation produces a synthetic distribution of satellites a lot more isotropic than the actual average inner halo. This drop in the signal around stacked haloes due to central torques is demonstrated by checking that at fixed halo mass, more massive centrals lead to a damped signal. This analysis is presented in Appendix~\ref{section:satmass}. 

Focusing on the highest mass bin and using the average $\Delta \xi_{\rm sat}$ in this mass bin (thick green dashed line),  an average halo minor-to-major axis ratio  $\langle c_{\rm 0}/a_{\rm 0}\rangle =0.54\pm 0.06$ is found for the uniform model. These values remain compatible with recent observations in the SDSS by \cite{Shin17} and inc RCS-2 and SpARCS by \cite{Just16}. Once again, estimating the average $c_{\rm 0}/a_{\rm 0}$ from stacked $\Delta \xi_{\rm sat}$  computed in the Virial shell (thick green dashed line) leads to an underestimation compared to the average value obtained from the inertia tensor of the dark matter material (i.e. overestimation of the ellipticity). Both values remain compatible within $1\sigma$ but the relative error on the mean remains around $10\%$.  As could be expected, not taking into account the increased isotropy of the inner halo slightly reduces the discrepancy between the expected degree of satellite alignment and the one actually observed in Horizon-AGN (i.e. including central torques), but results remain very comparable with those obtained using a NFW or a Einasto profile.

For individual haloes, the error soars quickly. The 16th to 84th percentile region for $\Delta \xi_{\rm sat}$ values computed for each individual halo with $M_{\rm 0}> 10^{13.5}\, M_{\odot}$ is highlighted in pink shade, with $c_{\rm 0}/a_{\rm 0}$ matching as dashed pink lines. The pink distribution is significantly shifted and skewed to lower values of $c_{\rm 0}/a_{\rm 0}$  compared to the yellow distribution:  axis ratios derived from $\Delta \xi_{\rm sat}$ seem to be {\it systematically} underestimated compared to values computed from the DM inertia tensor. haloes appear more triaxial than they actually are through their distribution of satellites. This model suggests that the relative error on the axis ratio measurement can reach  $30-40\%$ for individual haloes.

\section{Case study of the Milky Way: constraining the radial distribution of satellites}
\label{subsection:MWrad}

Let us now investigate the impact of a radially increasing halo ellipticity on alignments of satellites around the MW. Einasto and NFW profiles are notoriously poor choices to model the scarce radial density of satellites around the MW. Instead, let us set the number of satellites to 41 to match structures identified in \cite{Pawlowski15}, and using the 41 galacto-centric distances from \cite{Pawlowski15} as constrained radii. Using the relation $r_{\rm 1/2}=0.015\; R_{\rm vir}$ from \cite{Kravtsov13}, with $r_{\rm 1/2}\approx 4$ kpc the half mass radius of the Milky Way, we set $R_{\rm vir}=270$ kpc. This is also compatible with estimations in the Illustris simulation by \cite{Taylor16}.

The minor-to-major axis ratio of each synthetic ellipsoid is assumed to vary linearly between $0.1\; R_{\rm vir}$ and $1\;R_{\rm vir}$, with the constraint that:
\begin{eqnarray}
 c_{\rm 0}/a_{\rm 0}(0.1)=c_{\rm 0}/a_{\rm 0}(1)+0.1+\epsilon,
 \end{eqnarray}
 $\epsilon$ being randomly drawn from a gaussian distribution centred on 0.05 with standard deviation $\sigma=0.05$ for each ellipsoid sampled. This choice mimicks the evolution of the halo inertial axis ratio for haloes with $10^{11.8}\, M_{\odot}<M_{\rm 0}<10^{12.3}\, M_{\odot}$, with major axis aligned with the cosmic web ($\alpha_{\rm 1}<37^{\rm o}$), as the position of the MW in cosmic flows suggests it is. Indeed, haloes closer to the spine of their closest filament but not massive enough to lie in nodes of the cosmic web undergo stronger stretching and tidal torques along their filament, hence display slightly more elongated cores.
 
\begin{figure}
\center \includegraphics[width=0.9\columnwidth]{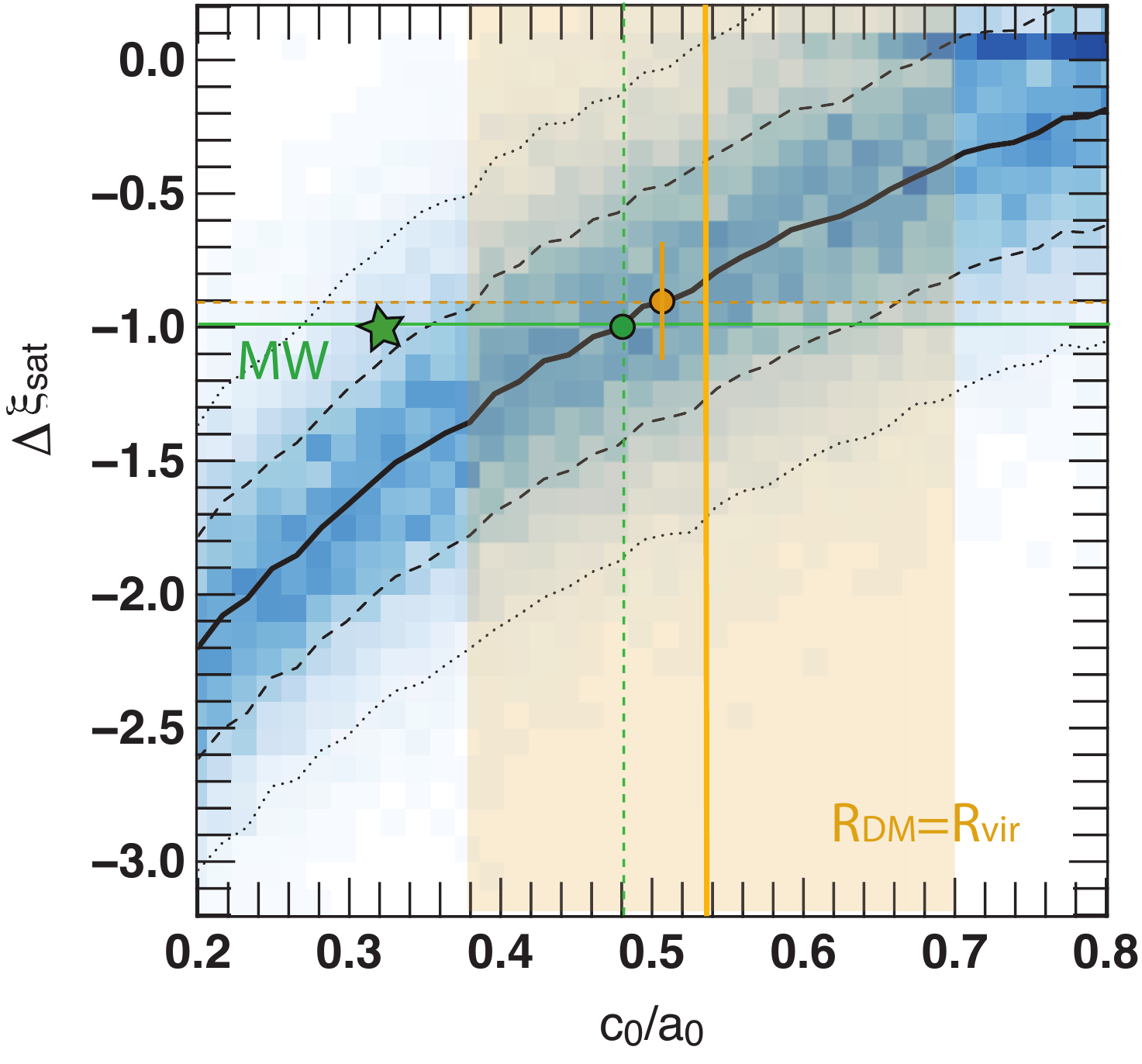}
 \caption{Same as Fig.~\ref{fig:samples} but using the 41 known radii of MW satellites in a halo model with radially increasing ellipticity. The solid horizontal green line indicates  $\Delta \xi_{\rm sat}^{\rm MW}$  for the Milky Way  equating the minor axis of the halo to that of the total system of satellites. The green star indicates the minor-to-major axis ratio of the system of satellites, the green dot the most likely halo minor-to-major axis ratio corresponding to $\Delta \xi_{\rm sat}^{\rm MW}$. The vertical orange line shows the average axis ratio for haloes with $10^{11.8}\, M_{\odot} <M_{\rm 0}<10^{12.3} \, M_{\odot}$ in Horizon-AGN, with the 16th to 84th percentile region shaded in orange. The dashed horizontal orange line (with poissonian error bars on the mean) shows the average stacked $\Delta \xi_{\rm sat}$ for the $50\%$ most elongated haloes with $10^{11.8}\, M_{\odot} <M_{\rm 0}<10^{12.3} \, M_{\odot}$ and with misalignment with their cosmic filament limited to $37^{\rm o}$
}
\label{fig:nfw2}
\end{figure}

One can see that results are not qualitatively different from what was observed for the uniform model: the degree of alignment of MW satellites is compatible with the distribution of halo ellipticities in the same mass range in Horizon-AGN. $35\%$ of haloes with $10^{11.8}\, M_{\odot}<M_{\rm 0}<10^{12.3}\, M_{\odot}$ have haloes more elongated than the one inferred for the MW halo from the distribution of its satellites. This rises up to $47\%$ for haloes in the same mass range with $\alpha_{\rm 1}<37^{\rm o}$. Note that the shape of the system of satellites in itself is less elongated than $8.5\%$ of haloes with $10^{11.8}\, M_{\odot}<M_{\rm 0}<10^{12.3}\, M_{\odot}$ in Horizon-AGN, and $13.5\%$ of these haloes with $\alpha_{\rm 1}<37^{\rm o}$. As a conclusion, even directly equating the axis ratio of the system of satellites to that of its underlying DM halo does not position the MW in a rare or extreme position among other haloes of the same mass in Horizon-AGN.


\section{Impact of halo shape and orientation.}
\label{section:haloeshape}

This section evaluates the impact of the halo shape and orientation on the tendency of satellites to align with the shape of the halo.

\begin{figure}
\center \includegraphics[width=0.9\columnwidth]{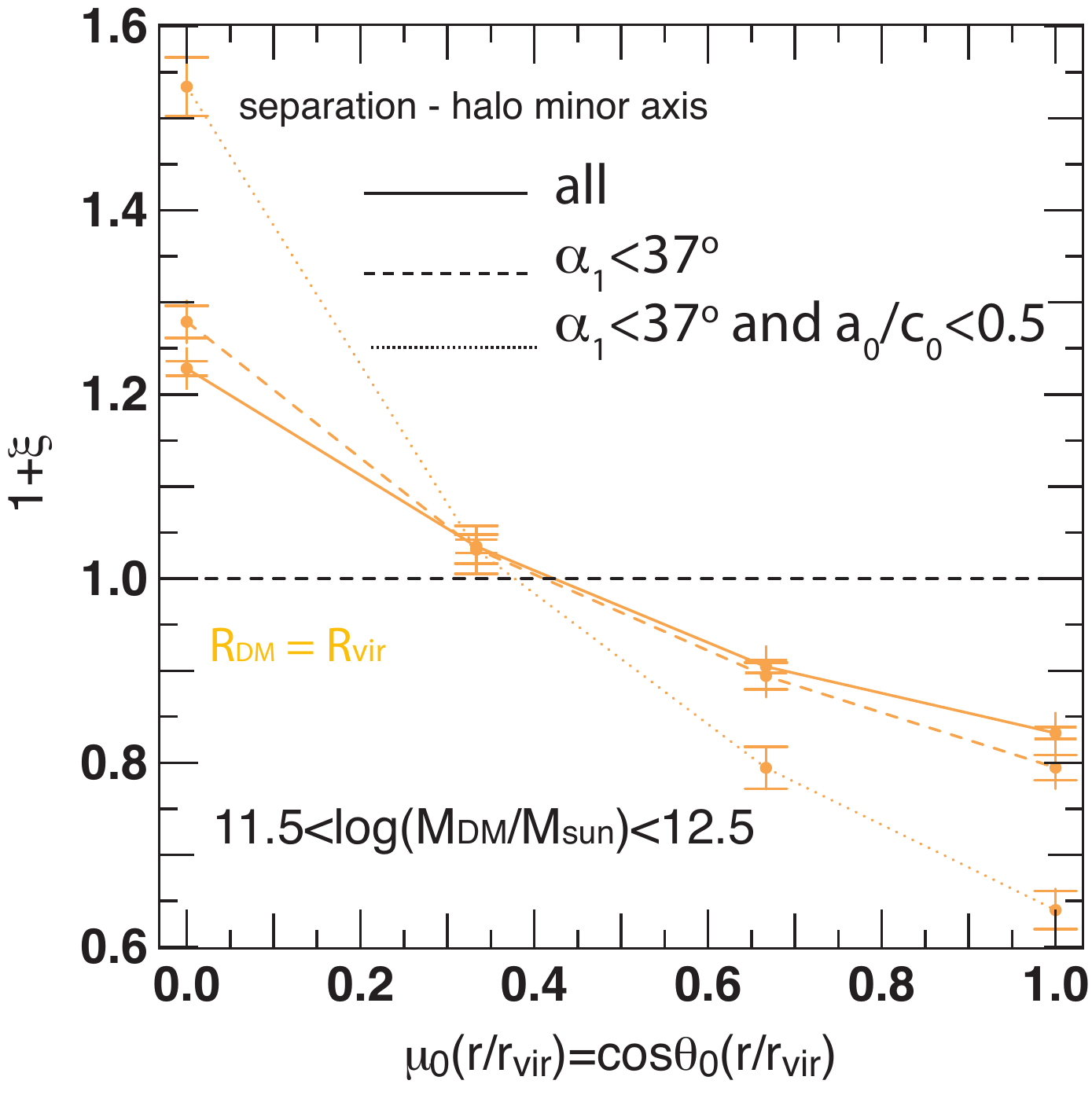}
 \caption{PDF of $\mu_{\rm0}$, the cosine of $\theta_{\rm 0}(r/r_{\rm vir})$, angle between the satellite separation vector and the minor axis of the halo DM material contained within $1\, R_{\rm vir}$ and for halo masses $10^{11.5}\, M_{\odot}<M_{\rm 0}<10^{12.5}\, M_{\odot}$ . Solid lines are used for the full sample, dashed lines for the $50\%$ most elongated haloes ($c_{\rm 0}/a_{\rm 0}<0.5$) and dotted lines for the most elongated haloes aligned with their cosmic filament ($\alpha_{1}<37^{\rm o}$).
}
\label{fig:hso}
\end{figure}

Fig.~\ref{fig:hso} displays the PDF of $\mu_{\rm 0}$, the cosine of $\theta_{\rm 0}(r/r_{\rm vir})$ the angle between the minor axis of the halo DM material contained  within $1\, R_{\rm vir}$ and for halo masses $10^{11.5}\, M_{\odot}<M_{\rm 0}<10^{12.5}\, M_{\odot}$. Solid lines are used for the full sample, dashed lines for the sample restricted to the $50\%$ most elongated haloes ($c_{\rm 0}/a_{\rm 0}<0.5$ and dotted lines for the most elongated haloes aligned with their cosmic filament ($\alpha_{1}<37^{\rm o}$ (major axis computed in the Virial radius.).
%
%
 Note that the tendency of satellites to lie in a plane orthogonal to their host's minor axis, i.e. to align with the shape of their halo, increases for more elongated haloes and even more for those better aligned with their nearest cosmic filament. In this mass range ($10^{11.5}\, M_{\odot}<M_{\rm 0}<10^{12.5}\, M_{\odot}$) compatible with measurements of the Milky Way halo, the stacked signal reaches $\Delta \xi_{\rm sat}=-0.9$ for haloes with $c_{\rm 0}/a_{\rm 0}<0.5$ and $\alpha_{1}<37^{\rm o}$.

\section{Impact of  central mass on alignments.}
\label{section:satmass}
This section evaluates the impact of the central-to-halo mass ratio on the tendency of satellites to align with the shape of the halo.
\begin{figure}
\center \includegraphics[width=0.9\columnwidth]{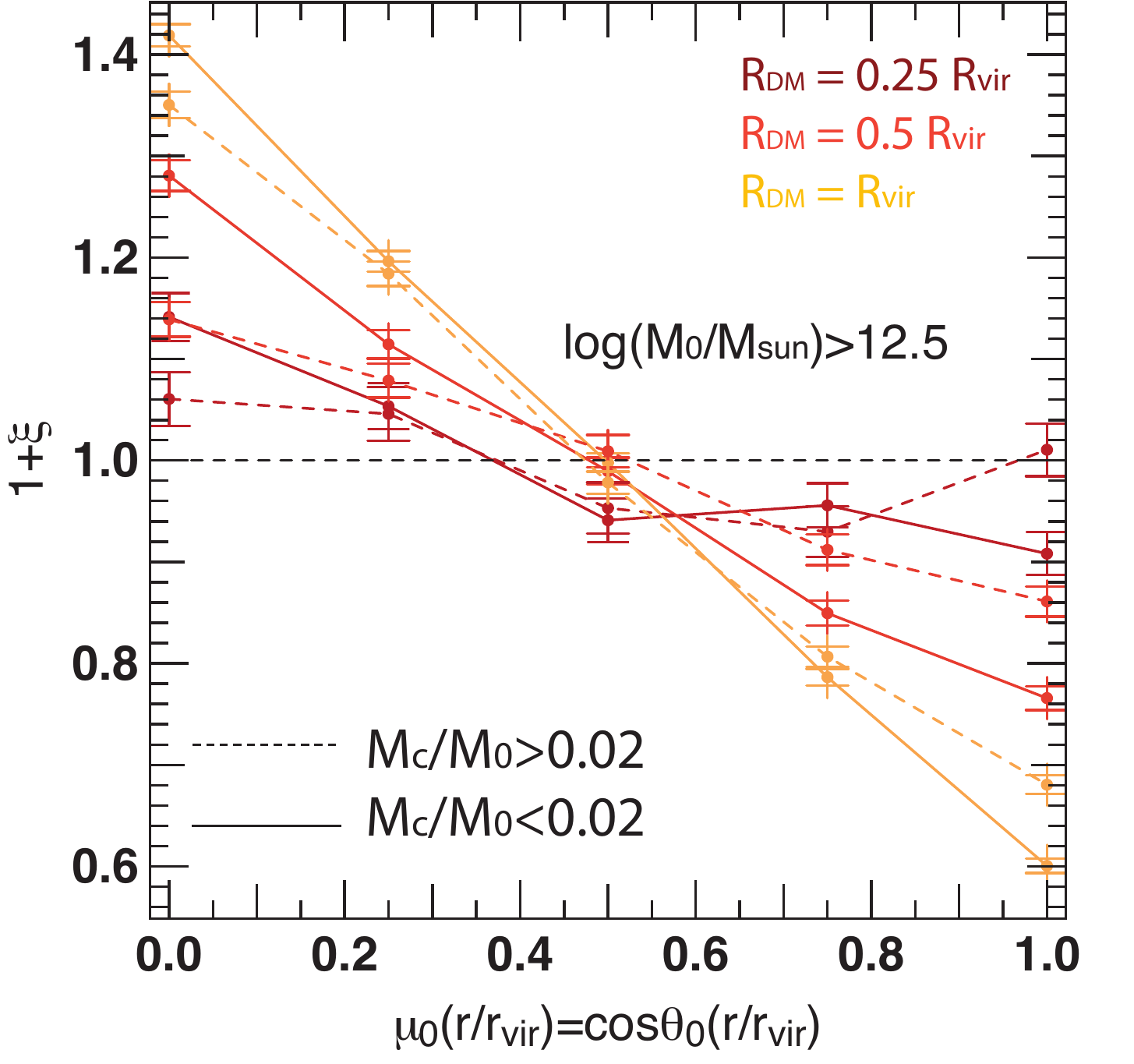}
 \caption{PDF of $\mu_{\rm0}$, the cosine of $\theta_{rm 0}(r/r_{\rm vir})$, angle between the the satellite separation vector and the minor axis of the halo DM material contained within radius $r/r_{\rm vir}$ , for three radial bins from $r/r_{\rm vir}< R_{\rm DM}/r_{\rm vir} = 0.25$ (dark red) to  $r/r_{\rm vir}<1$ (yellow), and for halo masses $M_{\rm 0}>10^{12.5}\, M_{\odot}$ . Dashed lines are used for central-to-halo mass ratios $M_{\rm c}/M_{\rm 0}>0.02$ and solid lines for $M_{\rm c}/M_{\rm 0}<0.02$. }
\label{fig:mccc}
\end{figure}
%

Fig.~\ref{fig:mccc} displays the PDF of $\mu_{\rm 0}$, the cosine of $\theta_{\rm 0}(r/r_{\rm vir})$ the angle between the minor axis of the halo DM material contained within radius $r/r_{\rm vir}$ and the satellite separation vector for the three innermost radial bins from $r/r_{\rm vir}< R_{\rm DM}/r_{\rm vir} = 0.25$ (in dark red) to  $r/r_{\rm vir}<1$ (in yellow).  The results are displayed for halo masses $M_{\rm 0}>10^{12.5}\, M_{\odot}$, and two different bins of central-to-halo mass ratios: $M_{\rm c}/M_{\rm 0}<0.02$ as solid lines and $M_{\rm c}/M_{\rm 0}>0.02$ as dashed lines. The halo mass bin are chosen to ensure a wide range of central-to-halo mass ratio is explored and to preserve good and similar statistics in all bins and on all scales.

Note that the tendency of satellites to lie orthogonally to their host's minor axis, i.e. to align with the shape of their halo, is decreased for high mass ratios. This is consistent with  the effect described in Section~\ref{section:ellipticity}: more massive centrals generate stronger torques on their satellites and bend them more efficiently in their galactic plane. Since in this mass range, centrals and their host haloes show strong misalignments on sub-Virial scales, this  damps the alignment signal around the halo by dragging the satellites away for their original distribution as mere tracers of the halo shape.

\end{document}